\documentclass[a4paper,11pt]{article}
\usepackage{jheppub} 
\usepackage{lmodern}
\usepackage{physics}
\usepackage{esint}

\usepackage{mathtools}
\newcounter{tableeqn}[table]

\newcounter{tablesubeqn}[tableeqn]

\usepackage{fancyhdr}
\usepackage[spanish,english]{babel}
\usepackage{lettrine} \usepackage[T1]{fontenc}
\usepackage{chngcntr} \counterwithout{footnote}{section}
\usepackage{enumerate}
\usepackage{listings} 
\usepackage{tikz} 
\usetikzlibrary{arrows.meta}
\usepackage{array} 
\usepackage{url}
\usepackage{rotating} 
\usepackage{simpler-wick} 
\usepackage{xcolor}
\usetikzlibrary{decorations.pathmorphing}
\usepackage{subcaption}
\usepackage{graphicx}
\usepackage{float}
\usepackage{wrapfig} 
\usepackage{graphics}
\usepackage{float}
\usepackage{pict2e}
\usepackage{epic}
\usepackage{eepic}
\usepackage{pstricks}
\usepackage{color}
\usepackage{xcolor}
\usepackage[absolute]{textpos} 
\usepackage{braket}
\usetikzlibrary{angles,quotes,calc,intersections}
\DeclareGraphicsExtensions{.png,.pdf,.jpg,.gif}
\providecommand{\abs}[1]{\lvert#1\rvert}

\definecolor{blueline}{RGB}{90,100,160}
\definecolor{tealline}{RGB}{110,190,180}
\definecolor{goldc}{RGB}{195,150,40}
\definecolor{connectorgray}{RGB}{90,90,90}

\newcommand{\circarc}[9]{%
  \pgfmathsetmacro{\dx}{#4-#2}
  \pgfmathsetmacro{\dy}{#5-#3}
  \pgfmathsetmacro{\L}{sqrt(\dx*\dx+\dy*\dy)}
  \pgfmathsetmacro{\ux}{\dx/\L}
  \pgfmathsetmacro{\uy}{\dy/\L}
  \pgfmathsetmacro{\px}{\uy}
  \pgfmathsetmacro{\py}{-\ux}
  \pgfmathsetmacro{\Mx}{(#2+#4)/2}
  \pgfmathsetmacro{\My}{(#3+#5)/2}
  \pgfmathsetmacro{\Cx}{\Mx+#6*\px}
  \pgfmathsetmacro{\Cy}{\My+#6*\py}
  \pgfmathsetmacro{\R}{sqrt((\Cx-#2)*(\Cx-#2)+(\Cy-#3)*(\Cy-#3))}
  \pgfmathsetmacro{\Astart}{atan2(#3-\Cy,#2-\Cx)}
  \pgfmathsetmacro{\Aend}{atan2(#5-\Cy,#4-\Cx)}
  \pgfmathsetmacro{\Amid}{\Astart-#7*(\Astart-\Aend)}
  \pgfmathsetmacro{\Atip}{\Aend-#8}
  \ifnum#9=1
    \draw[black,dashed,line width=0.8pt]
      ($(\Cx,\Cy)+(\Astart:\R)$) arc (\Astart:\Amid:\R);
  \fi
  \draw[black,line width=1.6pt]
    ($(\Cx,\Cy)+(\Amid:\R)$) arc (\Amid:\Aend:\R);
  \draw[black,dash pattern=on 7pt off 5pt,line width=1.6pt]
    ($(\Cx,\Cy)+(\Aend:\R)$) arc (\Aend:\Atip:\R);
  \pgfmathsetmacro{\splitX}{\Cx+\R*cos(\Amid)}
  \pgfmathsetmacro{\splitY}{\Cy+\R*sin(\Amid)}
  \global\let\lastSplitX\splitX
  \global\let\lastSplitY\splitY
}

\title{
Cusped Defects: Cusp Operator Expansion, Conformal Properties and Bootstrap Applications}

\author[1]{Lorenzo Bianchi,}
\author[1]{Andrea Cavaglià,}
\author[1]{A. David Guti\'errez,}
\author[1]{Stefanos R. Kousvos,}
\author[1]{and Marco Meineri}
\affiliation[1]{Dipartimento di Fisica, Università di Torino and INFN – Sezione di Torino, \\
Via P. Giuria 1, Torino 10125, Italy.}

\emailAdd{lorenzo.bianchi@unito.it, andrea.cavaglia@unito.it, abnerdavid.gutierrezromero@unito.it, stefanosrobert.kousvos@unito.it, marco.meineri@unito.it}

\abstract{We analyze general properties of cusped defect lines embedded in generic conformal field theories (CFTs). We prove that suitably defined cusp scaling operators transform like primary operators under general conformal transformations. We show that, via the cusp operator expansion (COE), defects of general shape can be expanded in a basis of cusped defects. These findings are checked and exemplified in the case of the localized magnetic field defect. Finally, we derive analytic constraints on dynamical cusp data by making use of a crossing equation, and discuss possible numerical bootstrap applications of our results.
}

\begin{document}
\maketitle
\flushbottom
\section{Introduction} 
\label{sec:intro}

Extended operators have proven to be physically important in Quantum Field Theory (QFT) for many reasons. They model the interaction between light and heavy degrees of freedom, such as impurities in a medium and heavy charged particles in a gauge theory. As such, they describe a plethora of 
tangible 
real-world phenomena, 
such as the logarithmic growth of resistivity in metals at low temperature \cite{kondo,WilsonKondo,Affleck:1995ge}, surface criticality \cite{Diehl:1996kd}, bremsstrahlung and infrared factorization in gauge theories \cite{Bauer:2001yt}. They are also essential in capturing features of QFT invisible to local operators, such as the breaking of generalized symmetries \cite{Gaiotto:2014kfa}---notably, at the deconfinement transition in pure gauge theories \cite{Wilson:1974sk,tHooft:1977nqb}---or the entanglement between spatially separated regions \cite{Calabrese:2004eu,Bianchi:2015liz}.   
In Conformal Field Theories (CFTs), one can consider defects that preserve a subgroup of the ambient conformal group, giving rise to a defect CFT \cite{Cardy:1984bb,McAvity:1995zd}, and this will be our focus.

Correlation functions involving defects have a rich dependence on their shape. The response of the CFT partition function to deformations of an extended operator is an active topic of research \cite{Graham:1999pm,Solodukhin:2008dh,Correa:2012at,Lewkowycz:2013laa,Allais:2014ata,Bianchi:2015liz,Billo:2016cpy,Cooke:2017qgm,Bianchi:2018zpb,Gabai:2025zcs,Drukker:2025dfm,Girault:2025kzt}.
In particular, one can consider extended operators supported on curves with singular loci, of which a simple example is shown in Figure \ref{fig:cusped}. The singular point in Figure \ref{fig:cusped} is known as a cusp.
Among many other applications, in Lorentzian signature time-like cusped line defects describe particles which receive a sudden acceleration \cite{Correa:2012at}, while in the light-like limit they are related to Sudakov double logarithms in gauge theories \cite{Sudakov:1956}. In Euclidean signature, cusped defects describe for instance the entanglement entropy of regions whose boundary has corners \cite{Bueno:2015rda}, and their small opening angle limit computes the static interquark potential, or more generally the potential between static sources in a CFT \cite{Correa:2012hh}. 

Early studies of cusped defects in gauge theories include \cite{Polyakov:1980ca}, \cite{Makeenko:1979pb} and  \cite{Brandt:1981kf}, where the latter showed that Wilson loops with cusps in non-abelian gauge theory with $SU(n)$ gauge group require additional local renormalization factors depending on the cusp angle, hinting at the additional physical information encapsulated in cusps. The associated anomalous dimension is the cusp anomalous dimension,  whose computation beyond leading order was later achieved by \cite{Korchemsky:1987wg} and then \cite{Grozin:2014hna}.\footnote{Cusped Wilson lines with operators inserted at the tip, whose renormalization leads to excited cusp anomalous dimensions such as the ones we will consider in this work, are also relevant in phenomenology, see e.g. \cite{Bruser:2018jnc,Falcioni:2019nxk}. }
Following these works, a plethora of cusped configurations in various contexts have been studied. In supersymmetric gauge theories, cusped loops have been approached from perturbation theory, scattering amplitudes, holography and integrability \cite{Drukker:2011za,Bykov:2012sc,Correa:2012nk,Correa:2012hh,Drukker:2012de,Gromov:2015dfa,Gromov:2016rrp,ladder,McGovern:2019sdd,Grabner:2020nis,Dorn:2015bfa,Dorn:2020meb,Dorn:2018srz,Dorn:2019yms,Chernikov:2026lcv,Griguolo:2012iq,Bonini:2016fnc,Bianchi:2018scb,Fiol:2015spa}. More recently, cusped defects have been considered from a CFT point of view, including 
 general line impurities \cite{Cuomo:2024psk}, fermionic CFTs \cite{Giombi:2025evu}, and holographic defect CFTs with corner contributions \cite{Sun:2024qhv}. Via dual conformal invariance, the spectrum of a cusped Wilson loop is related to Regge trajectories on the Coulomb branch of $\mathcal{N}=4$ Super Yang-Mills (SYM) \cite{Alday:2025pmg}. Light-like cusps in general CFTs where recently studied in \cite{Cuomo:2026mop}, inspired by their relation to Sudakov's double logarithm.

Recent results in explicit examples \cite{ladder,McGovern:2019sdd,Dorn:2020meb} have shown tantalizing similarities between cusps and ordinary local operators in CFT. In particular, expectation values of cusped line defects in (the ladder limit of) $\mathcal{N}$=4 SYM have covariance properties analogous to the ones of correlators of local primary operators. This is naively unexpected, since the symmetry group preserved by a cusp does not include special conformal transformations. Furthermore, it was shown that a pair of successive cusps along a contour could be replaced by a single one, at the price of deforming the contour appropriately and of summing over local insertions at the tip. 

These last developments motivate the present work. We show that cusps form a basis of a Hilbert space that comprises more general contours with specific features. From this fact, we derive the existence and the kinematical properties of the Cusp Operator Expansion (COE), that is, the expansion of these general contours into a basis. By defining non-perturbatively excited states on a cusp, we are also able to prove that cusps transform like local primaries operators in generic CFTs. We confirm these findings by simple examples, and we derive from them some analytic results on the spectral density of cusp operators, by exploiting crossing symmetry.

The rest of the paper is organized as follows. 
In Section \ref{sec:hilb}, we describe the construction of the Hilbert space and the state operator correspondence in the presence of a cusped defect. Then, we introduce 
the cusp operator expansion.
In Section \ref{sec:generic-cusp-correlators}, we prove transformation rules for correlation functions of cusp operators
under conformal transformations.
In Section \ref{sec:examples}, we study the magnetic line defect in the 4D free theory. We explicitly renormalize local cusp operators and check the expected transformation properties of their two- and three-point functions. Moreover we exemplify the COE by expanding a smooth contour into a basis of these cusped states. 
Section \ref{transformations-of-eigenstates} concerns itself with the transformation properties of NS Hamiltonian eigenstates in ordinary CFTs without defects. This detour serves to demystify the surprising transformation rules of scaling cusp states, by showing that they become specific bi-local operators with the same property, in the limit of a trivial defect.
In Section \ref{sec:bootsconf}, we use the general properties of the COE to study two distinct bootstrap configurations involving cusped defects. We show that these leads to analytic results on the asymptotics of certain spectral densities.  We finally conclude in Section \ref{sec:conclu} by describing some open questions as well as outlining a set of bootstrap setups that can be carried out numerically. 

\paragraph{Note added.} 
While this work was being completed we became aware of related work involving cusped defects in CFT. We  thank Ryan A. Lanzetta, Ian Moult and Yifan Wang for coordinating the submission of their works \cite{LanzettaMoultWang1,LanzettaMoultWang2} with us.

\section{Hilbert Space and Cusp Operator Expansion}  \label{sec:hilb}

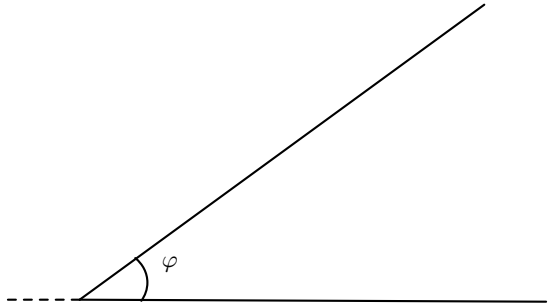
\begin{figure}[t]
    \centering

\begin{tikzpicture}[
    x=0.75pt,
    y=0.75pt,
    yscale=-1,
    xscale=1,
    line cap=round,
    line join=round
]


\draw[black, thick]
    (180,241) -- (414,242);

\draw[black, thick]
    (180,241) -- (383,93);

\draw[black, dashed, thick]
    (144,241) -- (180,241);

\draw[draw opacity=0]
    (207.9,220.11)
    .. controls (211.63,223.14) and (213.97,227.52) .. (213.98,232.39)
    .. controls (213.98,235.78) and (212.86,238.92) .. (210.93,241.55)
    -- (195.48,232.42) -- cycle;

\draw[black, thick]
    (207.9,220.11)
    .. controls (211.63,223.14) and (213.97,227.52) .. (213.98,232.39)
    .. controls (213.98,235.78) and (212.86,238.92) .. (210.93,241.55);

\draw (220,218.4)
    node[
        anchor=north west,
        inner sep=0.75pt,
        font=\footnotesize
    ] {$\varphi$};




\end{tikzpicture}

\caption{Line defect with a cusp at a finite point formed by the intersection of two straight lines extending to infinity. We will refer to the two lines meeting at the cusp as branches.}
\label{fig:cusped}

\end{figure}

In CFT, local operators are related to states via the state-operator correspondence. More precisely, a scaling operator inserted at the origin is indistinguishable from a state with the same quantum numbers prepared on a sphere around the origin, when probed at distances larger than the radius of the sphere. Said quantum numbers are the spin labels and the scaling dimension, i.e. the eigenvalues of the maximal set of commuting generators which leave the origin invariant. In the presence of an extended object, the symmetry is typically broken to a subgroup of the conformal group. For the present work, we will consider modifications of the Lagrangian of the theory along a submanifold, aka defects, which we define to possess the two following properties:
\begin{itemize}
    \item Their expectation value is invariant under any conformal transformation if the submanifold is smooth,
    \item They support a set of operators on any smooth part of the contour, whose transformation properties are local. 
\end{itemize}
Let us make a few remarks on these assumptions. Firstly, in this paper we use the term \emph{smooth}, in a non-standard fashion, to mean continuously differentiable ($C^1$). That this is the only condition required will become clear in the following.
Secondly, both assumptions concern any conformal transformation, not just the ones that preserve the contour. Furthermore, the term \emph{operators} in the second request is meant in the path-integral sense: modifications of the defect in a small neighborhood of a point. The  adjective \emph{local}, then, means that defect operators $\bar{\mathcal{O}}_a(p)$,\footnote{We reserve the notation $\hat{\mathcal{O}}$ for cusp operators, leveraging the fitting difference between the bar and hat accents.} change as $\bar{\mathcal{O}}_a(p) \to M_a^b(p) \bar{\mathcal{O}}_b(p)$, where $M$ only depends on the insertion point $p$. Beyond this, we are agnostic on the transformation properties of local defect operators, since in particular they can be subtle for transformations which do not preserve the contour \cite{Gabai:2025zcs}.

For concreteness, we limit our attention to one-dimensional conformal defects, although most of our results generalize.

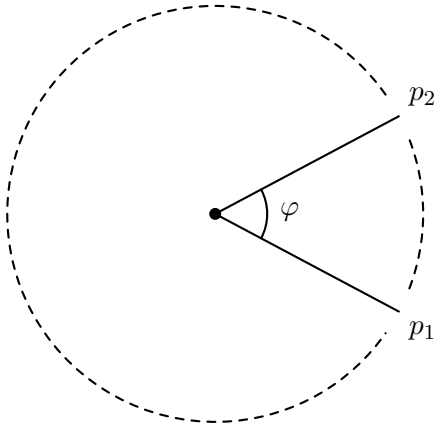
\begin{figure}[t]
    \centering
\centering
    \begin{tikzpicture}[scale=1.25, line cap=round, line join=round]

        \def\R{2.2}        
        \def\ang{28}       
        \def\gap{7}        

        \coordinate (C)  at (0,0);              
        \coordinate (P2) at (\ang:\R);          
        \coordinate (P1) at (-\ang:\R);         

        \draw[dashed, thick]
            ({\ang+\gap}:\R)
            arc[start angle=\ang+\gap, end angle=360-\ang-\gap, radius=\R];

        \draw[dashed, thick]
            ({360-\ang+\gap}:\R)
            arc[start angle=360-\ang+\gap, end angle=360+\ang-\gap, radius=\R];

        \draw[thick] (C) -- (P2);
        \draw[thick] (C) -- (P1);

        \fill (C) circle (1.8pt);

        \draw[thick]
            (\ang:0.55)
            arc[start angle=\ang, end angle=-\ang, radius=0.55];

        \node[right] at (0.58,0.03) {$\varphi$};

        \node[above right] at (P2) {$p_2$};
        \node[below right] at (P1) {$p_1$};



    \end{tikzpicture}
\caption{ A state is prepared by performing the path integral in the interior of the dashed sphere, with boundary conditions specified by the value of fields at the boundary of the UV ball (black).}
    \label{fig:punctured}
\end{figure}

Specifically, we will be interested in configurations such as the one in Figure \ref{fig:cusped}, which contain a cusp. The transformation property of the defect at the cusp is importantly not part of our assumptions. In fact, one of the aims of this paper is to derive it. The subgroup of the conformal group which leaves the contour in Figure \ref{fig:cusped} invariant is $\mathbb{R} \cross SO(d-2)\rtimes G_\text{discrete}$, the first factor representing dilatations and the second rotations orthogonal to the defect. The discrete symmetries $G_\text{discrete}$ include reflections in the transverse direction, a reflection with respect to the bisector of the angle $\varphi$, and inversion. The partition function of a conformal defect defined on the contour in Figure \ref{fig:cusped} is in general not invariant under all of these symmetries. We will be agnostic regarding discrete symmetries\footnote{Discrete symmetries play a role in classifying operators in the example considered in Section \ref{sec:examples}.}, but we will require a weaker property involving inversions below. More importantly, cusped defects are in general not invariant under dilatations, due to the cutoff dependence in the definition of the path integral at the cusp. As we explain below, the cutoff dependence contains universal information on the spectrum of local cusp operators, which correspond to states prepared on a sphere centered at the location of the cusp---see Figure \ref{fig:punctured}. The states are labeled by quantum numbers corresponding to the little group of the origin, among the symmetries preserving the configuration in Figure \ref{fig:cusped}: in particular, by their scaling dimension and $SO(d-2)$---rather than $SO(d)$---spin.\footnote{Rotations orthogonal to the defect serve as a global symmetry for observables fully defined in the plane of the cusp. Most of this paper is dedicated to such observables. When considering non-planar configurations, we will restrict to the scalar sector for simplicity}

\subsection{The Hilbert Space of a Cusp}
\label{sec:hilbert}

In this subsection, we discuss in some detail the Hilbert space in radial quantization around a cusp at the origin, as shown in Figure \ref{fig:punctured}. The method to prepare a state is the same as the one for local bulk operators  (see e.g. \cite{Simmons-Duffin:2016gjk}), with the obvious modification that the defect modifies the action. The state produced on the boundary of the ball depends in particular on the boundary conditions chosen at the location of the cusp. In concrete perturbative treatments, the latter will in turn depend on the regularization scheme, while we will later define a specific non-perturbative regulator, useful for our purposes. However, it is important that, as long as these boundary conditions, together with any other insertions, are defined away from the quantization surface, the set of states in Figure \ref{fig:punctured} are part of a Hilbert space that transforms in a representation of dilatations. By diagonalizing the dilatation operator, one finds the (scheme-independent) scaling dimension of the cusp operators. This is the perspective adopted here to define local cusp operators starting from the assumptions listed above on smooth defects: the eigenstates of the dilatation operators define the data at the cusp. They can be glued to piece-wise smooth defects in the path-integral to define correlation functions involving multiple cusps. Alternatively, the Cusp Operator Expansion defined below can be used to define such correlation functions starting from smooth loops. Given the importance of this point, we spend the rest of this subsection defining the cusp Hilbert space starting from path integrals containing only smooth loops.\footnote{The arguments below are also valid for radial quantization around smooth, and in particular flat, defects, which are a special case of this construction.} We also make some remarks on the existence of a Hermitian dilatation operator, but we will not try to be rigorous.

\begin{figure}[t]
\centering

\begin{subfigure}[t]{0.46\textwidth}
\centering

\begin{tikzpicture}[
    curve/.style={
        black,
        thick
    },
    seam/.style={
        black,
        dashed,
        thick
    },
    tangent/.style={
        orange!85!black,
        line width=1.15pt,
        -{Stealth[length=2.2mm]}
    },
    markedpoint/.style={
        circle,
        fill=black,
        inner sep=1.6pt
    }
]

\def\R{2.2}

\coordinate (P1) at (60:\R);
\coordinate (P2) at (-55:\R);
\coordinate (M)  at (-0.15,0.05);

\draw[seam] (0,0) circle (\R);

\draw[curve]
    (P1)
    .. controls +(-70:.85) and +(120:.90) .. (M)
    .. controls +(-60:.90) and +(80:.85) .. (P2)
    node[pos=.28, circle, fill=black, inner sep=1.3pt] {}
    node[pos=.72, circle, fill=black, inner sep=1.3pt] {};

\node at (0.4,0.72) {$\gamma$};

\node[markedpoint] at (P1) {};
\node[markedpoint] at (P2) {};

\node[above right=-1pt] at (P1) {$p_1$};
\node[below right=-1pt] at (P2) {$p_2$};

\draw[tangent]
    (P1) -- ++(110:1.00)
    node[pos=.72, left=1pt] {$\mathbf t_1$};

\draw[tangent]
    (P2) -- ++(-100:1.00)
    node[pos=.72, right=1pt] {$\mathbf t_2$};

\fill (-.85,.35) circle (1.4pt);
\fill (.05,-.95) circle (1.4pt);
\fill (.70,-.15) circle (1.4pt);

\end{tikzpicture}

\caption{The curve $\gamma$.}
\label{fig:inside}
\end{subfigure}
\hfill
\begin{subfigure}[t]{0.46\textwidth}
\centering

\begin{tikzpicture}[
    curve/.style={
        black,
        thick
    },
    seam/.style={
        black,
        dashed,
        thick
    },
    tangent/.style={
        orange!85!black,
        line width=1.15pt,
        -{Stealth[length=2.2mm]}
    },
    markedpoint/.style={
        circle,
        fill=black,
        inner sep=1.6pt
    }
]

\def\R{2.2}

\coordinate (P1) at (60:\R);
\coordinate (P2) at (-55:\R);

\draw[seam] (0,0) circle (\R);

\pgfmathsetmacro{\dtheta}{245*pi/180}

\pgfmathsetmacro{\drone}{
    \R*\dtheta/tan(50)
}
\pgfmathsetmacro{\drtwo}{
    \R*\dtheta/tan(135)
}

\pgfmathsetmacro{\C}{3.5}

\pgfmathsetmacro{\ua}{0.28}
\pgfmathsetmacro{\ub}{0.72}

\pgfmathsetmacro{\ra}{
    \R
    + \drone*\ua*(1-\ua)^4
    - \drtwo*\ua^4*(1-\ua)
    + \C*\ua^2*(1-\ua)^2*(1 + 0.80*sin(1080*\ua))
}
\pgfmathsetmacro{\rb}{
    \R
    + \drone*\ub*(1-\ub)^4
    - \drtwo*\ub^4*(1-\ub)
    + \C*\ub^2*(1-\ub)^2*(1 + 0.80*sin(1080*\ub))
}

\coordinate (Q1) at ({\ra*cos(60+245*\ua)},{\ra*sin(60+245*\ua)});
\coordinate (Q2) at ({\rb*cos(60+245*\ub)},{\rb*sin(60+245*\ub)});

\draw[curve]
    plot[
        domain=0:1,
        samples=180,
        variable=\u,
        smooth
    ]
    ({
        (
            \R
            + \drone*\u*(1-\u)^4
            - \drtwo*\u^4*(1-\u)
            + \C*\u^2*(1-\u)^2
                *(1 + 0.80*sin(1080*\u))
        )
        * cos(60+245*\u)
    },{
        (
            \R
            + \drone*\u*(1-\u)^4
            - \drtwo*\u^4*(1-\u)
            + \C*\u^2*(1-\u)^2
                *(1 + 0.80*sin(1080*\u))
        )
        * sin(60+245*\u)
    });

\node[circle, fill=black, inner sep=1.3pt] at (Q1) {};
\node[circle, fill=black, inner sep=1.3pt] at (Q2) {};

\node at (-2.50,1.25) {$\gamma'$};

\node[markedpoint] at (P1) {};
\node[markedpoint] at (P2) {};

\node[above right=-1pt] at (P1) {$p_1$};
\node[below right=-1pt] at (P2) {$p_2$};

\draw[tangent]
    (P1) -- ++(-70:1.00)
    node[pos=.72, left=1pt] {$-\mathbf t_1$};

\draw[tangent]
    (P2) -- ++(80:1.00)
    node[pos=.72, left=1pt] {$-\mathbf t_2$};

\fill (25:2.90)   circle (1.4pt);
\fill (-8:2.85)   circle (1.4pt);
\fill (195:2.55)  circle (1.4pt);

\end{tikzpicture}

\caption{The curve $\gamma'$.}
\label{fig:outside}
\end{subfigure}
\caption{On the left, a vector in the vector space obtained by path integrating inside a sphere punctured by a defect with fixed tangents $t_1$ and $t_2$. On the right, an element of the dual vector space.}
\label{fig:matching_pair}
\end{figure}
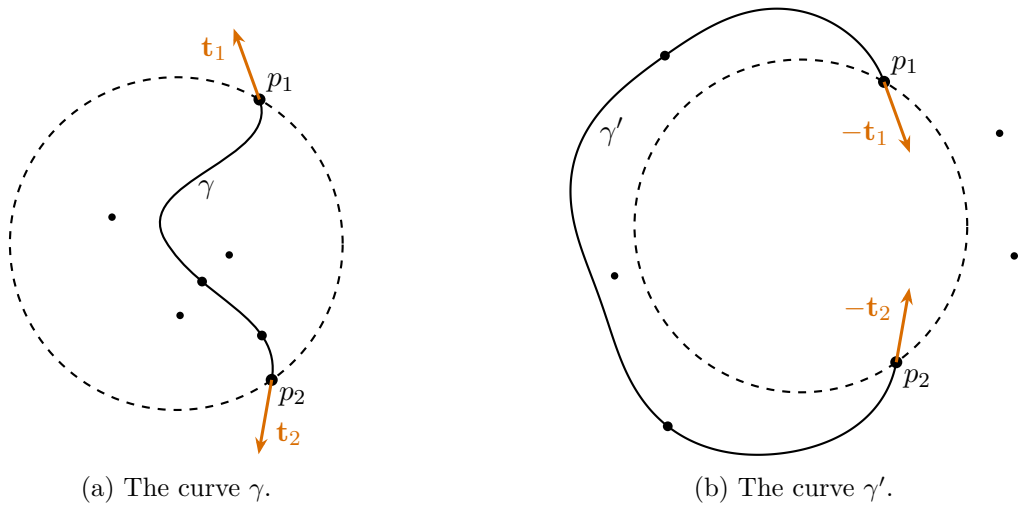
Let us therefore take a step back and start from a more general configuration than the one in Figure \ref{fig:punctured}.
There is a state  $|\Psi \rangle_{\gamma}$ for every smooth defect curve $\gamma$ lying within the quantization surface and extending between the punctures $p_1$ and $p_2$.\footnote{The defect can be decorated with defect-changing, or defect-ending, operators, cusps etc.} Call $t_i$ the vectors tangent to $\gamma$ at 
$p_i$ ($i=1,2$), see Figure \ref{fig:inside}.  
 Consider the vector space generated by all states of this form for tangents $t_i$, fixed but generic. 
 The dual space (i.e., the space of linear functionals acting on such vectors) is constructed by performing the path-integral in the exterior of the quantization surface, which is the punctured sphere. Thus, we can build a co-vector by taking a defect line $\gamma'$ which extends between $p_2$ and $p_1$ {outside the unit ball}, see Figure \ref{fig:outside}: 
 by construction, the action of the functional defined in this way on a state of the form $|\Psi \rangle_{\gamma}$ is obtained by gluing $\gamma$ (defining the state) with the curve $\gamma'$ outside the unit ball (defining the functional), and considering the expectation value of the resulting closed contour. 

Importantly, we require that the tangents of $\gamma$ and $\gamma'$ match at the points $p_1$ and $p_2$. Otherwise, after gluing the two curves, the new cusp at the punctures would introduce divergences, which require additional choices to define the pairing in the continuum limit. While these extra data do not obstruct linearity of the pairing, they would endanger positivity of the scalar product to be defined next.  For brevity, we call this the \emph{$C^1$ condition}. For the same reason, we require the defects on $\gamma$ and $\gamma'$ to be the same\footnote{This condition can be unambiguously defined, since our assumptions include the existence of defect operators on smooth defects: no operators are inserted at $p_1$ and $p_2$, meaning that the conformally transformed defect does not pick up factors depending on these two points.}
at the punctures---of course, the defect type at $p_1$ can be different from the one at $p_2$  if defect-changing operators are present along the curves---and we call this the \emph{defect-matching condition.}

We will now show that a subspace of this vector space admits the structure of a Hilbert space. To define a scalar product, one proceeds as usual \cite{Osterwalder:1973,Pappadopulo:2012jk} by defining an antilinear map between vectors and covectors. Such map must be compatible with the one of the bulk CFT, and therefore it acts on states in radial quantization by applying an inversion to all operator insertions:  
 \begin{equation}
     I(x^\mu)=\frac{R^2 x^\mu}{x^2}~,
 \end{equation}
where $R$ is the radius of the quantization surface. In particular, the Hermitian conjugate of a state $|\Psi \rangle_{\gamma}$, constructed as above, is defined by performing the path integral in the exterior region, in presence of the transformed curve $I(\gamma)$, which lies outside the quantization surface. Now, the conjugate state belongs to the dual space only if it satisfies the $C^1$ and the defect-matching conditions. 

The $C^1$ condition can be written as 
\begin{equation} \label{tan2}
    t_i^\mu=-\frac{\partial I^\mu}{\partial x^\nu}\Bigg|_{p_i}t_i^\nu~,
\end{equation}
which is easily seen to imply that $t_i$ \emph{is orthogonal to the quantization surface}.  The defect-matching condition, instead, requires the defect to be invariant under Hermitian conjugation. In concrete examples, this imposes constraints on the action of the defect. Consider the case of a Wilson loop in a (Abelian) gauge theory:
\begin{equation}
    S_\textup{defect}(\gamma) = \lambda \int_{\gamma} A_\mu dx^\mu~.
    \label{Swilson}
\end{equation}
Since the action is conformal invariant but odd under reversing the orientation of the curve $\gamma$, one can check that
\begin{equation}
    I\,S_\textup{defect}[\gamma(p_1,p_2)] = S_\textup{defect}[\gamma'(p_1,p_2)] =-S_\textup{defect}[\gamma'(p_2,p_1)]~,
    \label{InversionSwilson}
\end{equation}
where $I$ denotes the action of inversion and $\gamma(p_1,p_2)$ and $\gamma'(p_2,p_1)$ have opposite orientation, i.e. the sum of the two curves yields a loop with uniform orientation. Since Hermitian conjugation is antilinear, it follows that the defect-matching condition forces $\lambda$ to be pure imaginary.\footnote{Notice that one cannot choose the action of Hermitian conjugation on the defect action to include an extra sign, because the Hermitian conjugate of the gauge potential is fixed by reflection positivity of the bulk CFT.} It is also interesting to notice that this is compatible with reflection positivity \cite{Kravchuk:2024qoh,Witten:2025ayw}. Indeed, invariance under inversion implies that the path integral over the exterior of the ball of radius $R$, with the insertion of the Wilson line on the oriented contour $\gamma'(p_1,p_2)$, equals the path-integral over the interior, with the Wilson line inserted along $\gamma(p_1,p_2)$, if the fields on the boundary of the ball match (up to the same inversion). Hence, eq. \eqref{InversionSwilson} and $\lambda^*=-\lambda$ imply that reversing the orientation yields the complex conjugate of the path integral inside the ball. Therefore, the expectation value of the inversion-symmetric loop is positive.\footnote{We are assuming, as is the case here, that the cosmological constant counterterm needed to make the expectation value finite in the continuum is real. We are also being cavalier with the ambiguities of the boundary data on the quantization sphere up to gauge transformations. One may instead consider a line defect obtained by integrating a parity-odd scalar of dimension 1 in a parity-invariant CFT, to reach the same conclusions without the issue of gauge redundancy.}

To summarize, \emph{states belonging to the Hilbert space in radial quantization are those associated with defect curves that cross the quantization surface orthogonally, and the scalar product is defined by overlap with appropriately oriented defects with the same property}. The Hilbert space depends on the angle $\varphi \in (0,2\pi)$ defined by $p_1$ and $p_2$, and we simply denote it by $\mathcal{H}(\varphi)$. While, in general, it also depends on the type of defects at the two punctures, we will not explicitly denote this dependence in the following.

As implied by the terminology we use, we assume the defect to be reflection positive, hence the scalar product to be positive definite. The last remaining task is to show that a Hermitian dilatation operator acting on this Hilbert space exists. In a local theory, the dilatation operator is the integral of the stress tensor, possibly with appropriate subtractions at the location of the defect \cite{Meineri:2023mps,Lanzetta:2025xfw}, but we can also sketch a construction closer to the spirit of \cite{Osterwalder:1973} (see also \cite{Kravchuk:2021kwe} for a modern guide to the literature). In a translationally invariant theory,  one defines a positive operator that generates Euclidean time evolution. One can do the same here: the radial time evolution operator $T(\lambda)$ acts on any state in $\mathcal{H}(\varphi)$ by applying a dilatation that shrinks the state by a factor $\lambda>1$ to lie inside a smaller sphere, and attaches to endpoints of the defect, which now lie inside the quantization surface, two straight segments to join them to $p_1$ and $p_2$. This is where the existence (but not the knowledge) of the covariance properties of smooth defects and their local operators come in: the map $T(\lambda)$ is only defined if the latter are specified. Locality of the transformation law of defect operators is also used here, because it ensures that the action of $T(\lambda)$ can be determined without reference to the shape of the defect outside the quantization sphere. One then can show that $\braket{\Psi_1|T(\lambda)|\Psi_2}=\braket{\Psi_2|T(\lambda)|\Psi_1}^*$, because the latter overlap builds the same picture of the former up to a dilatation (of course the map $T(\lambda)$ is defined so that all the factors arising from the transformation of operators are included). Rigorously defining the time evolution, and the Hamiltonian from it, further requires showing that $T(\lambda)$ is densely defined, and bounded by reflection positivity, but we will not pursue this here, and assume that the remaining requirements are all met and a Hermitian dilatation operator $D$ exists.

As promised, we define a complete basis of eigenstates of the dilatation operator $D$:
\begin{equation}\label{eq:basis}
D |n\rangle = \Gamma_n(\varphi) |n \rangle ,
\end{equation}
where the spectrum (which we assume discrete for notational simplicity) continuously depends on the angle $\varphi$ defined by the position of the punctures $p_1$ and $p_2$. This dependence will sometimes be omitted, but always understood.
Using a dilatation to reduce the radius $R$ of the quantization surface, as usual in radial quantization, we conclude that energy eigenstates of $D$ are created by cusps with straight branches, as in Figure \ref{fig:punctured}.

If the spectrum is bounded from below, the ground state dimension $\Gamma_0(\varphi)$ controls the exponent of the leading divergence of a straight cusp as the regulator is removed, and is therefore customarily referred to as the cusp anomalous dimension.  Unitarity constrains the ground state dimension $\Gamma_0$  to be non-positive (vanishing only for topological defects or at $\varphi = \pi$ where the cusp straightens up), and to be a concave function of the angle~\cite{Cuomo:2024psk}.  
Excited eigenstates of $D$ are obtained by modifying the ground state only in an arbitrarily small neighborhood of the cusp, and therefore correspond to local cusp operators. In perturbative theories, one can construct them explicitly, as we will see in section \ref{sec:examples}. We will sometimes call eigenstates of $D$ (or more generally of time evolution along the cusp---see Subsection \ref{subsec:gener}) \emph{cusp eigenstates} or \emph{scaling cusp operators}.
Completeness of the eigenstates of a Hermitian operator implies that any element of the Hilbert space can be expanded in a linear combination of the states $\ket{n}$, as we explore in the next subsections.

\begin{figure}[t]
    \centering

\tikzset{
    every picture/.style={line width=0.75pt},
    qcircle/.style={
        black,
        line width=0.9pt,
        dash pattern=on 5pt off 4pt
    }
}

\begin{tikzpicture}[x=0.75pt,y=0.75pt,yscale=-1,xscale=1]

\draw [qcircle]
  (195,206.5) .. controls (195,157.07) and (235.07,117) .. (284.5,117) 
  .. controls (333.93,117) and (374,157.07) .. (374,206.5) 
  .. controls (374,255.93) and (333.93,296) .. (284.5,296) 
  .. controls (235.07,296) and (195,255.93) .. (195,206.5) -- cycle ;

\draw [line width=0.75]    (318.5,227.5) -- (370,177) ;

\draw [line width=0.75]    (318.5,227.5) -- (360.61,251.44) ;

\draw  [draw opacity=0] (333.57,235.94) .. controls (338.01,234.58) and (341.37,230.91) .. (342.05,225.95) .. controls (342.83,220.3) and (339.92,214.6) .. (335.12,211.36) -- (328.1,222.58) -- cycle ; 
\draw   (333.57,235.94) .. controls (338.01,234.58) and (341.37,230.91) .. (342.05,225.95) .. controls (342.83,220.3) and (339.92,214.6) .. (335.12,211.36) ;  

\draw (280,200) node [anchor=north west][inner sep=0.75pt]  [font=\footnotesize]  {$\bullet $};

\draw (343,218.4) node [anchor=north west][inner sep=0.75pt]  [font=\small]  {$\varphi $};

\draw (365,173.4) node [anchor=north west][inner sep=0.75pt]  [font=\footnotesize,color={rgb, 255:red, 255; green, 255; blue, 255 },opacity=1]  {$\bullet $};

\draw (357,247.4) node [anchor=north west][inner sep=0.75pt]  [font=\footnotesize,color={rgb, 255:red, 255; green, 255; blue, 255 },opacity=1]  {$\bullet $};

\draw (369,235.4) node [anchor=north west][inner sep=0.75pt]    {$p'_{1}$};

\draw (377,175.4) node [anchor=north west][inner sep=0.75pt]    {$p'_{2}$};

\end{tikzpicture}
\caption{Defect translated by $a^\mu$ while the quantization sphere remains fixed. The defect punctures the sphere at two different points $p'_1$ and $p'_2$. 
While this configuration defines a state on the sphere punctured at the new points, the tangent vectors at the punctures are not radial, meaning that this state does not belong to the Hilbert space. 
}
\label{fig:puncturedtrans}
\end{figure}
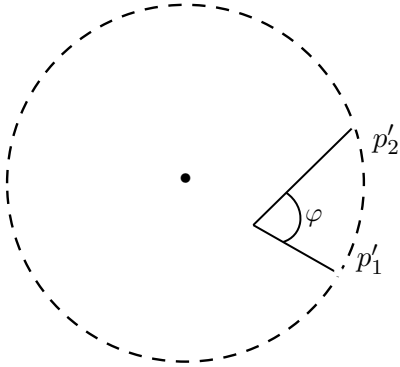

Since the cusped contour in Figure \ref{fig:cusped} is not invariant under translations, there is no translation generator $P_\mu$ (nor special conformal generator $K_\mu$) acting on the eigenstates $\ket{n}$, therefore the spectrum of cusp operators does not form integer spaced multiplets, and is not constrained by the unitarity bounds familiar from the conformal algebra. 
From the point of view of the present discussion, a finite translation by a vector $a_\mu$ is not an endomorphism of the Hilbert space:
\begin{equation}
e^{a\cdot P}:  \mathcal{H}(\varphi) \rightarrow  \mathcal{S}(\varphi; p'_1,p'_2).
\end{equation}
$\mathcal{S}(\varphi; p'_1,p'_2)$ is a set of states without a well-defined inner product, because the tangent vectors to the defect at the new punctures are not orthogonal to the quantization circle (see Figure \ref{fig:puncturedtrans}). Below, we will see that a contour formed by two arcs of circles meeting at two cusps---see Figure \ref{fig:NSalmond}---has an expectation value matching the two-point function of local operators. The remark above shows why, despite this fact, one cannot use a Taylor expansion to define operators with integer spaced dimensions, as in the case of local operators.\footnote{Instead, the action of an infinitesimal translation on eigenstates of the dilatation operator is computed by the insertion of the displacement operator integrated along the defect.} One practical consequence of this is that a conformal multiplet in a line defect-CFT, that would otherwise have integer-spaced scaling dimensions (a primary and its corresponding descendants), is broken once a cusp angle $\varphi$ is turned on.

Let us finally notice that the construction of this subsection easily generalizes to multiple punctures, with scaling eigenstates corresponding to junctions of multiple straight defects at the origin.

\subsection{The Cusp Operator Expansion (COE)}\label{sec:COE}

Consider a generic state $|\Psi\rangle$ belonging to the Hilbert space $\mathcal{H}(\varphi)$ constructed in the previous subsection.
By completeness of the Hilbert space, it can be written as a superposition of eigenstates of the dilation operator
\begin{equation} \label{generalCOE}
     |\Psi\rangle= \sum_n c_n(\varphi, \vec{r})  |n\rangle,
\end{equation}
where $\varphi$ is the cusp angle and $\vec{r}$ denotes a collection of parameters that determine the shape of the contour, such as characteristic length scales, radii of curvature, eccentricities, etc.
We shall call expression \eqref{generalCOE} above the Cusp Operator Expansion (COE). 

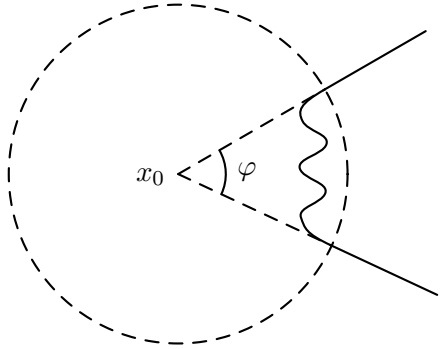
\begin{figure}[t]
  \centering
\begin{tikzpicture}[
    scale=0.70,
    line cap=round,
    line join=round
]

\def\R{3.2}       
\def\upperangle{30}
\def\lowerangle{-25}
\def\raylength{5.4}

\coordinate (O) at (0,0);
\coordinate (U) at (\upperangle:\R);
\coordinate (L) at (\lowerangle:\R);

\coordinate (Uout) at (\upperangle:\raylength);
\coordinate (Lout) at (\lowerangle:\raylength);

\node[anchor=east, xshift=-1pt, yshift=-1pt] at (O) {$x_0$};

\draw[
    black,
    thick,
    dash pattern=on 5pt off 4pt
]
(O) circle (\R);

\draw[
    black,
    thick,
    dash pattern=on 5pt off 4pt
]
(O) -- (U);

\draw[
    black,
    thick,
    dash pattern=on 5pt off 4pt
]
(O) -- (L);

\draw[
    black,
    thick
]
(O) ++(\lowerangle:9mm)
arc[
    start angle=\lowerangle,
    end angle=\upperangle,
    radius=9mm
];

\pgfmathsetmacro{\middleangle}{(\lowerangle+\upperangle)/2}

\node at (\middleangle:13mm) {$\varphi$};

\draw[
    black,
    thick
]
(U) -- (Uout);

\draw[
    black,
    thick
]
(L) -- (Lout);

%
\draw[
    black,
    thick
]
(U)
    .. controls (2.295,1.325) and (2.25,1.20)
    .. (2.35,1.00)

    .. controls (2.45,0.80) and (2.87,0.76)
    .. (2.80,0.56)

    .. controls (2.73,0.36) and (2.23,0.30)
    .. (2.30,0.10)

    .. controls (2.37,-0.10) and (2.85,-0.16)
    .. (2.78,-0.36)

    .. controls (2.71,-0.56) and (2.25,-0.62)
    .. (2.32,-0.82)

    .. controls (2.39,-1.02) and (2.40,-1.12)
    .. (L);
\end{tikzpicture}
\caption{ Generic line defect (solid line) defining a state in the cusped Hilbert space $\mathcal{H}(\varphi)$ by radial quantization.}
  \label{fig:sce2}
\end{figure}

When applied to a defect, it allows one to replace part of a contour by a sum over cusp eigenstates. For instance, the part of the defect inside the dashed circle in Figure \ref{fig:sce2} can be replaced by the dashed cusp with straight branches, with operator insertions at the tip. In the Section \ref{subsec:sce}, we will consider a simple but useful example where the form of the COE coefficients $c_n$ can be made more explicit. In Section \ref{sec:coeexample}, we will also see the COE at work, reproducing the expectation value of a defect in a specific CFT.

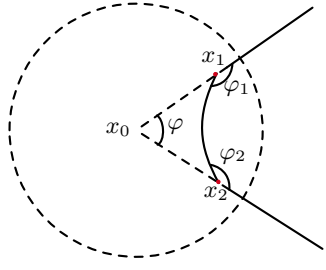
\begin{figure}[t]
  \centering

\begin{tikzpicture}[
    x=0.75pt,
    y=0.75pt,
    yscale=-1,
    xscale=1,
    line cap=round,
    line join=round
]

\draw[black,dashed,thick]
  (91,163.5) .. controls (91,128.43) and (119.43,100) .. (154.5,100)
  .. controls (189.57,100) and (218,128.43) .. (218,163.5)
  .. controls (218,198.57) and (189.57,227) .. (154.5,227)
  .. controls (119.43,227) and (91,198.57) .. (91,163.5) -- cycle ;

\draw[black,thick] (200,132) -- (243,102) ;
\draw[black,thick] (201.25,193.25) -- (248,223) ;

\draw[black,dashed,thick] (200,132) -- (154.5,163.5) ;
\draw[black,dashed,thick] (201.25,193.25) -- (154.5,163.5) ;

\draw [draw opacity=0]
  (164.97,156.07) .. controls (166.52,156.98) and (167.77,159.74) .. (167.98,163.07)
  .. controls (168.23,166.88) and (167.04,170.12) .. (165.25,170.76)
  -- (164.24,163.32) -- cycle ;

\draw[black,thick]
  (164.97,156.07) .. controls (166.52,156.98) and (167.77,159.74) .. (167.98,163.07)
  .. controls (168.23,166.88) and (167.04,170.12) .. (165.25,170.76) ;

\draw[black,thick]
  (195,135) .. controls (184,157) and (186,172) .. (196,190) ;

\draw [draw opacity=0]
  (192.04,180.89) .. controls (195.13,181.52) and (197.85,183.21) .. (199.57,185.88)
  .. controls (200.99,188.1) and (201.52,190.68) .. (201.25,193.25)
  -- (187.37,193.73) -- cycle ;

\draw[black,thick]
  (192.04,180.89) .. controls (195.13,181.52) and (197.85,183.21) .. (199.57,185.88)
  .. controls (200.99,188.1) and (201.52,190.68) .. (201.25,193.25) ;

\draw [draw opacity=0]
  (202.8,130.21) .. controls (202.69,133.37) and (201.47,136.33) .. (199.12,138.46)
  .. controls (197.17,140.24) and (194.71,141.18) .. (192.13,141.34)
  -- (189.37,127.73) -- cycle ;

\draw[black,thick]
  (202.8,130.21) .. controls (202.69,133.37) and (201.47,136.33) .. (199.12,138.46)
  .. controls (197.17,140.24) and (194.71,141.18) .. (192.13,141.34) ;

\draw (193,133.4) node [anchor=north west][inner sep=0.75pt]
  [font=\huge,color={rgb,255:red,208; green,2; blue,27},opacity=1] {$\dot{}$};

\draw (194.5,187.4) node [anchor=north west][inner sep=0.75pt]
  [font=\huge,color={rgb,255:red,208; green,2; blue,27},opacity=1] {$\dot{}$};

\draw (186,123) node [anchor=north west][inner sep=0.75pt] [font=\footnotesize] {$x_{1}$};
\draw (187.25,190.65) node [anchor=north west][inner sep=0.75pt] [font=\footnotesize] {$x_{2}$};
\draw (138,158.4) node [anchor=north west][inner sep=0.75pt] [font=\footnotesize] {$x_{0}$};
\draw (168.97,156.47) node [anchor=north west][inner sep=0.75pt] [font=\footnotesize] {$\varphi$};
\draw (197,137.4) node [anchor=north west][inner sep=0.75pt] [font=\footnotesize] {$\varphi_{1}$};
\draw (195,171.4) node [anchor=north west][inner sep=0.75pt] [font=\footnotesize] {$\varphi_{2}$};

\end{tikzpicture}
\caption{Cusp-cusp COE. The two cusps at $x_1$ and $x_2$ are expanded as a sum of exchanged cusp operators at $x_0$.}
  \label{fig:cuspcuspope0}
\end{figure}

In fact, nothing in the COE forces the piece of contour inside the quantization surface to be smooth, or even continuous. 
Consider the configuration in Figure 
\ref{fig:cuspcuspope0}, where the contour possesses two cusps. The COE replaces this piece of contour with an expansion over excited states of a cusp at the center of the quantization surface, with straight branches.
 
Such an expansion, which generalizes the notion of the OPE between two operators in a local theory to cusps, was first found in the context of the ladder limit of Wilson lines in $\mathcal{N}$=4 SYM~\cite{ladder}. We will elaborate on its kinematical properties (in the simplest possible setup) in Section \ref{sec:cuspcuspope}. 

 Finally, while we started by discussing radial quantization due to its illustrative simplicity, it is in fact not the only useful quantization scheme concerning spheres marked at two points. In Section \ref{subsec:gener}, we discuss the most general cusped line defect that is left invariant by a generator of the conformal group, the relation of this construction to North-South (NS) pole quantization, and the corresponding COE. A large portion of our explicit examples will indeed be carried out in the NS picture.

Before moving on, let us give a geometric diagnostic for the applicability of the COE to replace part of a defect. As discussed, it must be possible to draw a sphere which obeys the following:
\begin{itemize}
    \item the sphere intersects the defect at only two points $p_1$ and $p_2$;
    \item the tangent vectors $t_i$ to the defect at the two points are orthogonal to the sphere. 
\end{itemize}

It is not hard to check that the second condition is equivalent to the vectors $p_2-p_1$ and $t_2-t_1$ being parallel.

To obtain this constraint, the ambiguity in the direction of $t_i$ is solved by choosing 
\begin{equation}
   (p_2-p_1)\cdot t_1<0~,\quad (p_2-p_1)\cdot t_2>0~,
\end{equation}
which makes them outward normal to the sphere. Their constant of proportionality is then the radius $R$ of the quantization sphere:
\begin{equation}
    p_2-p_1= R\, (t_2-t_1) ,
    \label{parallel}
\end{equation}
and the center of the sphere $c$ is obviously fixed to $c=p_1-R\, t_1$.

\subsubsection{The COE of a Rounded Cusp, \emph{a.k.a.}
Squaring the Circle
} \label{subsec:sce}

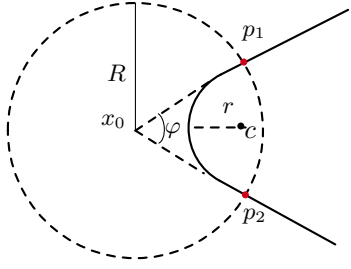
\begin{figure}[t]
  \centering

\begin{tikzpicture}[
    x=0.75pt,
    y=0.75pt,
    yscale=-1,
    xscale=1,
    line cap=round,
    line join=round
]

\draw[draw opacity=0]
    (192.6,188.56)
    .. controls (183.43,183.56) and (177.17,173.62) .. (177.17,162.17)
    .. controls (177.17,150.45) and (183.72,140.31) .. (193.25,135.43)
    -- (206.09,162.17) -- cycle ;

\draw[black,thick]
    (192.6,188.56)
    .. controls (183.43,183.56) and (177.17,173.62) .. (177.17,162.17)
    .. controls (177.17,150.45) and (183.72,140.31) .. (193.25,135.43) ;

\draw[black,dashed,thick]
    (86.5,163.5)
    .. controls (86.5,128.15) and (115.15,99.5) .. (150.5,99.5)
    .. controls (185.85,99.5) and (214.5,128.15) .. (214.5,163.5)
    .. controls (214.5,198.85) and (185.85,227.5) .. (150.5,227.5)
    .. controls (115.15,227.5) and (86.5,198.85) .. (86.5,163.5)
    -- cycle ;

\draw[black,dashed,thick]
    (150.5,163.5) -- (188.68,138.21) ;

\draw[black,dashed,thick]
    (150.5,163.5) -- (184.83,184.95) ;

\draw[draw opacity=0]
    (161.97,155.07)
    .. controls (163.52,155.98) and (164.77,158.74) .. (164.98,162.07)
    .. controls (165.23,165.88) and (164.04,169.12) .. (162.25,169.76)
    -- (161.24,162.32) -- cycle ;

\draw
    (161.97,155.07)
    .. controls (163.52,155.98) and (164.77,158.74) .. (164.98,162.07)
    .. controls (165.23,165.88) and (164.04,169.12) .. (162.25,169.76) ;

\draw[black,dashed,thick]
    (180.31,161.93) -- (205.81,161.93) ;

\draw
    (150.5,99.5) -- (150.5,163.5) ;

\draw[black,thick]
    (192.6,188.56) -- (236.65,212.83) -- (250.77,220.61) ;

\draw[black,thick]
    (193.25,135.43) -- (257.23,102.92) ;

\draw (199,121.4)
    node [anchor=north west][inner sep=0.75pt]
    [font=\Huge,color={rgb,255:red,208; green,2; blue,27},opacity=1]
    {$\cdot{}$};

\draw (199.8,188)
    node [anchor=north west][inner sep=0.75pt]
    [font=\Huge,color={rgb,255:red,208; green,2; blue,27},opacity=1]
    {$\cdot{}$};

\draw (203,109.4)
    node [anchor=north west][inner sep=0.75pt]
    [font=\footnotesize]
    {$p_{1}$};

\draw (203.25,201.65)
    node [anchor=north west][inner sep=0.75pt]
    [font=\footnotesize]
    {$p_{2}$};

\draw (132,154.4)
    node [anchor=north west][inner sep=0.75pt]
    [font=\footnotesize]
    {$x_{0}$};

\draw (163.97,158.47)
    node [anchor=north west][inner sep=0.75pt]
    [font=\footnotesize]
    {$\varphi$};

\draw (197.51,153.5)
    node [anchor=north west][inner sep=0.75pt]
    [font=\Huge]
    {$\cdot$};

\draw (193,148.4)
    node [anchor=north west][inner sep=0.75pt]
    [font=\footnotesize]
    {$r$};

\draw (204,160.4)
    node [anchor=north west][inner sep=0.75pt]
    [font=\small]
    {$c$};

\draw (135,129.4)
    node [anchor=north west][inner sep=0.75pt]
    [font=\footnotesize]
    {$R$};

\end{tikzpicture}

\caption{Smooth defect defining a state in the cusped Hilbert space. The smoothing circle is connected in a $C^1$ way to the rest of the defect (solid black).}
\label{fig:sce}
\end{figure}

As an example of the COE, let us consider the contour formed when
smoothing out the cusp in Figure \ref{fig:punctured} by capping it with a portion of a circle of radius $r$ (see Figure \ref{fig:sce}). The resulting contour has only one scale, so it provides a nice setup to study the COE explicitly. It also provides a useful regulator of a cusped contour, a feature that we will use in section \ref{sec:generic-cusp-correlators}. As shown in Figure \ref{fig:sce}, the arc of the circle is fixed such that its tangent joins continuously with the rest of the defect lines. In the figure, the dotted circle represents the quantization surface we will consider, with radius $R$, which cuts the defect at points $p_1$ and $p_2$. Notice that there is a maximal value of the smoothing radius $r=r_{max}$, beyond which the defect no longer punctures the (fixed) quantization sphere orthogonally.\footnote{In particular, $r_{\text{max}}=\tan\frac{\varphi}{2} R$ and $||a_{\text{max}}-x_0||=  \sec\frac{\varphi}{2} R$.}

\begin{figure}[t]
  \centering
\begin{tikzpicture}[scale=1.2, line cap=round, line join=round]

\def\R{1.4}      
\def\a{3.4}      

\pgfmathsetmacro{\theta}{acos(\R/\a)}

\coordinate (O) at (0,0);
\coordinate (A) at (\a,0);                  
\coordinate (Tup) at ({\R*cos(\theta)}, {\R*sin(\theta)});
\coordinate (Tdown) at ({\R*cos(\theta)}, {-\R*sin(\theta)});

\draw[thick]
  (Tup) arc[start angle=\theta, end angle=360-\theta, radius=\R]
  -- (A)
  -- cycle;

\node[right] at (A) {$|n\rangle$};
\end{tikzpicture}
\caption{The ice-cream contour computes the overlap $\langle \Psi_r | n\rangle$ between the smooth state and an excited cusp eigenstate. }
  \label{fig:icecream}
\end{figure}
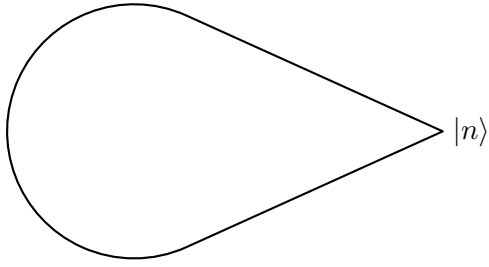

Considering $r\leq r_{\text{max}}$, radial quantization defines a  state $|\Psi_r\rangle$ in the Hilbert space $\mathcal{H}(\varphi)$. 
 By the COE, we can expand it in the basis of dilatation eigenstates
\begin{equation} \label{smoothstatea}   |\Psi_r\rangle= \sum_n \tilde{c}_n\left(\varphi,\frac{r}{R}\right)  |n\rangle .
\end{equation}
The coefficients in this expansion depend on the ratio of $r$ and $R$ by dimensional analysis, and therefore without loss of generality from now on we set $R=1$. They can be computed as 
\begin{equation}
    \langle n |\Psi_r \rangle = \tilde{c}_n(\varphi, r)
\end{equation}
 and 
 are equal to the expectation value of a one cusp contour as shown in Figure \ref{fig:icecream}.\footnote{Note that the expectation value is real when $d>2$, since the orientation of the defect can be switched by a rotation, and when $d=2$ if the bulk CFT preserves parity.} 
 It is simple to fix the exact dependence on the radius $r$. In fact,
considering the matrix element of the radial evolution operator $T(\lambda) \equiv e^{\log\lambda\hspace{2pt} D }$, we have
\begin{equation}
    \langle n | T(\lambda)|\Psi_r\rangle = \lambda^{\Gamma_n} \tilde{c}_n (\varphi,r).
\end{equation}
On the other hand, the action of the evolution operator on the smooth contour rescales $r$
\begin{equation}
     \langle n |T(\lambda)|\Psi_r \rangle = \langle n |\Psi_{\lambda r} \rangle =  \tilde{c}_n(\varphi, \lambda r).
\end{equation}
Comparing the last two equations yields 
\begin{equation}
    \tilde{c}_n(\varphi,r)= r^{\Gamma_n} c_n \left(\varphi\right).
\end{equation}
 Thus, the COE for the smoothed-out cusp state takes the form
\begin{equation} \label{smoothope}
    |\Psi_r\rangle= \sum_n c_n(\varphi) r^{\Gamma_n} |n\rangle ,
\end{equation}
where the dependence of $c_n$ and $\Gamma_n$ on the angle is not fixed by symmetry. 
Some of this data will be explicitly computed in an example in Subsection \ref{sec:coeexample}.  

As the regulator (i.e. the smoothing radius $r$) is taken to zero, the ground state dominates,
\begin{equation}
    |\Psi_r\rangle\overset{r\to0}{\sim} c_0(\varphi) r^{\Gamma_0}|0\rangle.
\end{equation}
Thus, the overlap of an $r$-regularized cusp with any smooth state $|\Phi \rangle$ will have the expected power-law divergence for small $r$:
\begin{equation} \label{powerlaw}
    \langle \Phi | \Psi_r\rangle \sim r^{ \Gamma_0} \cross  \text{finite},
\end{equation}
related to the ground state cusp anomalous dimension $\Gamma_0$.

\subsection{The Most General Symmetric Cusp} \label{subsec:gener}

While cusps with straight branches arise naturally in radial quantization, it is interesting and often convenient to work with defects that are not simply intersections of straight lines. 
With this in mind, we would now like to answer the following question: \emph{what is the most general cusped defect whose support is invariant under the action of at least one generator of the conformal algebra?} Of course, we are interested in the quotient of the stabilizer algebra by the pointwise stabilizer algebra, i.e. we want to find generators that act non-trivially on the points of the support. In practice, asking that one such generator exists already fixes the shape of the defect, hence we are looking for cusps with a conserved Hamiltonian. The result of this section is the intuitive one: the cusp with straight branches, as in Figure \ref{fig:cusped}, is the unique solution up to conformal transformations.

 To guide this construction, let us first understand what is special about configurations like those in Figure \ref{fig:cusped}. This defect is formed by two straight half-lines meeting at the cusp (which we take to be sitting at the origin) and extending to infinity. This configuration is a union of complete orbits of the flow in the space of coordinates generated by the dilatiation operator $D = x^{\mu} \partial_{\mu}$. In fact, the two straight lines are orbits of the form $e^{s D} \cdot x$, for $x$ any point on the line, and $s\in \mathbb{R}$. The origin and infinity are fixed points of dilatations, and thus can also be regarded as complete orbits.

For a more general Hamiltonian, associated to a transformation $H=v^\mu(x) \partial_\mu$ of the conformal group, we would like to do the same: construct a cusped defect as a union of two $H$-orbits meeting at one (or more)  fixed points of the flow, with distinct tangents at the fixed points. Notice that the orbits need to be complete in order for the full configuration to be invariant under $H$.

\usetikzlibrary{decorations.markings}
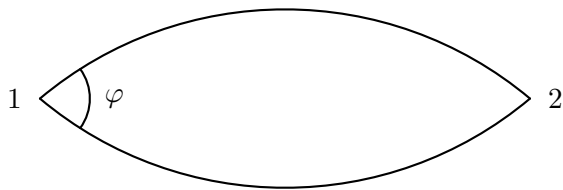
\begin{figure}[t]
  \centering

\begin{tikzpicture}[
    scale=1.2,
    line cap=round,
    line join=round
]


\def\a{2.7}

\def\phiang{80}

\def\anglerad{0.55}

\pgfmathsetmacro{\alpha}{(180-\phiang)/2}
\pgfmathsetmacro{\h}{\a*tan(\alpha)}
\pgfmathsetmacro{\Rc}{sqrt(\a*\a+\h*\h)}

\pgfmathsetmacro{\beta}{
    acos(\anglerad/(2*\Rc))-\alpha
}

\coordinate (C1) at (-\a,0);
\coordinate (C2) at ( \a,0);

\draw[black,thick]
    (C1)
    arc[
        start angle=180-\alpha,
        end angle=\alpha,
        radius=\Rc
    ];

\draw[black,thick]
    (C1)
    arc[
        start angle=180+\alpha,
        end angle=360-\alpha,
        radius=\Rc
    ];

\draw[black,thick]
    (C1) ++(-\beta:\anglerad)
    arc[
        start angle=-\beta,
        end angle=\beta,
        radius=\anglerad
    ];

\path (C1) ++(0.82,0) node {$\varphi$};

\node[left=3pt]  at (C1) {$1$};
\node[right=3pt] at (C2) {$2$};

\end{tikzpicture}

\caption{A defect with two cusps, invariant under a NS Hamiltonian, which we refer to as the almond.}
\label{fig:NSalmond}
\end{figure}

\begin{figure}[t]
\centering
\begin{tikzpicture}[
    scale=0.75,
    >=Latex,
    line cap=round,
    line join=round,
    flow/.style={
      red, thick,
      postaction={decorate},
      decoration={markings, mark=at position 0.55 with {\arrow{Latex}}}
    },
    slice/.style={
      black, thick,
      dash pattern=on 5pt off 4pt
    }
]

\pgfmathsetmacro{\a}{2.2} 

\coordinate (x1) at (-\a,0);
\coordinate (x2) at ( \a,0);


\draw[slice] (0,-3.0) -- (0,3.0);

\pgfmathsetmacro{\cA}{3.0}
\pgfmathsetmacro{\RA}{sqrt(\cA*\cA-\a*\a)}
\pgfmathsetmacro{\openA}{58}

\draw[slice]
  ({-\cA + \RA*cos(\openA)},{ \RA*sin(\openA)})
  arc[start angle=\openA, end angle=-\openA, radius=\RA];

\draw[slice]
  ({ \cA - \RA*cos(\openA)},{ \RA*sin(\openA)})
  arc[start angle=180-\openA, end angle=180+\openA, radius=\RA];

\pgfmathsetmacro{\cB}{4.2}
\pgfmathsetmacro{\RB}{sqrt(\cB*\cB-\a*\a)}
\pgfmathsetmacro{\openB}{52}

\draw[slice]
  ({-\cB + \RB*cos(\openB)},{ \RB*sin(\openB)})
  arc[start angle=\openB, end angle=-\openB, radius=\RB];

\draw[slice]
  ({ \cB - \RB*cos(\openB)},{ \RB*sin(\openB)})
  arc[start angle=180-\openB, end angle=180+\openB, radius=\RB];


\foreach \ang in {35,65}{
  \pgfmathsetmacro{\b}{\a/tan(\ang)}
  \pgfmathsetmacro{\R}{sqrt(\a*\a+\b*\b)}
  \draw[flow]
    plot[domain=-\a:\a, samples=150, variable=\x]
    (\x,{\b - sqrt(\R*\R - \x*\x)});
}

\draw[flow] (x1) -- (x2);

\foreach \ang in {35,65}{
  \pgfmathsetmacro{\b}{-\a/tan(\ang)}
  \pgfmathsetmacro{\R}{sqrt(\a*\a+\b*\b)}
  \draw[flow]
    plot[domain=-\a:\a, samples=150, variable=\x]
    (\x,{\b + sqrt(\R*\R - \x*\x)});
}

\draw[black, thick] (x1) circle (0.28);
\draw[black, thick] (x2) circle (0.28);

\fill[black] (x1) circle (1.5pt);
\fill[black] (x2) circle (1.5pt);

\node[left=6pt]  at (x1) {$w_1$};
\node[right=6pt] at (x2) {$w_2$};

\end{tikzpicture}
\caption{NS foliation with Hamiltonian $H$ as defined in \eqref{UDU}. The flow of $H$ (in red) takes us from $w_1$ to $w_2$. The foliation surfaces (in black) are constant-time slices and are circles orthogonal to the flow.}
\label{fig:NSq}
\end{figure}
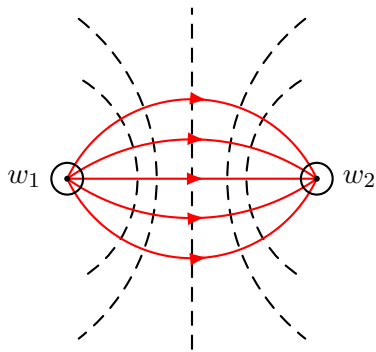

We first restrict our attention to configurations on a plane, we comment at the end about the general case. In two dimensions, $H$ should be the generator of M\"obius transformations. M\"obius transformations are classified into four types of conjugacy classes known as elliptic, hyperbolic, loxodromic, and parabolic 
(see, for instance, Chapter 3 of \cite{needham1997visual}). Each conjugacy class defines a  Hamiltonian up to the choice of a conformal frame.

\begin{figure}[t]
    \centering

\begin{tikzpicture}[
    x=0.75pt,
    y=0.75pt,
    yscale=-1,
    xscale=1,
    line cap=round,
    line join=round
]


\draw[draw opacity=0]
    (380.38,241.49)
    .. controls (373.55,254.35) and (347.71,263.82) .. (316.93,263.66)
    .. controls (284.29,263.49) and (257.32,252.55) .. (252.7,238.45)
    -- (317.08,234.42) -- cycle ;

\draw[black,thick]
    (380.38,241.49)
    .. controls (373.55,254.35) and (347.71,263.82) .. (316.93,263.66)
    .. controls (284.29,263.49) and (257.32,252.55) .. (252.7,238.45) ;

\draw[draw opacity=0]
    (252.83,238.91)
    .. controls (259.75,226.1) and (285.66,216.81) .. (316.44,217.19)
    .. controls (349.07,217.58) and (375.96,228.71) .. (380.48,242.84)
    -- (316.08,246.42) -- cycle ;

\draw[black,thick]
    (252.83,238.91)
    .. controls (259.75,226.1) and (285.66,216.81) .. (316.44,217.19)
    .. controls (349.07,217.58) and (375.96,228.71) .. (380.48,242.84) ;


\draw[draw opacity=0]
    (258.43,231.64)
    .. controls (261.15,233.81) and (262.86,237.12) .. (262.8,240.81)
    .. controls (262.75,243.74) and (261.59,246.39) .. (259.73,248.41)
    -- (250.93,240.62) -- cycle ;

\draw[black,thick]
    (258.43,231.64)
    .. controls (261.15,233.81) and (262.86,237.12) .. (262.8,240.81)
    .. controls (262.75,243.74) and (261.59,246.39) .. (259.73,248.41) ;


\draw[
    black,
    thick,
    dash pattern=on 5pt off 4pt
]
    (252.7,238.45)
    .. controls (292,144) and (380,205) .. (380.48,242.84) ;

\draw[
    black,
    thick,
    dash pattern=on 5pt off 4pt
]
    (253.7,237.45)
    .. controls (304.1,223.35) and (338.1,221.35) .. (381.48,241.84) ;


\draw (263.65,235.24)
    node [anchor=north west][inner sep=0.75pt]
    [font=\footnotesize]
    {$\varphi$};

\end{tikzpicture}

\caption{Almond with $\chi=0$ on the plane (solid) and an almond with $\chi\neq0$ on the same plane (dashed). Both almonds have the same cusp angle $\varphi$.}
\label{fig:nonzerochi}
\end{figure}
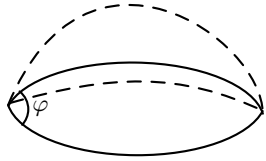

By studying their flow lines, it becomes apparent that only the hyperbolic maps have orbits that can form cusped defects (see  Appendix \ref{Conju}). One representative of the hyperbolic family is the dilatation generator $D$ itself, whose associated cusped defect is the one we have studied so far. Other representatives will include conjugations of $D$ (i.e., dilatations viewed in a conformally transformed frame):
\begin{equation}
H = U  D  U^{-1} ,
\label{UDU}
\end{equation}
with $U$ any conformal transformation. The most general $U \in PSL(2,\mathbb{C})$ which does not simply translate or rotate the cusp with straight branches is
 \begin{equation}\label{confmap}
    w=w_1 +\frac{(w_2-w_1)z}{z+a}~,
\end{equation}
where we introduced complex coordinates on the plane, and the transformation maps the cusps from $z=0$ and $z=\infty$ to $w_1$ and $w_2$. The complex parameter $a$ reflects the freedom of acting with a dilatation and a rotation in the $z$-complex plane, without altering the image of the cusps originally placed at the origin and infinity.
The associated cusped defect is formed by two arcs of circles flowing from one fixed point to the other.
 When  $\Im a=0$, the configuration is symmetric under reflection across the axis passing through the cusps---see Figure \ref{fig:NSalmond}. 
The phase of $a$ affects this shape by rotating each point along the dotted circles in Figure \ref{fig:NSq}---see Figure \ref{fig:nonzerochi}. We call an \emph{almond} any defect whose shape is obtained via the map \eqref{confmap}, independently of the phase of $a$. Since the generalization is straightforward, we mostly focus on the symmetric shape obtained by setting Im $a=0$ in the following.\footnote{In the context of  gauge theories,  cusped Wilson lines of this shape have often been called ``lens-shaped''. 
In fact, depending on the shape of the arcs, the dimension $d$---see below---and the reader's taste, the almonds might start to look more like bananas or potato wedges. We thank Nikolay Gromov for insightful correspondence on these matters. }

In higher dimensions, the result is the same: the most general cusped defect left invariant by a generator of the conformal algebra acting faithfully is a conformal transformation of the cusp with straight branches. The general shape is obtained by composing a rotation to the right of the map \eqref{confmap} (appropriately upgraded to a conformal transformation in $d$ dimensions). A non-planar shape is obtained when the rotation does not preserve the plane where the cusp with straight branches is contained, and where \eqref{confmap} acts. Since planes are mapped to spheres, the branches of non-planar almonds are arcs of circles fully contained in a 2-sphere, stretching between two---not necessarily antipodal---points.

\begin{figure}[t]
  \centering

\begin{tikzpicture}[
    x=0.75pt,
    y=0.75pt,
    yscale=-1,
    xscale=1,
    line cap=round,
    line join=round
]

\coordinate (Qc) at (150.5,159.3);
\def\QR{64}

\draw[black, thick, dash pattern=on 5pt off 4pt]
    (Qc) circle (\QR);

\draw[draw opacity=0]
    (182.22,152.35) .. controls (183.37,153.45) and (184.3,156.13) .. (184.5,159.31)
    .. controls (184.75,163.08) and (183.89,166.29) .. (182.53,167.03)
    -- (181.5,159.5) -- cycle ;

\draw[black, thick]
    (182.22,152.35) .. controls (183.37,153.45) and (184.3,156.13) .. (184.5,159.31)
    .. controls (184.75,163.08) and (183.89,166.29) .. (182.53,167.03) ;

\draw[black, thick, dash pattern=on 5pt off 4pt]
    (215.99,186.5) .. controls (208.94,186.72) and (199.25,182.85) .. (190.38,175.67)
    .. controls (184.11,170.59) and (179.37,164.75) .. (176.7,159.32) ;

\draw[black, thick, dash pattern=on 5pt off 4pt]
    (176.7,159.32) .. controls (179.21,152.98) and (185.42,145.81) .. (194.03,140.1)
    .. controls (200.26,135.96) and (206.72,133.27) .. (212.39,132.16) ;

\draw[black, thick]
    (207,134) .. controls (252,113) and (261,151) .. (287,122) ;

\draw[black, thick]
    (210,185) .. controls (248,194) and (276,164) .. (291,186) ;

\draw[black, thick]
    (207,134) .. controls (190,141) and (186,179) .. (210,185) ;

\draw (175,157.4) node [anchor=north west][inner sep=0.75pt]
    [font=\huge,color={rgb,255:red,0; green,0; blue,0},opacity=1] {$\dot{}$};

\draw (161,154.4) node [anchor=north west][inner sep=0.75pt]
    [font=\footnotesize] {$x_{0}$};

\draw (184.22,155.75) node [anchor=north west][inner sep=0.75pt]
    [font=\scriptsize] {$\varphi$};

\end{tikzpicture}

\caption{A smooth defect can be expanded in eigenstates of any NS Hamiltonian having orbits tangent to the defect on the quantization surface (which they pierce orthogonally). Given the smooth defect and quantization surface, there are several possible choices of the Hamiltonian. Here a particular choice is shown. The orbits touching the defect tangentially at the quantization surface are illustrated by dashed lines, and the smooth defect by the solid line.}
  \label{fig:sceg}
\end{figure}
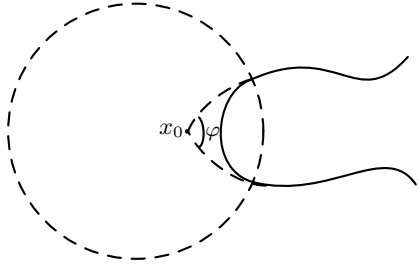

\subsubsection{The COE and NS Energy Eigenstates}
\label{sec:COENS}

The Hamiltonian $H$ defined by \eqref{UDU} with \eqref{confmap} generates time dilatation for states defined in the so called North-South pole (NS) quantization scheme. 
The quantization surfaces are nested, non-concentric circles whose centers shift continuously as time flows from one pole to the other, as shown in Figure \ref{fig:NSq}. Furthermore, since this scheme is related to radial quantization by a conformal transformation, the quantization circles are also orthogonal to the flow lines of the corresponding NS Hamiltonian. 

All the concepts we developed earlier can also be adapted to this scheme. In particular, the Hilbert space is constructed in the same way, changing the definition of the quantization surface. In fact, the conformal transformation $U$ acts as a map from the Hilbert space $\mathcal{H}(\varphi)$ (for the straight cusp at the origin) to the NS Hilbert space $\mathcal{H}^{\text{NS}}(\varphi)$. This map sends eigenstates of $D$ to eigenstates of $H$, which we denote as $|n \rangle_{\text{NS}}$, with the same eigenvalues:
\begin{equation}
    |n\rangle_{\text{NS}}= U |n\rangle_{\text{R}}~, \qquad
H | n\rangle_{\text{NS} }= \Gamma_n(\varphi) | n \rangle_{\text{NS}}~.
\end{equation}

As before, one can expand a section of a smooth contour into a linear combination of eigenstates of $H$, as depicted in Figure \ref{fig:sceg}. Explicitly, we write 
\begin{equation} \label{COENS}
     |\Psi_r\rangle_{\text{NS}}= \sum_n c^{\text{NS}}_n(\varphi, \vec{r})  |n\rangle_{\text{NS}}~ ,
\end{equation}

Above, $\varphi$ is the cusp angle 
and $\vec{r}$ schematically denotes a collection of parameters which define the section of the contour enclosed by the quantization surface.

As in Subsection \ref{subsec:sce}, let us consider again the example of an arc of circle of radius $r_\text{NS}$, which pierces the quantization circle orthogonally in two points. The COE in the NS quantization scheme \eqref{COENS} for this example can be obtained by means of a conformal transformation from the COE of \eqref{smoothope} given in radial quantization, and is expressed as

\begin{equation}\label{COEsimple} 
\Bigg| 
\begin{tikzpicture}[
    x=0.75pt,
    y=0.75pt,
    yscale=-1,
    xscale=1,
    baseline={(0,-5.54)},
    line cap=round,
    line join=round
]

\draw[draw opacity=0]
    (298.62,230.23)
    .. controls (309.8,227.77) and (318,219.64) .. (318,210)
    .. controls (318,200.04) and (309.25,191.7) .. (297.52,189.54)
    -- (291.5,210) -- cycle ;

\draw[black,thick]
    (298.62,230.23)
    .. controls (309.8,227.77) and (318,219.64) .. (318,210)
    .. controls (318,200.04) and (309.25,191.7) .. (297.52,189.54) ;

\end{tikzpicture} 
\Bigg\rangle_{\text{NS}}
=
\sum_{n=0} c_{n}(\varphi)
\left(r_\text{R}(r_\text{NS})\right)^{\Gamma_n}
|n\rangle_{\text{NS}} .
\end{equation} 
Here, $r_\text{R}$ is the image of $r_\text{NS}$ under the inverse of the map \eqref{confmap}. We set $w_1=0$ and $w_2=x_2$ real and positive. For simplicity, we also set  $\mathrm{Im}\, a=0$ there.
We further fix $|a|$ by demanding the image of the unit circle under \ref{confmap} to be the perpendicular bisector of the horizontal line between the origin and $x_2$. This choice of quantization surface is convenient because conjugation $\Theta$ is then just a reflection across this surface: 
\begin{equation}
    \Theta_{\text{NS}}(w) =
    x_2-\bar{w}~,
\end{equation}
where $\bar{w}$ denotes the complex conjugate of $w$.

Under the map \eqref{confmap}, for our specific choice of $w_1$ and $w_2$, the radii $r_\text{R}$ and $r_\text{NS}$ of the two smoothing circles are related by
\begin{equation} \label{rtrans}
    r_\text{R}=
\frac{
2\,\rho
}{
1-2\rho\csc\frac{\varphi}{2}
+
\sqrt{
1-4\rho\csc\frac{\varphi}{2}
+4\rho^2
}
},
\qquad
\rho\equiv\frac{r_\text{NS}}{x_2}~.
\end{equation}
In particular, at small $r_\text{NS}$, the COE \eqref{COEsimple} can be written as 
\begin{equation}\label{eq:nscoe}
\Bigg| 
\begin{tikzpicture}[
    x=0.75pt,
    y=0.75pt,
    yscale=-1,
    xscale=1,
    baseline={(0,-5.54)},
    line cap=round,
    line join=round
]

\draw[draw opacity=0]
    (298.62,230.23)
    .. controls (309.8,227.77) and (318,219.64) .. (318,210)
    .. controls (318,200.04) and (309.25,191.7) .. (297.52,189.54)
    -- (291.5,210) -- cycle ;

\draw[black,thick]
    (298.62,230.23)
    .. controls (309.8,227.77) and (318,219.64) .. (318,210)
    .. controls (318,200.04) and (309.25,191.7) .. (297.52,189.54) ;

\end{tikzpicture} 
\Bigg\rangle_{\text{NS}}
 \underset{r_{\text{NS}}\rightarrow 0}{=} \sum_{n=0} c^i_{n}(\varphi)\left( \frac{r_{\text{NS}}}{x_2}\right)^{\Gamma_n}|n\rangle_{\text{NS}}.
\end{equation}  \label{COEsimple2}

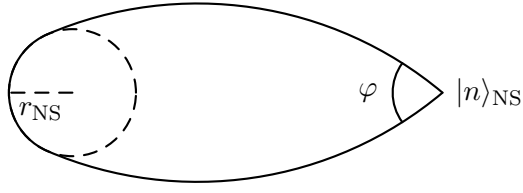
\begin{figure}[t]
  \centering

\begin{tikzpicture}[
    scale=1.2,
    line cap=round,
    line join=round
]


\def\a{2.7}
\def\phiang{80}
\def\anglerad{0.55}

\def\rprime{0.70}

\pgfmathsetmacro{\alpha}{(180-\phiang)/2}
\pgfmathsetmacro{\h}{\a*tan(\alpha)}
\pgfmathsetmacro{\Rc}{sqrt(\a*\a+\h*\h)}

\coordinate (C2) at (\a,0);

\pgfmathsetmacro{\d}{\Rc-\rprime}
\pgfmathsetmacro{\sx}{-sqrt(\d*\d-\h*\h)}
\coordinate (S) at (\sx,0);

\pgfmathsetmacro{\gamma}{atan(\h/(-\sx))}

\coordinate (Tup)   at ({\sx-\rprime*cos(\gamma)},{ \rprime*sin(\gamma)});
\coordinate (Tdown) at ({\sx-\rprime*cos(\gamma)},{-\rprime*sin(\gamma)});

\draw[black, thick]
    (Tup)
    arc[
        start angle={180-\gamma},
        end angle=\alpha,
        radius=\Rc
    ];

\draw[black, thick]
    (Tdown)
    arc[
        start angle={180+\gamma},
        end angle={360-\alpha},
        radius=\Rc
    ];

\draw[black, thick, dash pattern=on 5pt off 4pt]
    (S) circle (\rprime);

\draw[black, thick]
    (Tup)
    arc[
        start angle={180-\gamma},
        end angle={180+\gamma},
        radius=\rprime
    ];

\draw[black, thick, dash pattern=on 5pt off 4pt]
    (S) -- ++(180:\rprime);

\node[below] at ({\sx-0.5*\rprime},0) {$r_{\text{NS}}$};

\pgfmathsetmacro{\beta}{acos(\anglerad/(2*\Rc))-\alpha}

\draw[black, thick]
    (C2) ++({180-\beta}:\anglerad)
    arc[
        start angle={180-\beta},
        end angle={180+\beta},
        radius=\anglerad
    ];

\node at ({\a-0.82},0) {$\varphi$};

\node[right=1pt] at (C2) {$|n\rangle_{\text{NS}}$};

\end{tikzpicture}

\caption{Smoothed almond contour: the left cusp is regulated by a circular cap of radius $r_\text{NS}$, joined tangentially to the rest of the contour. The expectation value computes the coefficients of the COE \eqref{COEsimple}.}
\label{fig:smoothedalmond}
\end{figure}

The COE coefficients in \eqref{COEsimple} are computed by the conformally transformed ice cream contour of Figure \ref{fig:icecream}, shown in Figure \ref{fig:smoothedalmond}.
In Subsection \ref{sec:smooth}, we will compute the corresponding expectation value in an explicit example and check that the non-trivial function \eqref{rtrans} is reproduced.

\section{Cusp Operators Transform Like Primaries}\label{sec:generic-cusp-correlators}

In this section, we discuss correlation functions defined by properly renormalized cusped defects, with (potentially excited) scaling cusp operators prepared at the cusps. We prove a remarkable property: these correlators transform like those of primary local operators in a CFT (provided one of course keeps track also of the changing shape of the connecting arcs under the conformal map). 
This was first observed in the context of gauge theory, in the case of the ladders limit of Wilson lines in  $\mathcal{N}$=4 SYM~\cite{ladder}, for contours consisting of arcs of circles, and was further elaborated on in \cite{Cavaglia:2020hdb}. For the full $\mathcal{N}$=4 SYM theory and ground state cusps on the same type of piecewise-circular contour, it was proved using Ward identities in \cite{Dorn:2020meb}. 
Inspired by these results, here we aim to clarify these findings and present an argument valid for generic conformal defects. We concentrate on piecewise circular contours, offering some comments on more general contours at the end.

\subsection{Cusp Operators and their Correlation Functions}

\paragraph{Defining Cusp Correlation Functions. }

In the previous sections, we have defined scaling cusp operators, associated to cusps with circular branches.
 We can also give a natural definition of `cusp correlation functions'. Consider a contour with $m$ cusps, made by the arcs connecting them. 
The pair of arcs emanating from each cusp uniquely defines a family of quantization surfaces around the cusp. That is, the surfaces associated to the NS Hamiltonian that leaves the two arcs invariant. 
Then, we can define correlation functions where we prepare a state $ | k_i \rangle_{\text{NS}}$ in the appropriate Hilbert space around each cusp. Notice that we need to declare on which quantization surface the state is normalized to $1$.\footnote{In fact, for two quantization surfaces $\Sigma_1$, $\Sigma_2$ related by NS evolution as $\Sigma_2 = \lambda^{H}\circ \Sigma_1$, preparing a state $|\Psi \rangle$ on $\Sigma_1$ creates the state $\lambda^{-H} |\Psi \rangle$ on $\Sigma_2$. } The natural choice, inspired by creating a local excitation at the cusp points, is to prepare state $ | k_i \rangle$ on a quantization surface which is, roughly, at distance $\epsilon$ from the  cusp.\footnote{The NS quantization surface is generically not a sphere centered at the cusp point. We can define, for instance, $\epsilon$ as the minimal distance between the surface and the cusp point. For small $\epsilon \sim 0$, this becomes the radius. } This defines a correlator with a small-$\epsilon$ behavior  controlled by the cusp anomalous dimensions of the states at the cusps (as we will see). Then, we send $\epsilon \to 0$ after dividing out the appropriate power of $\epsilon$. 
The resulting finite quantity is what we refer to as the correlation function of cusp operators.

Notice that we could introduce the $\epsilon$ regularized quantity in many ways, modifying the original cusped contour on a scale $\epsilon$ around each cusp. We will make a concrete example below. As will be demonstrated, the result is universal and does not depend on the details of the scheme.

\paragraph{Covariance of Correlation Functions. }
Correlation functions constructed in this way exhibit covariance under  conformal transformations, at least in the case of a contour made from a sequence of arcs, which is what we will focus on.
 
In this case, under a conformal map $M: x \rightarrow x'$, the correlation function transforms as 

\begin{equation} \label{trans}
\left\langle
\begin{tikzpicture}[
    scale=0.85,
    baseline=-0.5ex,
    line cap=round,
    line join=miter
]

\coordinate (x1) at (0,0);
\coordinate (x2) at (0.90,1.25);
\coordinate (x3) at (2.50,1.15);
\coordinate (xn) at (1.45,-1.20);

\draw[black,thick]
    (x1)
    arc[
        start angle=297.9358,
        delta angle=52.6205,
        radius=1.73757
    ]
    arc[
        start angle=241.1107,
        delta angle=50.6259,
        radius=1.87472
    ]
    arc[
        start angle=131.3756,
        delta angle=49.0978,
        radius=3.09759
    ]
    arc[
        start angle=20.6349,
        delta angle=59.5088,
        radius=1.89625
    ];

\node[left=3pt] at (x1) {$x_1$};
\node[above left=2pt] at (x2) {$x_2$};
\node[above=2pt] at (x3) {$x_3$};
\node[below=3pt] at (xn) {$x_n$};

\node at (1.99,0.10) [rotate=-110] {$\cdots$};

\end{tikzpicture}
\right\rangle
=
\left\langle
\begin{tikzpicture}[
    scale=0.85,
    baseline=-0.5ex,
    line cap=round,
    line join=miter
]

\begin{scope}[
    rotate around={20:(1.25,0)},
    transform shape
]

\coordinate (xp1) at (0,0);
\coordinate (xp2) at (0.90,1.25);
\coordinate (xp3) at (2.50,1.15);
\coordinate (xpn) at (1.45,-1.20);

\draw[black,thick]
    (xp1)
    arc[
        start angle=297.9358,
        delta angle=52.6205,
        radius=1.73757
    ]
    arc[
        start angle=241.1107,
        delta angle=50.6259,
        radius=1.87472
    ]
    arc[
        start angle=131.3756,
        delta angle=49.0978,
        radius=3.09759
    ]
    arc[
        start angle=20.6349,
        delta angle=59.5088,
        radius=1.89625
    ];

\node[left=3pt] at (xp1) [rotate=-15]{$x'_1$};
\node[above left=2pt] at (xp2)  [rotate=-15]{$x'_2$};
\node[above=2pt] at (xp3)  [rotate=-15]{$x'_3$};
\node[below=3pt] at (xpn)  [rotate=-15]{$x'_n$};

\node at (1.99,0.10) [rotate=-110] {$\cdots$};

\end{scope}

\end{tikzpicture}
\right\rangle
\prod_{i=1}^{n}
\abs{\frac{\partial x'_i}{\partial x_i}}^{
\Gamma_{k_i}(\varphi_i)/d
}\, 
\end{equation}
where $\Gamma_{k_i}(\varphi_i)$ is the dimension of the state we prepare at each cusp. 
Notice that the transformation law is always the one above, even when involving excited states. Namely, the rescaling factors are the same ones that we would obtain for conformal \emph{primaries} in the case of local operators. 

In the next two sections we present a simple argument establishing \eqref{trans}, in the case of a piecewise arc-like contour. We will furthermore address briefly the case where the cusped contour is more general.

\subsubsection{Correlators of Leading Cusp Operators}
We start by considering cusp operators in the ground state. 
We will give a simple proof based on a regularization scheme where the cusped contour is softened into a $C^1$ curve. 

To prove the covariance, we will use the following property: if $ \mathcal{C}_1$ and $\mathcal{C}_2$ are two $C^1$ defect contours related by a conformal transformation, then 
    \begin{equation} \label{transmooth}
        \left \langle  W[\mathcal{C}_1]\right\rangle = \left\langle W[\mathcal{C}_2 ] \right\rangle,
\end{equation}  
where $W[\mathcal{C}]$ represents an extended operator defined on the  contour $\mathcal{C}$. This is a standard assumption in defect-CFT literature, dating back to \cite{McAvity:1993ue,McAvity:1995zd}. We expect it to be valid when the extended operators are properly renormalized, in the sense that coincident-point UV singularities have been taken care of. 

\paragraph{Regularizing and Renormalizing. }  
Let us specify a regularization scheme: for every cusp, we can choose a sufficiently small quantization surface, and replace the corresponding cusp with a piece of circle inside the surface, keeping the full contour $C^1$. This is precisely the `rounded cusp' configuration we have considered in Sections \ref{subsec:sce} and \ref{sec:COENS}. 

Let us denote the original cusped contour by $\mathcal{C}(x_1,...,x_n)$, and its regularization with rounded-off cusps by $\mathcal{C}(\{x_1,\epsilon_1\},..., \{x_n,\epsilon_n\}) $, where $\epsilon_i$ denotes the rounding radius at the $i$-th cusp. The COE  \eqref{COEsimple} now shows that as we take these cutoffs to zero $\epsilon_{i} \rightarrow 0^+$, we have \begin{equation} \label{asu2}
        \langle W[\mathcal{C}(\{x_1,\epsilon_1\},..., \{x_n,\epsilon_n\}) ] \rangle \sim \epsilon_1^{\Gamma_0(\varphi_1)} \cdots \epsilon_n^{\Gamma_0(\varphi_n)} \; \texttt{finite} + \texttt{subleading},
    \end{equation} 
which comes from using the leading term of the COE around each cusp.
The finite part above does not depend on the rate at which the various cutoffs are taken to zero (so we can take them all equal for simplicity), and defines our renormalized correlator:
\begin{equation} \label{renloop}
    \langle W_{\text{ren}}[\mathcal{C}(x_1,...,x_n)] \rangle:= \lim_{\epsilon \to 0^+} \langle W[ \mathcal{C}(\{x_1,\epsilon\},..., \{x_n,\epsilon\}) ] \rangle \; 
 \epsilon^{-\sum_{i=1}^n \Gamma_0(\varphi_i)}.
\end{equation}

\paragraph{Proof of Covariance. }
Consider a conformal transformation $M: x \rightarrow x'$. This sends the cusped contour $\mathcal{C}$ into a new cusped contour $\mathcal{C}'$, where the cusps have the same angles $\varphi_i$. The map also sends the regularized version of the contour $\mathcal{C}$ into a similarly regularized version of the new contour $\mathcal{C}'$: the only difference is that the regularization circles now have different radii $\epsilon_i'$. 
 Since the Jacobian of the map $M$ gives a local dilatation factor, in the limit $\epsilon_i \rightarrow 0^+$ they are related as
\begin{equation} \label{radit}
    \epsilon'_i=\epsilon_i \abs{\frac{\partial x'_i}{\partial x_i}}^{\frac{1}{d}} + \sum_i\mathcal{O}(\epsilon_i^2).
\end{equation}

On the other hand, the regularized contours are $C^1$ curves, and thus should be conformally invariant by the assumption made earlier:
\begin{equation} \label{last}
    \langle W[\mathcal{C} (\{x_1,\epsilon_1\},...,\{x_n,\epsilon_n\}) ]\rangle =\langle W[\mathcal{C}'(\{x'_1,\epsilon'_1\},...,\{x'_n,\epsilon'_n\})] \rangle.
\end{equation}
Plugging \eqref{last} into the RHS of \eqref{renloop} before taking the limit, we then obtain
\begin{eqnarray}
     \langle W_{\text{ren}}[\mathcal{C}(x_1,...,x_n)]\rangle &=& \lim_{\epsilon \to 0} \langle W[\mathcal{C}'(\{x'_1,\epsilon'_1\},...,\{x'_n,\epsilon'_n\})] \rangle \epsilon^{-\sum_{i=1}^n \Gamma_0(\varphi_i)}. 
\end{eqnarray}
Comparing the RHS with the definition of the renormalized correlator for the transformed contour $\mathcal{C}'$, and recalling \eqref{radit}, we arrive precisely at the transformation rule \eqref{trans}:
\begin{equation} \label{trans2}
    \langle W_{\text{ren}}[\mathcal{C}(x_1,...,x_n)] \rangle= \langle W_{\text{ren}}[\mathcal{C}'(x'_1,...,x'_n) ] \rangle \prod_{i=1}^n \abs{\frac{\partial x'_i}{\partial x_i}}^{\Gamma_0(\varphi_i)/d}. 
\end{equation}

\paragraph{Comments on Universality.} Above, we have used a specific regularization scheme. This is related to the idea introduced in the previous section of preparing the ground state $| 0 \rangle_{\text{NS}}$ at scale $\epsilon$ around each cusp. In particular, for small $\epsilon$, these two prescriptions differ only by an overall frame-independent normalization factor. 
 
To see this, let us consider the contour regularized as  described above, rounded off with a circle of radius $\epsilon$ around each cusp at a given point $x_i$. This procedure defines a state on the quantization surface $\Sigma_{\epsilon_i}$, defined at the intersection between the arcs and the `rounding' circle.  
Comparison with \eqref{smoothstatea} shows that the state defined on this surface is
\begin{equation}\label{eq:state}
\sum_{k=0}^{\infty} \tilde{c}_k(\varphi_i) | k \rangle_{\text{NS}} ,   
 \end{equation}
where the state does not depend on $\epsilon_i$, but is defined on a surface that varies with the cutoff. 
 The state defined on any \emph{fixed} quantization surface $\Sigma$ is obtained  acting with the appropriate evolution operator, in particular if $\Sigma = \lambda^{H} (\Sigma_{\epsilon_i})$ the state evolves with $\lambda^{-H}$. Since 
$\lambda \propto \epsilon_i^{-1}$ for  fixed $\Sigma$, the  leading behavior will be $\epsilon_i^{\Gamma_0(\varphi_i)}$, determined by 
evolution of the ground state in  
\eqref{eq:state} (where we assume $\tilde{c}_0(\varphi_i) \neq 0$). 
This means that multiplicative  renormalization of the correlation function projects to the ground state component, washing away the regularization details associated to the presence of excited states in \eqref{eq:state}.  
 
Thus, we see that the details of the scheme we are using simply amount to  a rescaling of cusp operators. A first rescaling comes from the factor $\tilde{c}_0(\varphi_i)$ in \eqref{eq:state}. Further rescaling factors may come from the details of how we define the cutoffs $\epsilon_i$. For instance, in the argument given above we used the cutoff $\epsilon_i^{\text{rounding}}$, defined as the radius of curvature of the rounding circle. We could have used as an alternative cutoff the minimal distance between the quantization surface attached to this circle and the cusp point, $\epsilon_i^{\text{distance}}$. In the limit where the cutoffs are taken to be small these two prescriptions differ only by a (frame independent) constant. Concretely, $\epsilon_i^{\text{distance}} \sim \epsilon_i^{\text{rounding}} \; \cot(\varphi_i/2)$. 
 
Therefore, scheme dependence amounts to simply changing the normalization of the cusp operators in a frame-independent way. Hence, scheme dependence does not affect the covariance properties, which, as we have seen, come from the way a cutoff transforms between two frames, cf. \eqref{radit}.  

Finally, remember that we could have repeated the same discussion for a different smoothing of the cusp (different from a circle). In a good regularization scheme the smoothing curve converges to a fixed shape (up to rescaling) in the small cutoff limit. The shape of the curve is then fixed by dimensionless moduli (again, up to rescaling). Then, choosing a scheme characterized by such a smooth curve would have simply changed the form of the coefficients $\tilde{c}_n$, which would depend on these moduli. Provided the scheme is adopted consistently, we always obtain the same set of correlation functions, with the scheme dependence boiling down to an overall  normalization.

\subsubsection{Excited States}
\label{sec:transfExcited}
Let us now consider correlators involving excited states. In this case, we simply want to project to a possibly generic excited state 
 of the relevant NS Hamiltonian 
 on the quantization surface surrounding each cusp.

Let us describe a concrete, very precise regularization scheme. We will then argue again that the result is universal. Starting from a cusped contour $\mathcal{C}$ with cusps at points $x_i$, let $$W[\mathcal{C}(\{x_1,\epsilon_1, k_1\},..., \{x_n,\epsilon_n, k_n\}) ]$$ denote the regularized configuration where on the quantization surface $\Sigma_{i,\epsilon_i}$ around the cusp at $x_i$ the state is prepared as exactly $| k_i \rangle_{\text{NS}}$. In the exterior of the $n$ quantization surfaces, the defect lines are left unchanged. 

The expectation value of this configuration corresponds to the path integral computed in this exterior region, in presence of the defect lines, with boundary conditions specified by the assigned states. Under a conformal transformation, this expectation value transforms as
\begin{equation}
\langle W[\mathcal{C}(\{x_1,\epsilon_1, k_1\},..., \{x_n,\epsilon_n, k_n\}) ] \rangle = \langle W[\mathcal{C}'(\{x_1',\epsilon_1', k_1\},..., \{x_n',\epsilon_n', k_n\}) ] \rangle , \label{eq:invarianceexc}
\end{equation}
where on the RHS we transform the shape of the arcs, as well as the quantization surfaces, and we prepare the  excited states  using the NS Hamiltonian transformed to the new frame. 
 Notice the importance of defining the states using the NS Hamiltonian adapted to the arcs of the contour around each cusp: this provides a universal description of the regularization such that the above conformal invariance of the regularized correlators is guaranteed.

The above equation \eqref{eq:invarianceexc} should be seen as analogous to \eqref{last} in our previous argument. Now the argument for the covariance property of such correlators is the same as before. Generalizing the previous discussion, we see that the regularized correlator will now behave, for small cutoffs, as
\begin{equation} \label{asu3}
        \langle W[\mathcal{C}(\{x_1,\epsilon_1, k_1\},..., \{x_n,\epsilon_n, k_n\}) ] \rangle \sim \epsilon_1^{\Gamma_{k_1}(\varphi_1)} \cdots \epsilon_n^{\Gamma_{k_n}(\varphi_n)} \; \texttt{finite} + \texttt{subleading},
   \end{equation} 
    where the scaling is, again, determined by the state with the lowest cusp dimension present on the quantization surfaces around each cusp. The finite part in \eqref{asu3} defines the renormalized correlator of excited cusps. Again, keeping track of the transformation of the scales $\epsilon_i \rightarrow \epsilon_i'$ under the conformal transformation, i.e.  \eqref{radit}, we get the covariance property: 
\begin{equation} \label{trans2b}
    \langle W_{\text{ren}}[\mathcal{C}(x_1,...,x_n)] \rangle= \langle W_{\text{ren}}[\mathcal{C}'(x'_1,...,x'_n) ] \rangle \prod_{i=1}^n  \abs{\frac{\partial x'_i}{\partial x_i}}^{\Gamma_{k_i}(\varphi_i)/d}. 
\end{equation}
Just as before, notice that we would obtain exactly the same correlation functions adopting   different kinds of regularization procedures\footnote{This includes schemes where the state on $\Sigma_{\epsilon_i}$ is prepared by a $C^1$ regularization of the cusp: it is not difficult to see that such smooth states can be combined to obtain a combination like \eqref{eq:state} but starting from a certain excited state.}, which should have the following crucial feature in common: the state defined on  the quantization surface traced at scale $\epsilon_i$ around the cusp at $x_i$ should be, at small enough $\epsilon_i$, a linear combination of $| k_i \rangle_{\text{NS}}$ defined with the NS Hamiltonian adapted to the arcs, and states with higher dimensions, all with coefficients which are $O(1)$ in $\epsilon$. Again, the presence of higher-dimensional states is completely irrelevant, as it will be the coefficients of the lowest-dimensional operators around each cusp that determine the \texttt{finite} part in \eqref{asu3}. Thus, all such schemes determine the same set of covariant correlation functions.

Examples of calculations done to construct excited states in a concrete regularization scheme may be found in \cite{ladder}, and in Section  \ref{sec:examples} below, as well as Appendix \ref{app:chopped}, in the case of the magnetic (pinning) defect.

\subsubsection{More General Contours}

Les us now consider a cusped contour which is not necessarily made of arcs and see if renormalized correlation functions can be constructed in a similar way.

A case where the situation simplifies is when the lines forming the contour are not necessarily arcs all the way, but are shaped like arcs in a finite neighbourhood of each cusp. Then, clearly, we can repeat the construction presented above, since the presence of a precise NS quantization scheme close to each cusp allows us to repeat our argument. In this case, we would have correlation functions that still transform like \eqref{trans2} under conformal transformations. 

Finally, Let us consider a contour made of generic smooth lines  forming a number of cusps. Here, by `cusp' we mean, precisely, a point where two smooth lines meet with well defined tangent vectors, forming a finite angle. We call the angles of such cusps $\varphi_i |_{\text{cusp}}$. 
The key difference with the previous cases is that it is now more complicated to give a universal recipe to prepare states around the cusps. 
Each cusp is now surrounded by two  generic smooth defect lines. Suppose that we can  still construct uniquely 
 (at least in a finite neighbourhood of the cusp) a family of quantization surfaces, surrounding the cusp and converging to it.\footnote{For example, we can consider a point on one of the two lines and trace a sphere orthogonal to the tangent vector at this point. If the defect lines are not too wild, in a finite neighbourhood of the cusp there will be a unique value of the radius such that the sphere crosses a point on the other line orthogonally. We are restricting the discussion to cases where this procedure works. } On each of these surfaces, we can still define a Hilbert space picture. However, we no longer have a canonical choice, which before was the NS Hamiltonian leaving the arcs invariant. This choice   ensured that on each surface lives a isomorphic Hilbert space. 
 Now, we just have a sequence of spheres with two punctures: each of them could be viewed as a `constant-time' slice for a different (and not uniquely defined) NS Hamiltonian. In general, we can \emph{choose} a sequence of Hamiltonians depending on the cutoff scale $\epsilon_i$ -- each will  
define a defect Hilbert space on each surface (where now notice that even the angle  characterizing the Hilbert space becomes a function of the cutoff). 
 This dependence on the cutoff is not completely arbitrary, because to quadratic order the lines must approach arcs. Therefore the geometry forces us to have
 \begin{equation}
\varphi_i(\epsilon_i) \sim \varphi_i |_{\text{cusp} }+ \mathcal{O}(\epsilon_i), \;\;\;  H_i(\epsilon_i) \sim H_i^{\text{NS}} |_{\text{arcs}}+ \mathcal{O}(\epsilon_i).
 \end{equation}
That is, for small cutoffs the Hamiltonians will converge to the one determined by the shape of arcs approximating the two lines to second order, and correspondingly the angle will converge to the angle at the cusp.

A question that arises given these conditions is whether we may still define a renormalized correlation function by demanding, for example, that the state defined on the quantization surface at scale $\epsilon$ is the ground state of the corresponding Hamiltonian. While we leave a definitive answer to future work, we are doubtful that this can be done in a way that the resulting correlator would still be conformally covariant. The reason is that, even if we expect that the regularized correlator would still have a divergence of the form \eqref{asu2}, ruled by the cusp dimensions that determined purely by the opening angle~\footnote{While we do not have a rigorous argument, this is generally expected in gauge theory, as the leading UV singularities should be determined only by an infinitesimal neighborhood of the cusp. 
A check at strong coupling that the rate of divergence depends only on the angle for a cusp formed by smooth lines in $\mathcal{N}$=4 SYM was performed in \cite{Dorn:2015bfa}. However, naturally the finite part in front of this divergence is much more difficult to define unambiguously. 
} the \texttt{finite part} could now become dependent on regularization details (including the precise sequence of Hamiltonians chosen) 
 in a way that potentially spoils the covariance transformation properties of the correlator. 

\subsection{Piecewise Circular Correlators}\label{sec:ncusps} 
In this section, we review some important consequences of the above covariance properties for the case of cusped contours made of arcs. 

 Notice that the consequences of the covariance property are not immediately the same as for the correlation functions of primary operators in a CFT, due to the fact that in \eqref{trans} one should keep track also of the transformation of the shape of the arcs under a conformal map. 
 However, there is a notable simplification when the piecewise circular contour lies on a 2D plane. In fact, in this case the shape of the $n$ arcs is  very constrained once we specify the points $x_i$, $i=1,\dots, n$ where the cusps sit, and the cusp angles $\varphi_i$, $i=1,\dots, n$. In particular, in the case of 2 and 3-point functions made of planar arcs, covariance becomes as constraining as in the case of correlators of local primary operators in 2D CFT, with the additional feature that dynamical data need to depend on the cusp angles.

 In the rest of this section, we discuss in turn the case of 2-, 3-, and higher-point functions of cusps, mostly focusing on the case of configurations on the plane. The contents of this section are mostly a review of results from \cite{ladder,Cavaglia:2020hdb,Dorn:2020meb,Dorn:2020vzj}. The case of non-coplanar piecewise circular defects is more complicated by the presence of additional conformal invariants describing the shape of the arcs, on which correlation functions will depend. While we do not discuss such cases in detail, the relevant conformal kinematics were studied thoroughly in \cite{Dorn:2020vzj,Dorn:2023qbj}.

\subsubsection{Two-Cusp Functions for Coplanar Arcs} \label{subsec:twocuspf}
In Section \ref{subsec:gener}, we introduced `almond' configurations, which are obtained as a conformal transformation of the cusp with straight branches in Figure \ref{fig:punctured}.

When the arcs lie on a plane, the almond has a shape which can be conveniently parametrized in complex coordinates~\cite{ladder}, by exploiting the map \eqref{confmap}. We denote the location of the cusps as $x_i=(\text{Re}(w_i), \text{Im}(w_i),0,\dots,0)$, ($i=1,2$), and parameterize the two arcs as
\begin{equation}\label{xpm}
     x_{\pm}(s)=(\text{Re}(\zeta_\pm(s)), \text{Im} (\zeta_\pm(s)),0,0),
\end{equation} 
where
\begin{equation} \label{zeta}    \zeta_\pm(s)=w_1+\frac{w_2-w_1}{1+ e ^{\mp s +i(\chi \mp \varphi)/2}}, \qquad s\in (-\infty,\infty).
\end{equation}
Here, $\chi$ labels a one-parameter family of almonds with the same cusp angle $\varphi$ and cusp points (see Figure \ref{fig:nonzerochi}). The relation between the parameters in \eqref{zeta} and the ones in \eqref{confmap} is written in Appendix \ref{sec:conf} for convenience. In particular, $\chi$ is linearly related to the phase of $a$ in \eqref{confmap} and can be changed via a rotation.
Since rotations have have unit determinant, the covariance property shows that the expectation value of an almond on the plane cannot depend on $\chi$.

Thus, the planar two-cusp function can only depend on the cusp points and the cusp scaling dimensions through the covariance property, and the standard arguments give us the usual kinematical dependence:
\begin{equation}\label{eq:cusp2pt}
\langle W[ \mathcal{C}(x_1, x_2)] \rangle \propto \frac{\delta_{ij}}{|x_1 - x_2|^{ 2 \; \Gamma_i(\varphi)}},
\end{equation}
where the (potentially excited) states $i$ and $j$ are considered at the cusps.

\begin{figure}[t]
\centering

\begin{tikzpicture}[
    scale=1.35,
    contour/.style={
        black,
        thick,
        line cap=round,
        line join=round
    },
    angle mark/.style={
        draw,
        line width=0.4pt
    }
]

\coordinate (z1) at (0,0);
\coordinate (z2) at (1.5,2);
\coordinate (z3) at (3,0);

\draw[
    contour,
    name path=curve12
]
    (z1)
    .. controls (0.10,0.90) and (0.70,1.80) ..
    (z2);

\draw[
    contour,
    name path=curve23
]
    (z2)
    .. controls (2.00,1.80) and (2.90,0.80) ..
    (z3);

\draw[
    contour,
    name path=curve31
]
    (z3)
    .. controls (2.25,-0.55) and (0.80,-0.40) ..
    (z1);

\path[name path=circle1] (z1) circle[radius=0.42];
\path[name path=circle2] (z2) circle[radius=0.34];
\path[name path=circle3] (z3) circle[radius=0.42];

\path[
    name intersections={of=curve12 and circle1, by=p12}
];
\path[
    name intersections={of=curve31 and circle1, by=p13}
];

\path[
    name intersections={of=curve12 and circle2, by=p21}
];
\path[
    name intersections={of=curve23 and circle2, by=p23}
];

\path[
    name intersections={of=curve23 and circle3, by=p32}
];
\path[
    name intersections={of=curve31 and circle3, by=p31}
];

\pic[
    angle mark,
    angle radius=4.2mm
] {angle=p13--z1--p12};

\node[font=\footnotesize]
    at ($(z1)+(0.56,0.29)$)
    {$\varphi_1$};

\pic[
    angle mark,
    angle radius=3.4mm
] {angle=p21--z2--p23};

\node[font=\footnotesize]
    at ($(z2)+(0,-0.55)$)
    {$\varphi_2$};

\pic[
    angle mark,
    angle radius=4.2mm
] {angle=p32--z3--p31};

\node[font=\footnotesize]
    at ($(z3)+(-0.57,0.27)$)
    {$\varphi_3$};

\node[left=3pt]  at (z1) {$1$};
\node[above=3pt] at (z2) {$2$};
\node[right=3pt] at (z3) {$3$};

\end{tikzpicture}

\caption{Three-cusp contour.}
\label{fig:three}
\end{figure}
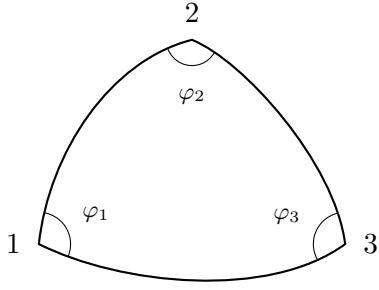

\subsubsection{Three-Cusp Functions for Coplanar Arcs and the Cusp-Cusp OPE}\label{sec:cuspcuspope}
The simplest 3-cusp function is made of a circular triangle, formed by the intersection of three circles, as shown in Figure \ref{fig:three}, where we now have three cusp angles $\varphi_i$ for $i=1,2,3$. 
An explicit parametrization of this configuration is given in \cite{ladder}. The key feature is that the shape of the arcs is now \emph{completely fixed} once we specify the cusp angles and the points $x_1$, $x_2$, $x_3$ where the cusps sit. 

Then, we can use a unique conformal transformation to map any such 3-cusp configuration to a reference one with cusps at three chosen points. The covariance transformation rule then fixes completely the kinematics of this correlator, which is the same as for local primary operators:
\begin{equation}\label{eq:cusp3pt}
\langle W[ \mathcal{C}(x_1, x_2, x_3)] \rangle  =    \frac{C_{n_1 n_2  n_3}(\varphi_1, \varphi_2, \varphi_3 )}{x_{12}^{\Gamma_{123}}x_{23}^{\Gamma_{231}} x_{31}^{\Gamma_{312}}} ,
\end{equation}
with
\begin{equation}
   \Gamma_{ijk} \equiv \Gamma_{n_i}(\varphi_i)+\Gamma_{n_j}(\varphi_j) -\Gamma_{n_k}(\varphi_k) ,
\end{equation}
where the (potentially excited) state $n_i$ is considered at the $i$-th cusp. Above, the coefficient $ C_{n_1 n_2  n_3}(\varphi_1, \varphi_2, \varphi_3 )$ is uniquely defined if we normalize canonically the 2-point functions. 

These  coefficients are the COE coefficients corresponding to the expansion of two cusps in the Hilbert space in terms of a third cusp (as in Figure \ref{fig:cuspcuspope0}). To draw an explicit link with the Hilbert space picture, consider the defect $W[ \{x_1, \varphi_1 , n_1 \} , \{ x_2, \varphi_2 , n_2 \} ]$, defined by three planar arcs forming two cusps at which the operators $n_i$ $(i=1,2$) are prepared. 
Then, we choose a quantization surface associated to the NS Hamiltonian
 leaving the first and third arc invariant. 
The state created by the defect on this surface is
\begin{equation}\label{eq:cuspcuspope}
{\Big |} W[ \{x_1, \varphi_1 , n_1\} , \{ x_2, \varphi_2 , n_2 \} ]  {\Big\rangle }=  \sum_n  C_{n_1 n_2  n}(\varphi_1, \varphi_2, \varphi_0 ) \times  \frac{ x_{0 \bar{0} }^{2 \Gamma_n(\varphi_0)}}{x_{12}^{\Gamma_{120}}x_{2\bar{0}}^{\Gamma_{201}} x_{\bar{0}1}^{\Gamma_{012}}} \times  W[ \{x_0, \varphi_0 , n \} ]  {\Big\rangle },
 \end{equation}
 where the points $x_0$ are cusp points of the almond defined by the first and third arc ($x_0$ inside the quantization surface, and $x_{\bar{0}}$ outside), and the basis of states chosen on the RHS are the ones created by local cusp operators at $x_0$ with canonical 2-point function. 
 
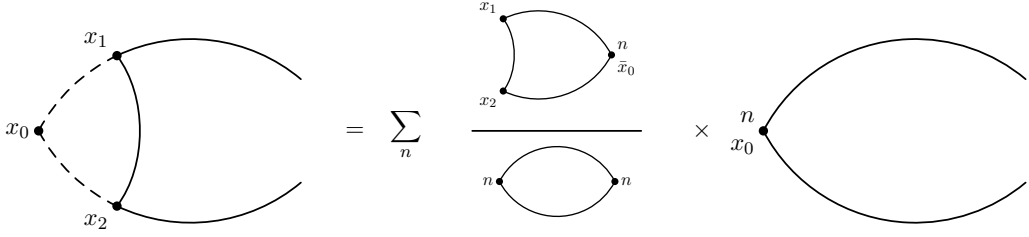
\begin{figure}[t]
\centering
\resizebox{0.90\linewidth}{!}{%
\begin{tikzpicture}[
    line cap=round,
    line join=round,
    contour/.style={black,thick},
    small contour/.style={black,semithick},
    aux/.style={black,thick,dash pattern=on 5pt off 4pt}
]

\def\almondradius{2.7112}
\def\upperstart{152.294}
\def\uppersplit{115.400}
\def\upperend{50.000}
\def\upperfullend{27.706}
\def\lowerstart{-152.294}
\def\lowersplit{-115.400}
\def\lowerend{-50.000}
\def\lowerfullend{-27.706}


\coordinate (A0) at (-2.4,0);

\draw[aux]
    (A0) arc[start angle=\upperstart,
             end angle=\uppersplit,
             radius=\almondradius]
    coordinate (A1);
\draw[contour]
    (A1) arc[start angle=\uppersplit,
             end angle=\upperend,
             radius=\almondradius];

\draw[aux]
    (A0) arc[start angle=\lowerstart,
             end angle=\lowersplit,
             radius=\almondradius]
    coordinate (A2);
\draw[contour]
    (A2) arc[start angle=\lowersplit,
             end angle=\lowerend,
             radius=\almondradius];

\draw[contour]
    (A1) .. controls (-0.68,0.55) and (-0.68,-0.55) .. (A2);

\foreach \p in {A0,A1,A2}
    \fill (\p) circle (2.1pt);

\node[left]       at (A0) {$x_0$};
\node[above left] at (A1) {$x_1$};
\node[below left] at (A2) {$x_2$};


\node at (2.60,0) {$=$};
\node at (3.40,-0.05) {$\displaystyle\sum_n$};


\def\numeratorscale{0.48}
\pgfmathsetmacro{\numeratorradius}{\almondradius*\numeratorscale}

\coordinate (Nbar0) at (6.65,1.20);

\draw[small contour]
    (Nbar0) arc[start angle=\upperfullend,
             end angle=\uppersplit,
             radius=\numeratorradius]
    coordinate (N1);
\draw[small contour]
    (Nbar0) arc[start angle=\lowerfullend,
             end angle=\lowersplit,
             radius=\numeratorradius]
    coordinate (N2);

\draw[small contour]
    (N1) .. controls (5.172,1.464) and (5.172,0.936) .. (N2);

\foreach \p in {Nbar0,N1,N2}
    \fill (\p) circle (1.6pt);

\node[above left,scale=0.72] at (N1) {$x_1$};
\node[below left,scale=0.72] at (N2) {$x_2$};
\node[above right,scale=0.72] at (Nbar0) {$n$};
\node[below right,scale=0.72] at (Nbar0) {$\bar{x}_0$};

\draw[black,line width=0.8pt] (4.45,0)--(7.12,0);

\coordinate (D0) at (4.88,-0.80);
\def\denominatorscale{0.38}
\pgfmathsetmacro{\denominatorradius}{\almondradius*\denominatorscale}

\draw[small contour]
    (D0) arc[start angle=\upperstart,
             end angle=\upperfullend,
             radius=\denominatorradius]
    coordinate (D1);
\draw[small contour]
    (D0) arc[start angle=\lowerstart,
             end angle=\lowerfullend,
             radius=\denominatorradius];

\fill (D0) circle (1.6pt);
\fill (D1) circle (1.6pt);
\node[left,scale=0.72]  at (D0) {$n$};
\node[right,scale=0.72] at (D1) {$n$};

\node at (8.08,0) {$\times$};


\begin{scope}[shift={(11.45,0)}]

\coordinate (B0) at (-2.4,0);

\draw[contour]
    (B0) arc[start angle=\upperstart,
             end angle=\upperend,
             radius=\almondradius];
\draw[contour]
    (B0) arc[start angle=\lowerstart,
             end angle=\lowerend,
             radius=\almondradius];

\fill (B0) circle (2.1pt);

\node[below left] at (B0) {$x_0$};
\node[above left] at (B0) {$n$};

\end{scope}

\end{tikzpicture}%
}
\caption{
Two local cusp operators connected by arcs can be expanded in terms of excited cusp operators, as in equation \eqref{eq:cuspcuspope}.   
}
\label{fig:completedalmondexpansion}
\end{figure}

This type of COE first appeared in \cite{ladder} and is illustrated in Figure \ref{fig:completedalmondexpansion}.

\subsubsection{Multiple Cusp Configurations} \label{subsec:3loop}
To understand what degrees of freedom remain after conformal transformations, let us reexamine the case of three point functions. We have a configuration defined by the intersection of three circles. 
 The number of free parameters $N$ is given by the difference between the total number of parameters and the number of parameters we can fix by conformal symmetry
\begin{align}
    N&= \underbrace{ 3}_{\text{ 1 radius per circle}} + 3 \times \underbrace{ (2)}_{\text{Parameters for each center}} - \underbrace{6}_{\text{Parameters fixed by conformal symmetry}}\nonumber \\
    &= 3 ,
\end{align}
which correspond to the three cusp angles.  Therefore, if we fix the angles, there are no degrees of freedom left. 
This mirrors what also happens in the case of a three-point function of local operators in a CFT, where there is no left over kinematical variable after exhausting the full freedom allowed by conformal transformations.

\paragraph{Cusps Connected by Planar Arcs. }
The formula above can easily be generalized to $n$ cusps obtained from the intersection of $n$ circles in the plane, to give
\begin{equation}
    N= 3(n-2).
\end{equation} 
Note that we can write $N$ as $N= (2n -6) + n $, where the second term on the RHS corresponds to the $n$ cusp angles of the loop and the term in parentheses corresponds to the number of conformal invariants of $n$ points in $d=2$ spacetime dimensions. Thus, if we fix the $n$ angles, the number of conformal invariants of an $n$-cusped loop is the same as the number of conformal invariants of $n$ arbitrary points on a plane. Thus, for such planar configurations we would have correlation functions with the same kinematics as for local primary operators in a CFT.  

Despite this similarity, notice that the fact that dynamical data such as COE coefficients depend on the cusp angles introduces important differences. 
 In particular, notice that for a four-point correlator of cusps, $n=4$, we have $N=6$ which corresponds to the sum of the four cusp angles plus two parameters. We can either view these parameters as the cross ratios of the 4-point function, or as the angles $\phi_s$ and $\phi_t$ of the two `virtual' cusps that we can use to decompose the correlator using the Cusp-Cusp expansion in two channels (see Figures \ref{fig:sboot} and \ref{fig:tboot}). 
In other words, after accounting for such angles on which the COE coefficients depend, there are no free parameters left. This complicates the task of bootstrapping  the dynamical data. 

\paragraph{Non-Planar Arcs.} 
For configurations of non-coplanar points, one has to take into account the subgroup of the conformal group that stabilizes  the configuration of $2n$ points $\{x_i, z_i\}$ in $d$ dimensions, where $x_i$ are the cusp positions and $z_i$ are the centers of the circles forming the contour. This is because, unlike in the coplanar case, specifying the $n$ cusp positions does not fix the contour; however, specifying the centers and the cusp positions does. The number of parameters in this case is then given by 
\begin{equation}
    N= 2 \hspace{1pt}n\hspace{1pt }d-n -\frac{(d+2)(d+1)}{2} + \frac{(d+1-m)(d+2-m)}{2}
\end{equation}
where $m=\text{min}(2\hspace{1pt}n, d+2)$. 
This result was first derived in \cite{Dorn:2020meb} (Appendix B). In particular, for trivial stabilizers, $N$ can be rewritten as 
\begin{equation}
    N= N_{n \,\text{points}} + n(d-2)
\end{equation}
where $N_{n\, \text{points}}$ represent the number of free parameters for $n$ points in $d$ dimensions.
For $d>2$, there are additional conformal invariants for the cusped contours. Geometrically, these are associated with the relative orientation of the arc planes.

\section{Worked Examples}  \label{sec:examples}
In this section, we consider explicit examples of the general features we discussed in Sections \ref{sec:hilb} and \ref{sec:generic-cusp-correlators}. We choose the pinning defect in the theory of a single free scalar field as our setting.
We first  
compute the expectation values of contours with two or three cusps in a plane, confirming that they transform in the same way as two- and three-point functions of operators in a CFT do, as discussed in Section \ref{sec:ncusps}.  
Then, in Section \ref{sec:spectrum} we proceed to explicitly construct the excited states on a cusped line defect diagonalizing the action of the NS Hamiltonian, performing several checks on the resulting basis of scaling operators. Lastly, in Section \ref{sec:coeexample}, we check that the COE correctly reproduce the expectation value of a smooth defect. 

\subsection{Two-Point Function} \label{subsec:cuspedline}
Consider the contour $\gamma$ shown in Figure \ref{fig:cusped}, over which we will integrate a scalar defect known as the pinning (magnetic line) defect, defined as 
\begin{equation} \label{dp}
    W= \exp\left(-\lambda \int_\gamma \mathrm{d}\tau \abs{\frac{\mathrm{d} x}{\mathrm{d}\tau}} \phi(x(\tau))\right)
\end{equation}
where $\phi(x)$ is a scalar free field in $d=4$ spacetime dimensions. The expectation value $\langle W \rangle $ has been computed in \cite{Cuomo:2024psk}, yielding  
\begin{equation} \label{expeflat}
    \langle W \rangle \propto\left(\frac{\epsilon}{\Lambda_{\text{IR}}}\right)^{ \Gamma},
\end{equation} 
with $\epsilon$ and $\Lambda_{\text{IR}}$ being UV and IR cutoffs, respectively. As discussed above, the specific form of the cutoff is not important, but our choice is specified below. 
The cusp anomalous dimension turns is
\begin{equation} \label{gammac}
    \Gamma= \lambda^2 \left(1- \frac{\pi-\varphi}{\sin\varphi}\right).
\end{equation} 
 One may now apply a conformal transformation to obtain a more generic contour with two cusps. We concentrate on the case related to the straight defect lines by a conformal map in the plane, discussed in detail in Appendix \ref{sec:conf}. Such a transformation leads to an almond-shaped curve, such as the one shown in Figure \ref{fig:NSalmond}. We parametrize the upper ($+$) and lower ($-$) circular arcs as in equations \eqref{xpm} and \eqref{zeta}.
The expectation value $\langle W \rangle$ then takes the form 
\begin{align} \label{almond}
    &\left \langle \exp{-\lambda \int_1^2 \mathrm{d}\tau \abs{\frac{\mathrm{d} x}{\mathrm{d}\tau}}\phi(x(\tau)) - \lambda \int_2^1 \mathrm{d}\tau \abs{\frac{\mathrm{d} x}{\mathrm{d}\tau}}\phi(x(\tau)) } \right \rangle\\ \nonumber
    &= 1+\underbrace{\lambda^2 \int_1^2 \mathrm{d}\tau_1 \int_{1}^2\mathrm{d}\tau_2 \abs{\frac{\mathrm{d} x}{\mathrm{d}\tau_1}} \abs{\frac{\mathrm{d} x}{\mathrm{d}\tau_2}}\langle \phi(x(\tau_1)) \phi(x(\tau_2)) \rangle}_{A}  \\ \nonumber
    &+ \underbrace{\lambda^2 \int_2^1 \mathrm{d}\tau_2 \int_1^2 \mathrm{d}\tau_1 \abs{\frac{\mathrm{d} x}{\mathrm{d}\tau_1}} \abs{\frac{\mathrm{d} x}{\mathrm{d}\tau_2}} \langle \phi(x(\tau_1)) \phi(x(\tau_2)) \rangle}_{B} + \cdots .
\end{align}
where $1$ and $2$ stand for the parameters representing the positions of the two cusps $x_1$ and $x_2$ respectively. 
In the free scalar theory, one can show that these two diagrams are all we need since the answer resums into an exponential of the sum of these two integrals; see e.g Section 2 of  \cite{Soderberg:2021kne}. Let us proceed with the first integral, connecting propagators to the same line. Using the parametrization $\tau_{1,2} =s_{1,2}$ introduced in \eqref{xpm},  the first integral $(A)$ in \eqref{almond} is computed as follows 
\begin{align} \label{first}
   A & = \lambda^2 \int_{-\Lambda}^{\Lambda} \mathrm{d}s_2 \int_{-\Lambda}^\Lambda\mathrm{d}s_1 \frac{\abs{\dot{x}_{\pm}(s_1)} \abs{\dot{x}_{\pm}(s_2)}}{\abs{ x_{\pm}(s_1) -  x_{\pm}(s_2)}^2}\\ \nonumber
    &= -\lambda^2\int_{-\Lambda}^{\Lambda} \mathrm{d}s_2= -2\Lambda \lambda^2,
\end{align}  
where $\Lambda$ here is a dimensionless cutoff that satisfies 
\begin{equation} 
    \begin{split}
        x_1&=\lim_{\Lambda\rightarrow \infty}x_\pm(\mp \Lambda) \\
        x_2&=\lim_{\Lambda\rightarrow \infty}x_\pm(\pm \Lambda).
    \end{split}
\end{equation}
 and should not be confused with the IR cutoff in \eqref{expeflat}. The latter will always be written as $\Lambda_{\text{IR}}$. In equation \eqref{first}, we have also renormalized a coincident point singularity  by a cosmological constant counterterm, i.e. a divergence proportional to the perimeter.

The second integral ($B$), which connects points in the upper and lower arcs, follows analogously 
\begin{equation} \label{second}
\begin{split}
   B &= \lambda^2\int_{-\Lambda}^{\Lambda} \mathrm{d}s_2 \int_{-\Lambda}^\Lambda\mathrm{d}s_1 \frac{\abs{\dot{x}_\pm(s_1)} \abs{\dot{x}_\mp(s_2)}}{\abs{ x_\pm(s_1) -  x_\mp(s_2)}^2} \\
    &=\frac{\lambda^2}{2}  \int_{-\Lambda}^{\Lambda} \mathrm{d}s_2 \int_{-\Lambda}^\Lambda\mathrm{d}s_1 \frac{1}{\cosh (s_1+s_2) -\cos \varphi} \\
    &= \lambda^2 \left(\frac{\pi-\varphi}{\sin\varphi} \Lambda -\frac{2}{\sin\varphi} \operatorname{Im} \operatorname{Li}_2(e^{i \varphi}) \right).
\end{split}
\end{equation}
Introducing a dimensionful cut-off $\epsilon$ by imposing that $\abs{\zeta_+(-\Lambda) -w_1}= \abs{\zeta_-(\Lambda) -w_1}=\epsilon$ in the parametrization \eqref{zeta}, we find, for small $\epsilon$,
\begin{equation} \label{biglambda}
    \Lambda=\log\left( \frac{x_{12}}{\epsilon}\right)
\end{equation} 
where $x_{12}= \abs{x_1-x_2}$. Lastly, by adding \eqref{first} and \eqref{second} and exponentiating the result, we arrive at 
\begin{equation} \label{expealm}
    \langle W \rangle = \exp(A+B)= e^{\frac{\lambda^2}{2} F_{\text{almond}}(\varphi)}\left(\frac{\epsilon}{x_{12}}\right)^{2 \Gamma}
\end{equation} 
with $\Gamma $ given as in \eqref{gammac}, and 
\begin{equation} \label{falm}
    F_{\text{almond}}(\varphi)=-\frac{4}{\sin\varphi}\operatorname{Im} \operatorname{Li}_2(e^{i \varphi}).
\end{equation}
This result is equivalent to that of equation \eqref{expeflat} by a conformal transformation\footnote{The overall factor is the same.} 
\begin{equation}
    z(w)= \frac{w_{12}(w-w_1)}{w_2-w}~.
\end{equation}
Considering the two points on the arcs such that $|w - w_2 | = \epsilon$, 
 we have  $ \abs{z(w)}= \Lambda_{\text{IR}}=\frac{x_{12}^2}{\epsilon}+O(\epsilon)$, which maps \eqref{expealm}  to \eqref{expeflat}. 
When appropriately normalized by $\epsilon$ factors, the expectation value of the defect $W$ in equation \eqref{expealm} is exactly the same as that of the $2-$pt function of two identical scaling operators with scaling dimension $\Gamma$. In particular,

\begin{equation}\label{eq:2ptfunc}
    \langle O(x_1) O(x_2) \rangle \sim \frac{\langle W \rangle}{\epsilon^{2\Gamma}} \propto \frac{1}{x_{12}^{2\Gamma}}.
\end{equation}
Notice that this last step of dividing by the appropriate power of $\epsilon$, while trivial, is crucial in defining states and correlation functions that remain finite in the limit where all regulators are removed.

\subsection{Three-Point Function} 

Consider the configuration in Figure \ref{fig:three}, where the circular
arcs intersect at three distinct cusps. We denote the internal angle at the
$i$-th cusp by $\varphi_i$. These angles satisfy \cite{ladder}
\begin{equation}
    0<\varphi_i<\pi,
    \qquad
    \varphi_1+\varphi_2-\varphi_3<\pi,
    \qquad
    \varphi_2+\varphi_3-\varphi_1<\pi,
    \qquad
    \varphi_3+\varphi_1-\varphi_2<\pi ,
\end{equation}
guaranteeing that virtual intersections  of the arcs (occurring when extending all the arcs) lie outside our loop of interest.

The three-cusp correlator has the following decomposition 
\begin{align} \label{3cusp}
    &\left \langle \exp \left( -\lambda \int_1^2 \mathrm{d}\tau \left| \frac{\mathrm{d}x_{12}}{\mathrm{d}\tau} \right| \phi(x_{12}(\tau)) - \lambda \int_2^3 \mathrm{d}\tau \left| \frac{\mathrm{d}x_{23}}{\mathrm{d}\tau} \right| \phi(x_{23}(\tau)) - \lambda \int_3^1 \mathrm{d}\tau \left| \frac{\mathrm{d}x_{31}}{\mathrm{d}\tau} \right| \phi(x_{31}(\tau)) \right) \right \rangle   \nonumber \\ \nonumber
    & =1+\frac{\lambda^2}{2} \biggl[\int_1^2 \mathrm{d}\tau_1 \mathrm{d}\tau_2 \left| \frac{\mathrm{d}x_{12}}{\mathrm{d}\tau_1} \right| \left| \frac{\mathrm{d}x_{12}}{\mathrm{d}\tau_2} \right| \langle \phi(x_{12}(\tau_1)) \phi(x_{12}(\tau_2)) \rangle \nonumber \\
    &+ \int_2^3 \mathrm{d}\tau_1 \mathrm{d}\tau_2 \left| \frac{\mathrm{d}x_{23}}{\mathrm{d}\tau_1} \right| \left| \frac{\mathrm{d}x_{23}}{\mathrm{d}\tau_2} \right| \langle \phi(x_{23}(\tau_1)) \phi(x_{23}(\tau_2)) \rangle \\ \nonumber
    &+ \int_3^1 \mathrm{d}\tau_1 \mathrm{d}\tau_2 \left| \frac{\mathrm{d}x_{31}}{\mathrm{d}\tau_1} \right| \left| \frac{\mathrm{d}x_{31}}{\mathrm{d}\tau_2} \right| \langle \phi(x_{31}(\tau_1)) \phi(x_{31}(\tau_2)) \rangle \biggr] \\ \nonumber
    &+ \lambda^2 \biggl[ \int_1^2 \mathrm{d}\tau_{1} \left| \frac{\mathrm{d}x_{12}}{\mathrm{d}\tau_{1}} \right| \int_3^1 \mathrm{d}\tau_{2} \left| \frac{\mathrm{d}x_{31}}{\mathrm{d}\tau_{2}} \right|  \langle \phi(x_{12}(\tau_{1})) \phi(x_{31}(\tau_{2})) \rangle + \text{cyclic permutations} \biggr] + \ldots ,
\end{align}
where $x_{ij}(\tau)$ parametrizes the arc going from the point $i$ to $j$.  The first three integrals above have been computed in \eqref{first}, so we focus on the contributions from the propagators joining different arcs. We only compute the first contribution as the others can be obtained by cyclic permutations. As before, we consider a planar configuration and we use complex coordinates with cusp points identified by $x_i=(\Re(w_i),\Im(w_i),0,0)$ for $i=1,2,3.$  Following \cite{ladder}, we parametrize the arcs
$x_{12}$ and $x_{13}$ as
\begin{align}
    x_{12}(s)
    &=
    \bigl(
        \Re\zeta_{12}(s),
        \Im\zeta_{12}(s),
        0,0
    \bigr),
    \\
    x_{13}(t)
    &=
    \bigl(
        \Re\zeta_{13}(t),
        \Im\zeta_{13}(t),
        0,0
    \bigr),
\end{align}
where
\begin{align}
\label{12}
\zeta_{12}(s)
&=
w_1-
\frac{w_{12}w_{13}e^s}{
e^s w_{13}
+
\frac{i}{2\sin\varphi_1}
w_{23}(1-e^s)
\left(
e^{-i\varphi_1}
+
e^{i(\varphi_2-\varphi_3)}
\right)
},
\\
\label{13}
\zeta_{13}(t)
&=
z_1-
\frac{w_{12}w_{13}e^t}{
e^t w_{12}
+
\frac{i}{2\sin\varphi_1}
w_{23}(1-e^t)
\left(
e^{i\varphi_1}
+
e^{i(\varphi_3-\varphi_2)}
\right)
},
\end{align}
with $\zeta_{12}(0)=w_2, \hspace{2pt} \zeta_{13}(0)=w_3$ and $\zeta_{12}(-\infty)= \zeta_{13}(-\infty)=w_1$ and $w_{ij}=w_i-w_j$. We regularize the integration at $w_1$ by introducing the cut-offs
\begin{equation}
\label{lam1}
\Lambda_s
=
\ln\left[
\frac{
x_{12}x_{13}\sin\varphi_1
}{
x_{23}\epsilon
\cos\left(
\frac{\varphi_1-\varphi_2+\varphi_3}{2}
\right)
}
\right],
\qquad
\Lambda_t
=
\ln\left[
\frac{
x_{12}x_{13}\sin\varphi_1
}{
x_{23}\epsilon
\cos\left(
\frac{\varphi_1+\varphi_2-\varphi_3}{2}
\right)
}
\right].
\end{equation}
which are obtained, as before, by imposing that $|\zeta_{12}(-\Lambda_s)-w_1|=|\zeta_{13}(-\Lambda_t)-w_1|=\epsilon$. The ranges of the parameters $s$ and $t$ are
\begin{equation}
    s\in[-\Lambda_{s},0],
    \qquad
    t\in[-\Lambda_{t},0].
\end{equation}
Thus, for terms connecting different lines, we have
\begin{equation}
\label{cross-integral}
\begin{aligned}
&
\int_1^2 \mathrm{d}\tau_1
\left|
\frac{\mathrm{d}x_{12}}{\mathrm{d}\tau_1}
\right|
\int_3^1 \mathrm{d}\tau_2
\left|
\frac{\mathrm{d}x_{31}}{\mathrm{d}\tau_2}
\right|
\left\langle
\phi\bigl(x_{12}(\tau_1)\bigr)
\phi\bigl(x_{31}(\tau_2)\bigr)
\right\rangle
\\[2mm]
&\qquad =
\frac{1}{2}
\int_{-\Lambda_s}^{0}\mathrm{d}s
\int_{-\Lambda_t}^{0}\mathrm{d}t\,
\frac{1}{
\cosh\bigl(s-t-\delta x_1\bigr)
-\cos\varphi_1
}
\\[2mm]
&\qquad =
\frac{\pi-\varphi_1}{\sin\varphi_1}
\log\left(
\frac{x_{12}x_{13}}
{x_{23} \epsilon }
\right)
+
f(\varphi_1,\varphi_2,\varphi_3)
+
\mathcal{O}(\epsilon).
\end{aligned}
\end{equation}
where we have used the notation
\begin{equation}
\delta x_1
=
\log\left[
\frac{
\cos\left(
\frac{\varphi_1-\varphi_2+\varphi_3}{2}
\right)
}{
\cos\left(
\frac{\varphi_1+\varphi_2-\varphi_3}{2}
\right)
}
\right].
\end{equation}
The constant $f(\varphi_1,\varphi_2,\varphi_3)$ denotes a finite, angle-dependent piece that contributes to the COE coefficient of the $3-$pt function, and it is given by
\begin{align}\label{f}
f(\varphi_1,\varphi_2,\varphi_3)
&= \frac{1}{\sin\varphi_1}
\left[
(\pi-\varphi_1)
\left(
\log\frac{1}
{L_{123}}
-\frac{|\delta x_1|}{2}
\right)
-\operatorname{Im}\operatorname{Li}_2\!\left(
e^{-|\delta x_1|+i\varphi_1}
\right)
-\operatorname{Im}\operatorname{Li}_2\!\left(
e^{i\varphi_1}
\right)
\right]~,  \\ 
L_{123}
&=
\frac{
\sqrt{
\cos\left(
\frac{\varphi_1+\varphi_2-\varphi_3}{2}
\right)
\cos\left(
\frac{\varphi_1-\varphi_2+\varphi_3}{2}
\right)
}
}{
\sin\varphi_1
}~.
\end{align}
Combining these results with the other contributions in \eqref{3cusp}, we get
\begin{equation}
\begin{split}
    &\left \langle \exp{-\lambda \int_1^2 \mathrm{d}\tau \phi(x(\tau)) - \lambda \int_2^3 \mathrm{d}\tau \phi(x(\tau)) - \lambda \int_3^1 \mathrm{d}\tau \phi(x(\tau))} \right \rangle=  \\
    &\exp\left(- \underbrace{\lambda^2\left(1-\frac{\pi-\varphi_1}{\sin\varphi_1}\right)}_{\Gamma(\varphi_1)}\log\left( \frac{x_{12} x_{13}}{x_{23} \epsilon} \right)+ {\lambda^2 f_1(\varphi_1,\varphi_2,\varphi_3) + \text{cyclic} }   \right),
\end{split}
\end{equation}
where $\Gamma(\varphi_1)$ is the cusp anomalous dimension \eqref{gammac}. Substituting all the cyclic permutations above, we arrive at 
\begin{multline}
    \left \langle \exp\left(-\lambda \int_1^2 \mathrm{d}\tau \phi(y(\tau)) - \lambda \int_2^3 \mathrm{d}\tau \phi(y(\tau)) - \lambda \int_3^1 \mathrm{d}\tau \phi(y(\tau))\right) \right \rangle=  \\
    \frac{{e^F}}{x_{12}^{\Gamma_1+\Gamma_2- \Gamma_3}x_{23}^{\Gamma_2+\Gamma_3-\Gamma_1}x_{31}^{\Gamma_1+\Gamma_3-\Gamma_2}} \epsilon^{\Gamma_1+\Gamma_2+\Gamma_3}
\end{multline} 
where ${F=\lambda^2 (f(\varphi_1,\varphi_2,\varphi_3) + \text{cyclic})}$ and $\Gamma_i\equiv \Gamma(\varphi_i)$. As expected from our transformation rule \eqref{trans}, the expression above has the same kinematic structure as a CFT three-point function of primary operators with dimensions $\Gamma_{1,2,3}$. 

Note that similar observations, a number of which inspired our present analysis, were also made in the case of $\mathcal{N}=4$ SYM in \cite{Dorn:2020vzj,ladder}. In case \cite{ladder}, a double scaling limit in an internal parameter 
(
known as the ladder limit ) was taken for the cusps. In that limit, the expectation value for a configuration such as the one in Figure \ref{fig:three} can be computed exactly for any excited states at the cusps and was shown to satisfy the form expected for a CFT $3$-pt function. 

\subsection{Spectrum of Cusp Operators} \label{sec:spectrum}\label{sec:CompositeOps}

Now, we would like to understand how to systematically construct excited states localized at the cusp (according to the general ideas explained in Section \ref{sec:generic-cusp-correlators}) in an explicit example. While here we provide a construction that very closely mimics the renormalization of composite operators in standard perturbative field theory \cite{Brezin:1974zr,Brown:1979pq,Collins:1984xc}, it should be mentioned that an alternative regularization scheme to generate the excited states was employed  in \cite{ladder}, and we pursue it in Appendix \ref{app:chopped}. This provides a useful cross-check for results derived in this section.

Suppose we have scaling operators $\hat{\mathcal{O}}_n(x)$ on the cusp. We define another set of operators $\hat{\Psi}_n(x)$ which are not scaling operators but do have an engineering dimension $\Gamma^e_n$. To define these operators, we introduce a dimensionful cutoff $\epsilon$ which has units of length,
such that correlation functions are invariant under 
\begin{equation}\label{eq:rescalings}
    x\rightarrow \alpha x, \hspace{1cm} \epsilon\rightarrow \alpha \epsilon, \hspace{1cm } \hat{\Psi}_{n} \rightarrow \alpha^{-\Gamma_n^e} \hat{\Psi}_n
\end{equation} 
the $\hat{\Psi}$s are usually called bare operators. On the other hand, the correlators of the scaling operators are invariant under 
\begin{equation}
    x\rightarrow \alpha x, \hspace{1cm} \hat{\mathcal{O}}_n \rightarrow \alpha^{-\Gamma_n}\hat{\mathcal{O}}_n.
\end{equation}
with fixed cut-off.
Here we notice the difference between the engineering dimension ($\Gamma^e_n$), which is the response to rescaling the unit of measure and the ruler, and the scaling dimension $\Gamma_n$, which is the one that controls the large distance decay of a correlator. We can write the bare operators as 
\begin{equation} \label{bare}
    \hat{\Psi}_{n}= \sum_{m} \epsilon^{\Gamma_n-\Gamma_n^e}  R_{n}^{m} \hat{\mathcal{O}}_{m}.
\end{equation}
For small $\epsilon$, the equation above tells us that the $\hat{\Psi}_{n}$ only mix with scaling operators for which $\Gamma_n\leq \Gamma^e_n$. 

Our strategy to build the scaling operators $\hat{\mathcal{O}}_{n}$ on the cusp is as follows: we consider a basis consisting of bare operators that can be constructed up to engineering dimension $\Gamma^e_n=2$, multiplied by factors of $\epsilon$ to make them all dimensionless\footnote{While somewhat unconventional we find this to be useful for bookeeping. This also makes comparison to the differential operator approach of Appendix \ref{app:chopped} straightforward.}
\begin{equation} \label{basisop}
 \{1, \epsilon \phi, \epsilon^2 \partial_y \phi, \epsilon^2 \partial_x \phi, \epsilon^2 \phi^2\},
\end{equation} 
where here the local operators will be inserted at the cusps, as we discuss in detail below. We remind the reader that the final physical correlators should be normalized by appropriate powers of $\epsilon$, as described around \eqref{eq:2ptfunc}, in order to have an appropriate limit as $\epsilon \rightarrow0$. As a consequence of the basis choice, the resulting renormalized operators will thus also carry additional powers of $\epsilon$.
In building this basis, we have ignored derivatives orthogonal to the plane of the defect, since, due to the $SO(d-2)$ symmetry, they do not mix with the rest of the operators up to $\Gamma^e_n=2$.
One may wonder whether the basis above is complete, or, maybe, we have unknowingly left out operators that cannot be written in terms of bulk operators. This can be straightforwardly answered by character-counting techniques, or in more pedestrian terms, by putting the theory on the finite temperature cylinder. Recall that, in general, the coefficients $\Omega(E)$ of  the partition function of a system at finite temperature 
\begin{equation} \label{zcount}
    Z(\beta)=\mathrm{Tr}_{\mathcal H} e^{-\beta H}
=\sum_E \Omega(E)e^{-\beta E},
\end{equation}
encode the number of states at a given energy $E$ \cite{Cardy:2008jc,Ginsparg:1988ui}. The partition function $Z_{W}$ of the cusped defect can be computed in the geometry $S^1\cross S^3$ with the periodicity in Euclidean time taken to be $\tau \sim \tau+\beta$. The result of this computation (see Appendix \ref{sec:parti}) tells us that the number of degeneracies at each level is the same as in the bulk free theory, albeit with the scaling dimensions shifted by the vacuum cusp anomalous dimension, indicating that our basis \eqref{basisop} is indeed complete. 

Let us now consider the matrix $G$ of two-point functions between all the aforementioned operators (inserted at the cusps) and diagonalize it. The elements $G_{nm}$ of the this matrix are defined as follows 
\begin{equation} \label{gmn}
   G_{mn}= \langle  \hat{\Psi}_m(x_2)\hat{\Psi}_n(0)  W\rangle ,
\end{equation}
where in this section we will adopt the notation $\langle O_1  \dots O_n W \rangle$ to indicate a correlation function of operators inserted on a cusped contour, whose shape will be clear by the context. Here for instance, we consider the almond defect lines connecting the points $x_1$ and $x_2$, and bare operators in the matrix elements are inserted at the cusp points. 
For simplicity, we have set $x_1=0$ and $x_2$ to lie at the $y=0$ line.  In the computation of these matrix elements, we will set $\chi=0$, which corresponds to the $ y$-reflection-symmetric almond. This corresponds to working in a  (particularly convenient) choice of conformal frame. 
For our purposes it suffices to consider $G_{nm}$ up to order $\epsilon^{2(\Gamma+2)}$ (more generally, $\epsilon^{[\hat{\Psi}_n]+[\hat{\Psi}_m]}$).

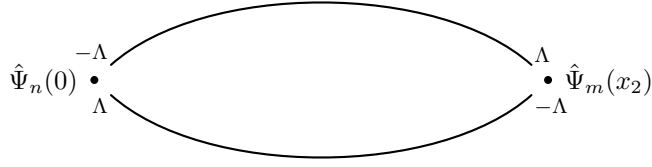
\begin{figure}[t]
    \centering

    \begin{tikzpicture}[
        contour/.style={
            black,
            thick,
            line cap=round,
            line join=round
        },
        cutoff/.style={
            font=\scriptsize,
            inner sep=1pt
        },
        operator/.style={
            font=\small,
            inner sep=2pt
        }
    ]

        \coordinate (L) at (-3,0);
        \coordinate (R) at ( 3,0);

        \coordinate (LT) at (-2.78, 0.20);
        \coordinate (LB) at (-2.78,-0.20);
        \coordinate (RT) at ( 2.78, 0.20);
        \coordinate (RB) at ( 2.78,-0.20);

        \draw[contour]
            (LT) .. controls (-1.55,1.30) and (1.55,1.30) .. (RT);

        \draw[contour]
            (LB) .. controls (-1.55,-1.30) and (1.55,-1.30) .. (RB);

        \fill (L) circle (1.5pt);
        \fill (R) circle (1.5pt);

        \node[operator, anchor=east] at (-3.14,0)
            {$\hat\Psi_n(0)$};

        \node[operator, anchor=west] at (3.14,0)
            {$\hat\Psi_m(x_2)$};

        \node[cutoff, above left]  at (LT) {$-\Lambda$};
        \node[cutoff, below left]  at (LB) {$\Lambda$};
        \node[cutoff, above right] at (RT) {$\Lambda$};
        \node[cutoff, below right] at (RB) {$-\Lambda$};

    \end{tikzpicture}

    \caption{Chopped almond with operators inserted at the two cusps.}
    \label{fig:cutalmwithops}
\end{figure}

We regularize the two-point functions by inserting the local operators $\hat{\Psi}_n$ directly at the cusp and chopping the contour with the cut-off $\Lambda$ \eqref{biglambda} as in Figure \ref{fig:cutalmwithops}. We then take the limit $\epsilon\to 0$. The explicit form of the matrix $G$ in this scheme is spelled out in Appendix \ref{sec:matrix}. 

The matrix $G$ is not quite the overlap of states in NS quantization. The latter is obtained by conjugating, say, the operators placed at $x_2$.  If we choose the quantization surface as the infinite-radius circle bisecting the almond, conjugation amounts to a reflection across this line. Therefore, the matrix of scalar products is $R\,G$, with $R=\operatorname{diag}(1,1,1,-1,1)$. This is symmetric and has positive eigenvalues.  
The basis $\{ \widehat{\mathcal{O}}_{n\geq 0}\}$ that diagonalizes the matrix $R\,G$ up to order four in $\epsilon $ is 
\begin{equation} \label{diagx1}
    B_{\text{diagonal}, x_1=0}= \left \{1, \phi_R, \epsilon^2 \partial_y \phi, \epsilon^2 \left(\partial_x \phi - \frac{2 }{x_2} \phi \right) +c_2, (\phi_R)^2\right\},
\end{equation} 
with the following definitions 
\begin{align} 
    \phi_R(0)&= \epsilon \phi +c_1= \epsilon \phi 
   +2\lambda~,  \\
    c_2&=2 \lambda \cos\varphi/2~.
\end{align} 
In this basis, the matrix of scalar products reads
\begin{equation}
    (R\,G)_{\text{diagonal}}= e^{\frac{\lambda^2}{2} F_{\text{almond}}} \left(\dfrac{\epsilon}{x_2}\right)^{2\Gamma} \text{diag}\left(1,\left(\dfrac{\epsilon}{x_2}\right)^2,2\left(\dfrac{\epsilon}{x_2}\right)^4,2\left(\dfrac{\epsilon}{x_2}\right)^4 ,2\left(\dfrac{\epsilon}{x_2}\right)^4\right)~,
\end{equation} 
with $F_{\text{almond}}$ given as in \eqref{falm}.
Let us make a few comments. First, as a sanity check, $\partial_y \phi$ does not mix with other operators, due to the reflection symmetry of the configuration. The specific mixing coefficients $c_{1,2}$ of the other operators, however, are scheme-dependent, as is pointed out in Appendix \ref{sec:matrix}. A different scheme is presented in appendix \ref{app:chopped}, and one can check that, for instance, the value of $c_1$ is different there---see \eqref{oc1}.

Furthermore, notice we have put an $x_1$ subscript for the diagonal basis. To obtain the basis of local operators that, inserted at $x_2$, have diagonal correlations with the set $B_{\text{diagonal}, x_1=0}$, one must act with the reflection $R$, which, for each operator, produces the Hermitian conjugate. The only change is of course in the sign of $\partial_x\phi$:
\begin{equation} \label{b2}
    B_{\text{diagonal}, x_2}= (B_{\text{diagonal}, x_1=0})^\dagger=\left \{1, \phi_R, \epsilon^2 \partial_y \phi, \epsilon^2 \left(-\partial_x \phi - \frac{2 }{x_2} \phi \right) +c_2, (\phi_R)^2\right\}~.
\end{equation}

More consequential than the previous observations is the fact that the basis is coordinate dependent, something unusual when constructing a diagonal operator basis and which would be considered off-limits for local operators. 
To illustrate this point, consider an example, first without the defect. For the local operators $\{\phi, \partial_x \phi\}$, the matrix of 2-point functions for the points $x_1=0$ and $x_2$  takes the form (on the line)
\begin{equation}
    \left(\begin{array}{cc}
        \frac{1}{x_2^2}& -\frac{2}{x_2^3} \\
        \frac{2}{x_2^3} & -\frac{6}{x_2^4}
    \end{array}\right),
\end{equation}
which cannot be diagonalized with a constant (i.e., position-independent) matrix. This is, of course, explained by the fact that the two-point function in this example involves an operator and its descendant on the line. 

When we turn on the defect, this feature persists, and diagonalizing the matrix is only possible if we allow coefficients depending on $x_2$. Naively, thus, the diagonalization has forced us to consider non-local operators, where the basis of operators at $x_1$, i.e. \eqref{diagx1}, depends on position $x_2$ of the other cusp operator. 
Notice, however, that $x_2$ is also determined by local\footnote{Local in the sense that they can be inferred  by looking at the contour in an arbitrarily close neighborhood around the cusp $x_1$.} geometric data of the almond. In fact, the shape of the arcs forming the almond is uniquely fixed by their tangent and curvature close to $x_1$, and these data are determined by $\chi$ and $x_2$. For the symmetric almond case $\chi=0$ of our computation, the radius of curvature of the arcs at $x_1$ is given by
 $ r_c=\frac{\abs{x_2-x_1}}{2 \cos\varphi/2}$.
This indicates that the 
distance $\abs{x_{12}}$ represents local data of the defect. 
In this sense, the operator $O_3(0)=\epsilon^2 \left(\partial_x \phi - \frac{2 }{x_2} \phi \right) +c_2$ is indeed a local cusp operator.
In Section \ref{subsec:Hamiltonians} we will further elaborate on this structure by confirming that these operators are eigenstates of the NS Hamiltonian and that they transform like primary operators. 
Before that, we perform a few simple computations that exemplify the general results of this paper.

\subsubsection{A Three-point Function Involving Scaling Cusp Operators}\label{sec:transfpropop}

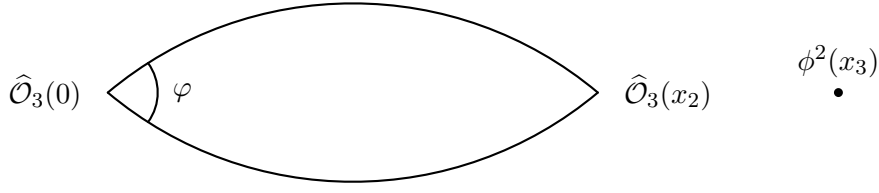
\begin{figure}[t]
\centering
 \begin{tikzpicture}[
    scale=1.2,
    line cap=round,
    line join=round
]


\def\a{2.7}
\def\phiang{80}
\def\anglerad{0.55}

\pgfmathsetmacro{\alpha}{(180-\phiang)/2}
\pgfmathsetmacro{\h}{\a*tan(\alpha)}
\pgfmathsetmacro{\Rc}{sqrt(\a*\a+\h*\h)}

\pgfmathsetmacro{\beta}{
    acos(\anglerad/(2*\Rc))-\alpha
}

\coordinate (C1) at (-\a,0);
\coordinate (C2) at ( \a,0);

\draw[black,thick]
    (C1)
    arc[
        start angle=180-\alpha,
        end angle=\alpha,
        radius=\Rc
    ];

\draw[black,thick]
    (C1)
    arc[
        start angle=180+\alpha,
        end angle=360-\alpha,
        radius=\Rc
    ];

\draw[black,thick]
    (C1) ++(-\beta:\anglerad)
    arc[
        start angle=-\beta,
        end angle=\beta,
        radius=\anglerad
    ];

\node at ($(C1)+(0.82,0)$) {$\varphi$};

\node[left=6pt]  at (C1) {$\widehat{\mathcal O}_3(0)$};
\node[right=6pt] at (C2) {$\widehat{\mathcal O}_3(x_2)$};

\coordinate (B) at (5.35,0);
\fill (B) circle (1.4pt);
\node[above=2pt] at (B) {$\phi^2(x_3)$};

\end{tikzpicture}
\caption{Three-point function involving an excited operator at each cusp and one bulk operator.}
\label{fig:3pcusp}
\end{figure}

Consistency with the results of Section \ref{sec:generic-cusp-correlators} requires that the correlation functions involving the basis elements \eqref{diagx1} obey the same constraints as correlators of primary operators. To test this, we place the unconventional-looking operators $\hat{\mathcal{O}}_3$ defined by 
\begin{equation}\label{eq:O3}
    \hat{\mathcal{O}}_3(0)= \epsilon^2 \left(\partial_x \phi(0) - \frac{2 }{x_2} \phi(0) \right) +c_2~,\qquad
    \hat{\mathcal{O}}_3(x_2)= \epsilon^2 \left(-\partial_x \phi(x_2) - \frac{2 }{x_2} \phi(x_2) \right) +c_2.
\end{equation}
on the two cusps, 
and a third primary bulk operator, such as $\phi^2(x_3)$, at some collinear point $x_3$, as shown in Figure \ref{fig:3pcusp}. As expected, we find this $3-$pt function to be the same as that of primaries with dimensions $\{ \Gamma+2, \Gamma+2, 2\}$:
\begin{equation}\label{eq:O3op}
   \langle \hat{\mathcal{O}}_3(0) \hat{\mathcal{O}}_3(x_2) \phi^2(x_3) W \rangle=\left(\frac{\epsilon}{x_2} \right)^{2\Gamma}\frac{2\epsilon^4 }{x_2^2(x_3-x_2)^2x_3^2} \left[4+\lambda^2\frac{(\varphi)^2}{\sin^2(\varphi/2)} \right].
\end{equation}

Notice that the form of the $3-$pt function is preserved at $\lambda=0$, meaning that even in the absence of the defect, the operator $\hat{\mathcal{O}}_3$ still behaves like a primary, at least as far as the $3-$point function is concerned. At $x_1=0$, when $\lambda=0$
\begin{equation} \label{lzero}
    \hat{\mathcal{O}}_3(0)_{\lambda=0}= \epsilon^2 \partial_x \phi(0)-\frac{2\epsilon^2}{x_2} \phi(0).
\end{equation} 
In Section \ref{subsec:Hamiltonians}, we will understand the explicit form of the operator \eqref{lzero} in terms of the diagonalization of the NS Hamiltonian, which leaves the almond invariant.

\subsubsection{Overlap with a Smooth Contour}\label{sec:smooth}

As a further check on our basis, we show that the operators we constructed appear in the COE of a smooth contour.

The overlaps of the states created by the operators in \eqref{diagx1} with a circle of radius $r_{NS}$ are computed by the expectation value of the pinning defect on the contour in Figure \ref{fig:smoothedalmond} with the corresponding 
 operator insertion (defined as we explained above at $x_2$), which, as matrix elements, correspond to the bra  $\langle \widehat{\mathcal{O}}_n(x_1) |$ overlapped with the state created by the circle arc of radius $r_{\text{NS}}$. This computation gives 
\begin{equation} \label{smwc}
\begin{split}
\Big \langle  \hat{\mathcal{O}}_0(x_1) \Big|
\begin{tikzpicture}[x=0.75pt,y=0.75pt,yscale=-0.5,xscale=0.5,baseline={(0,-2.9)},line cap=round,line join=round]
\draw[draw opacity=0]
    (298.62,230.23)
    .. controls (309.8,227.77) and (318,219.64) .. (318,210)
    .. controls (318,200.04) and (309.25,191.7) .. (297.52,189.54)
    -- (291.5,210) -- cycle ;
\draw[black,thick]
    (298.62,230.23)
    .. controls (309.8,227.77) and (318,219.64) .. (318,210)
    .. controls (318,200.04) and (309.25,191.7) .. (297.52,189.54) ;
\end{tikzpicture} \Big\rangle&= e^{\frac{\lambda^2}{2} F(\varphi)}\left( \frac{\epsilon \hspace{2pt} r_R(r_{\text{NS}})}{x_2} \right)^\Gamma \\ 
  \Big \langle \hat{\mathcal{O}}_1(x_1) \Big| 
\begin{tikzpicture}[x=0.75pt,y=0.75pt,yscale=-0.5,xscale=0.5,baseline={(0,-2.9)},line cap=round,line join=round]
\draw[draw opacity=0]
    (298.62,230.23)
    .. controls (309.8,227.77) and (318,219.64) .. (318,210)
    .. controls (318,200.04) and (309.25,191.7) .. (297.52,189.54)
    -- (291.5,210) -- cycle ;
\draw[black,thick]
    (298.62,230.23)
    .. controls (309.8,227.77) and (318,219.64) .. (318,210)
    .. controls (318,200.04) and (309.25,191.7) .. (297.52,189.54) ;
\end{tikzpicture} \Big\rangle&= e^{\frac{\lambda^2}{2} F(\varphi)}\left( \frac{\epsilon r_R(r_{\text{NS}})}{x_2} \right)^{\Gamma+1} (-\lambda)\left(\pi -\varphi-2  \cot(\varphi/2) \right)\\ 
   \Big \langle \hat{\mathcal{O}}_2(x_1) \Big| 
\begin{tikzpicture}[x=0.75pt,y=0.75pt,yscale=-0.5,xscale=0.5,baseline={(0,-2.9)},line cap=round,line join=round]
\draw[draw opacity=0]
    (298.62,230.23)
    .. controls (309.8,227.77) and (318,219.64) .. (318,210)
    .. controls (318,200.04) and (309.25,191.7) .. (297.52,189.54)
    -- (291.5,210) -- cycle ;
\draw[black,thick]
    (298.62,230.23)
    .. controls (309.8,227.77) and (318,219.64) .. (318,210)
    .. controls (318,200.04) and (309.25,191.7) .. (297.52,189.54) ;
\end{tikzpicture} \Big\rangle&= 0 \\
   \Big \langle \hat{\mathcal{O}}_3(x_1) \Big| 
\begin{tikzpicture}[x=0.75pt,y=0.75pt,yscale=-0.5,xscale=0.5,baseline={(0,-2.9)},line cap=round,line join=round]
\draw[draw opacity=0]
    (298.62,230.23)
    .. controls (309.8,227.77) and (318,219.64) .. (318,210)
    .. controls (318,200.04) and (309.25,191.7) .. (297.52,189.54)
    -- (291.5,210) -- cycle ;
\draw[black,thick]
    (298.62,230.23)
    .. controls (309.8,227.77) and (318,219.64) .. (318,210)
    .. controls (318,200.04) and (309.25,191.7) .. (297.52,189.54) ;
\end{tikzpicture}\Big\rangle&= e^{\frac{\lambda^2}{2} F(\varphi)}\left( \frac{\epsilon r_R(r_{\text{NS}})}{x_2} \right)^{\Gamma+2} 2 \lambda \csc\frac{\varphi}{2}\left(-\pi +\varphi +\cot\frac{\varphi}{2}+\frac{1}{2} \sin\varphi\right) \\
\Big \langle \hat{\mathcal{O}}_4(x_1) \Big| 
\begin{tikzpicture}[x=0.75pt,y=0.75pt,yscale=-0.5,xscale=0.5,baseline={(0,-2.9)},line cap=round,line join=round]
\draw[draw opacity=0]
    (298.62,230.23)
    .. controls (309.8,227.77) and (318,219.64) .. (318,210)
    .. controls (318,200.04) and (309.25,191.7) .. (297.52,189.54)
    -- (291.5,210) -- cycle ;
\draw[black,thick]
    (298.62,230.23)
    .. controls (309.8,227.77) and (318,219.64) .. (318,210)
    .. controls (318,200.04) and (309.25,191.7) .. (297.52,189.54) ;
\end{tikzpicture}\Big\rangle&= e^{\frac{\lambda^2}{2} F(\varphi)}\left( \frac{\epsilon r_R(r_{\text{NS}})}{x_2} \right)^{\Gamma+2} \lambda^2\left(\pi -\varphi-2  \cot(\varphi/2) \right)^2,
\end{split}
\end{equation} 
where
\begin{align}
{F}(\varphi)
&=
\frac{2(\pi-\varphi)}{\sin\varphi}
\log\!\left(\tan\frac{\varphi}{2}\right)
+
2\log\!\left(\frac{\cos(\varphi/2)}{2}\right)
\nonumber\\
&\quad
-
\frac{4}{\sin\varphi}
\operatorname{Im}\!\left[
\operatorname{Li}_2\!\left(e^{i\varphi}\right)
\right]
-2(\pi-\varphi)\tan\!\left(\frac{\varphi}{2}\right),
\end{align} 
with $r_R(r_{\text{NS}})$ given precisely as in  equation \eqref{rtrans}.  
The relation between these results and the COE coefficients $c_n(\varphi)$ in \eqref{COEsimple} is 
\begin{equation} \label{presc}
 c_n(\varphi) \, \left(r_R(r_{\text{NS}}) \right)^{\Gamma_n}
 = \frac{ \Big\langle \hat{\mathcal{O}}_n(x_1) \Big|
\begin{tikzpicture}[x=0.75pt,y=0.75pt,yscale=-0.5,xscale=0.5,baseline={(0,-2.9)},line cap=round,line join=round]
\draw[draw opacity=0]
    (298.62,230.23)
    .. controls (309.8,227.77) and (318,219.64) .. (318,210)
    .. controls (318,200.04) and (309.25,191.7) .. (297.52,189.54)
    -- (291.5,210) -- cycle ;
\draw[black,thick]
    (298.62,230.23)
    .. controls (309.8,227.77) and (318,219.64) .. (318,210)
    .. controls (318,200.04) and (309.25,191.7) .. (297.52,189.54) ;
\end{tikzpicture} \Big\rangle}{\sqrt{\Big\langle \hat{\mathcal{O}}_n(x_1) | \hat{\mathcal{O}}_n(x_1)\Big\rangle }}   ,
\end{equation}
which is regularization-independent. Using this relation, we find 
\begin{equation} \label{COEcoeff}
\begin{split}
c_0(\varphi) &=\frac{ e^{\frac{\lambda^2}{2} F(\varphi)} }{e^{\frac{\lambda^2}{4} F_{\text{almond}}} } \\
c_1(\varphi)&= c_0(\varphi)(-\lambda)\left(\pi -\varphi-2  \cot(\varphi/2) \right) \\ 
c_2(\varphi)&=0 \\
c_3(\varphi)&=c_0(\varphi) \sqrt{2}\, \lambda \csc\frac{\varphi}{2}\left(-\pi +\varphi +\cot\frac{\varphi}{2}+\frac{1}{2} \sin\varphi\right) \\
c_4(\varphi)&=c_0(\varphi)\frac{\lambda^2}{\sqrt{2}} \left(\pi -\varphi-2  \cot(\varphi/2) \right)^2 .
\end{split}
\end{equation}
with $F_{\text{almond}}$ given as \eqref{falm}.
The vanishing of the COE coefficient for $\hat{\mathcal{O}}_2$ is due to the $y$ reflection symmetry of the contour.

\subsection{Explicit Check of the COE in Radial Quantization}\label{sec:coeexample}

\begin{figure}[t]
  \centering

\begin{tikzpicture}[
    scale=1.25,
    line cap=round,
    line join=round
]

\def\th{28}      
\def\R{2.0}      
\def\ac{1.05}    
\def\bc{4.00}    

\pgfmathsetmacro{\rs}{\ac*sin(\th)}
\pgfmathsetmacro{\rb}{\bc*sin(\th)}

\pgfmathsetmacro{\angup}{90+\th}
\pgfmathsetmacro{\angdown}{270-\th}

\coordinate (O)  at (0,0);
\coordinate (Cs) at (\ac,0);   
\coordinate (Cb) at (\bc,0);   

\coordinate (Pp) at ({\R*cos(\th)},{ \R*sin(\th)});
\coordinate (Pm) at ({\R*cos(\th)},{-\R*sin(\th)});

\coordinate (Su) at ({\ac*cos(\th)^2},{ \ac*cos(\th)*sin(\th)});
\coordinate (Sd) at ({\ac*cos(\th)^2},{-\ac*cos(\th)*sin(\th)});

\coordinate (Bu) at ({\bc*cos(\th)^2},{ \bc*cos(\th)*sin(\th)});
\coordinate (Bd) at ({\bc*cos(\th)^2},{-\bc*cos(\th)*sin(\th)});

\draw[black,thick,dash pattern=on 5pt off 4pt]
    (O) circle (\R);

\draw[black,thick,dash pattern=on 5pt off 4pt] (O) -- (Pp);
\draw[black,thick,dash pattern=on 5pt off 4pt] (O) -- (Pm);

\draw[black,thick,dash pattern=on 5pt off 4pt]
    ({\ac-\rs},0) -- (Cs);

\draw[black,dotted] (O) -- (0,\R);


\draw[black,thick]
    (Su)
    arc[start angle=\angup, end angle=\angdown, radius=\rs];

\draw[black,thick] (Su) -- (Bu);

\draw[black,thick]
    (Bu)
    arc[start angle=\angup, end angle={-\angup}, radius=\rb];

\draw[black,thick] (Bd) -- (Sd);

\draw[black,thick]
    (\th:0.28)
    arc[
        start angle=\th,
        end angle=-\th,
        radius=0.28
    ];

\node at (0.4,0.02) {$\varphi$};

\fill (Cs) circle (1.5pt);

\node[left] at (0,1.0) {$R$};
\node[above] at ({\ac-0.5*\rs},0) {$r$};
\node[below right=-1pt] at (Cs) {$a$};


\node[left] at (O) {$0$};

\end{tikzpicture}

\caption{Smooth contour representing the norm of the state defined by the path integral inside the quantization circle (dashed).}
\label{fig:normcurve}
\end{figure}
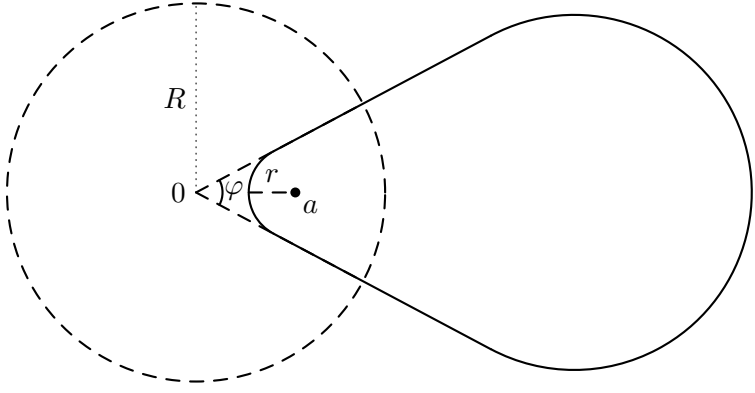
Let us now consider an explicit computation of the COE in the free scalar theory as one more check of the general features discussed in Section \ref{sec:hilb}. In particular, we consider Figure \ref{fig:normcurve}, and we compute the expectation value along that contour in two different ways. The first method consists in expanding the circular segments into cusp states using the COE, then taking the inner products between the resulting in and out states. The alternative is to directly compute the contour in field theory, and expand the results in the small radius ($r$) limit, in order to recover the COE. Here we show how these two methods produce identical results.

Take a smooth defect as shown in Figure \ref{fig:sce}, which consists of an arc of a circle centered at $a$
\begin{equation}
    C(\theta)= a +r e^{i\theta}, \hspace{1cm} \frac{\pi}{2}+\varphi/2\leq \theta \leq \frac{3 \pi}{2}-\varphi/2
\end{equation}
that joins smoothly to two $y-$symmetric radial rays through the origin $l_{\pm}=s e^{\pm i \varphi/2}$ with $s\in [0,\infty)$ and $\theta $ and $\varphi/2$ measured with respect to the real horizontal axis. The tangential union yields the center $a$ to be
\begin{equation}
    a=(r \csc\frac{\varphi}{2}, 0).
\end{equation}  
The points where the rays and the arc coincide are  parametrized by $s'=r \cot \varphi/2$. As prescribed in \eqref{presc}, the explicit form of the coefficients in the COE \eqref{smoothope} can be computed as the overlaps with unit-norm dilatation eigenstates
\begin{equation}
    \frac{\langle n | \hat{\Psi}_r \rangle }{\sqrt{\langle n|n\rangle}}= c_n(\varphi) r^{\Gamma_n}.
\end{equation}
As explained in Subsection \ref{sec:COENS}, the COE in radial quantization is related to that in the NS picture \eqref{COEsimple} by a conformal transformation, which doesn't change the angle-dependent coefficients, a fact that can also be checked explicitly. Therefore, with the coefficients \eqref{COEcoeff} computed above, the COE \eqref{smoothope} is given by
\begin{equation}
    |\hat{\Psi}_{r_R}\rangle= c_0(\varphi) r_R^\Gamma|0\rangle+ c_1(\varphi) r_R^{\Gamma+1} |1\rangle+\ldots,
\end{equation}
from which we easily obtain the norm:
\begin{equation} \label{expnorm}
    \langle \hat{\Psi}_r| \hat{\Psi}_r\rangle= r_R^{2 \Gamma}\left(c_0^2(\varphi)+c_1^2(\varphi) r_R^2 +\ldots\right).
\end{equation} 
On the other hand, this norm also corresponds to computing the expectation value of $W_\gamma$, for 
\begin{equation}
    W_\gamma= \exp\left(-\lambda \int_\gamma \mathrm{d}l \abs{\frac{\mathrm{d}x}{\mathrm{d}l}} \phi(x(l)) \right)
\end{equation}
with $\gamma$ the contour depicted in Figure \ref{fig:normcurve}.
The in-contour (inside the quantization circle) is given by the union of the following contours 
\begin{align}
    C(\theta)&=a +r e^{i\theta}, \hspace{1cm} \frac{\pi}{2}+\frac{\varphi}{2}\leq \theta \leq  \frac{3\pi}{2}-\varphi/2  \\
    U_{\text{in}}(s)&= s e^{i\varphi/2}, \hspace{2.3cm} s'\leq s\leq R \\
    L_{\text{in}}(s)&= s e^{-i\varphi/2}, \hspace{2.1cm} s'\leq s\leq R
\end{align}
with $s'=r \cot\varphi/2<R$, where $U$ and $L$ stand for the upper and lower ray respectively. The out-contour is constructed by radial reflection $I_R(z):= \frac{R^2}{\bar{z}}$ and it can be parametrized as 
\begin{align}
    C_*(\theta)&=a_* +r_* e^{i\theta}, \hspace{0.5cm} -\frac{\pi}{2}-\frac{\varphi}{2}\leq \theta \leq  \frac{\pi}{2}+\varphi/2  \\
    U_{\text{out}}(s)&= s e^{i\varphi/2}, \hspace{2.5cm} R\leq s\leq s'_* \\
    L_{\text{out}}(s)&= s e^{-i\varphi/2}, \hspace{2.3cm} R\leq s\leq s'_*
\end{align} 
with 
\begin{equation}
    \begin{split}
        s'_*&=\frac{R^2 \tan\varphi/2}{r}\\
        a_*&= (s'_* \sec\varphi/2,0) \\
        r_*&= s'_* \tan\varphi/2.
    \end{split}
\end{equation}

Defining $U=U_{\text{in}}\cup U_{\text{out}}$ and $L=L_{\text{in}}\cup L_{\text{out}}$, the whole contour $\gamma$ is given by $\gamma=C\cup U\cup L \cup \ C_*$, and Wick contractions in the free field theory yield 
\begin{equation}
    \langle W_\gamma \rangle= \exp\left( \frac{\lambda^2}{2}\left[  2 I_{UU}+I_{CC} + I_{C_*C_*}+2 I_{UL}+4 I_{UC}+ 4 I_{U C_*} +2 I_{C C_*}\right]\right),
\end{equation}
where we have also used the fact that, by symmetry, $I_{UC}=I_{LC}$ and $I_{UC_*}=I_{LC_*}$. For instance, $I_{UL}$ is given by 
\begin{equation}
    \begin{split}
    I_{UL}&=\int_{s'}^{s'_*} \mathrm{d}s \int_{s'}^{s'_*} \mathrm{d}t \frac{1}{\abs{se^{i\varphi/2}-t e^{-i\varphi/2}}^2} \\
    &=\frac{2(\pi-\varphi)}{\sin\varphi} \log\left(\frac{\tan\varphi/2}{q}\right) -\frac{2}{\sin\varphi} \Im\left[ \operatorname{Li}_2(e^{i\varphi}) -\operatorname{Li}_2(q^2 \cot^2(\varphi/2) e^{i\varphi}) \right] \\
    & \underset{q\rightarrow0}{=} -\frac{2(\pi-\varphi)}{\sin \varphi} \log q +\frac{2(\pi-\varphi)}{\sin\varphi} \log(\tan\varphi/2) -\frac{2}{\sin\varphi} \Im \operatorname{Li}_2(e^{i\varphi})+2q^2 \cot^2(\varphi/2),
    \end{split} 
\end{equation}
where $q=r_R/R$. In addition, as we have always done above, when evaluating diagrams with propagators ending on the same portion of the contour, such as $I_{UU}$, we subtract the perimeter divergence.
All in all, we find 
\begin{equation}
    \langle W_\gamma \rangle=c_0^2(\varphi) \left(\frac{r_R}{R} \right)^{2 \Gamma}\left [1+\lambda^2 (\pi-\varphi -2 \cot \varphi/2)^2 \left( \frac{r_R}{R}\right)^2 +\cdots\right]
\end{equation}
and setting $R=1$ we find agreement with \eqref{expnorm}, providing an explicit check of the COE \eqref{smoothope}.

\section{Covariance properties of NS Hamiltonian Eigenstates in CFT}\label{transformations-of-eigenstates}\label{subsec:Hamiltonians}

Eigenstates of NS quantization played an important role in defining the cusp operators with the right transformation law in Section \ref{sec:generic-cusp-correlators}. However, the existence of the defect itself did not: the trivial defect is in particular conformal, and the argument given in \ref{sec:transfExcited} goes through. This leaves us with an apparent puzzle: operators in a CFT certainly do not all transform like primaries. This question was essentially addressed in Section \ref{sec:examples}, where we noticed that scaling cusp operators are explicitly coordinate dependent in the limit where the defect trivializes. In this section, we consider vanilla translational invariant CFTs and elaborate on this observation. We show that the results in previous sections are consistent, and we demystify them to some extent. In this section, we work in one spacetime dimension for simplicity.

In an ordinary CFT, consider a conformal transformations $U_f$ acting as $f(x)$ on spacetime points $x$.
Then, given a primary operator (scalar, for simplicity) $\mathcal{O}(x)$, 
\begin{equation}
    U_f \mathcal{O}(x) U_f^{-1}
    =
    (b_f(x))^{\Delta}\mathcal{O}(f(0))~.
\end{equation}
for a scalar primary of dimension $\Delta$, with
\begin{equation}
    b_f(x)
    =
    \left|
    \frac{\partial f(x)}{\partial x}
    \right|
\end{equation}
We also require that
\begin{equation}
    b_f(0)=1~,
    \label{bconstraint}
\end{equation}
for reasons that will become clear in a moment.
Let us now define the following family of operators:
  
\begin{equation}\label{OnDef}
    \mathcal{O}^{(n)}(f(0)|f(\infty))
    =
    U_f P^n \mathcal{O}(0) U_f^{-1}~,
\end{equation}
where $P^n \mathcal{O}$ denotes the nested commutator. This is just the set of transformed descendants. 
The fields contained in $\mathcal{O}_f^{(n)}$ are evaluated at $f(0),$ but explicit dependence on  $f(\infty)$ also arises through the operator $U_f$, unless the transformation is a pure translation or rotation. In fact, $U_f$ is fixed by specifying $f(0),\, f(\infty)$ and \eqref{bconstraint},\footnote{In $d>1$, the action of the transverse $SO(d)$ must be specified as well. This is the only complication when repeating the arguments of this section in higher $d$. For instance, everything goes through unchanged for scalar operators in all dimensions.} which justifies the notation. We can therefore interpret \eqref{OnDef} as defining a bilocal operator: given any pair of points $x_1$ and $x_2$, $O^{(n)}(x_1|x_2)$ is unique.
Explicitly, the map is obtained by specifying $a=x_2-x_1=x_{21}$ in (the real slice of) \eqref{confmap}, in order to satisfy \eqref{bconstraint}:
\begin{equation} \label{condg}
    f(x)= x_1+\frac{x_{21} x}{x+x_{21}}~.
\end{equation}

The first obvious but important feature of the operators \eqref{OnDef} is that they create eigenstates of the Hamiltonian $H_f=U_f D U_f^{-1}$. In particular, if both $f(0)$ and $f(\infty)$ are finite, this is the NS quantization Hamiltonian evolving from one point to the other. 

The condition \eqref{bconstraint} can now be understood as a constraint on the norm of these eigenstates. Consider the norm of $P^n O(0) \ket{0}$ on a sphere of radius $\epsilon$ centered at the origin. Conjugation by $U_f$ preserves this norm computed on the transformed quantization sphere, i.e. when the adjoint is computed using $U_f I(\epsilon) U_f^{-1}$, $I(\epsilon)$ being the inversion the preserves the original sphere. The condition \eqref{bconstraint} ensures that a quantization sphere of radius $\epsilon$ is sent to a sphere of equal radius in the limit $\epsilon \to 0$, and so the norm of
$\mathcal{O}^{(n)}(x_1|x_2)\ket{0}$ on a fixed small sphere surrounding $x_1$ is independent of $f$ (in fact, restricting $f$ so that $b_f(0)$ is any $f$-independent value would work equally well, and in particular one could take it to scale with $\epsilon$ to give the state a finite norm in the limit $\epsilon \to 0$).

In sum, equation \eqref{OnDef} precisely defines the local operators used in the proof of covariance of Section \eqref{sec:generic-cusp-correlators}, for the case of the trivial defect: they create eigenstates of NS quantization, with a position independent norm on a vanishing quantization sphere. Correspondingly, these are the operators that diagonalize the expectation value on the almond in the example of the pinning field defect. For instance, using \eqref{condg} one gets
\begin{equation}\label{eq:PcommaO}
    \mathcal{O}^{(1)}(x_1|x_2)= \left(\partial_{x_1} -\frac{2 \Delta}{x_{21}} \right )\mathcal{O}(x_1),
\end{equation}  
which matches \eqref{lzero}. 

It is now easy to check that indeed all of the $O^{(n)}(x_1|x_2)$ transform like primaries. 
We apply another conformal transformation $U_g$, this time arbitrary:
\begin{equation} \label{Ugf}
    U_g \mathcal{O}^{(n)}(x_1|x_2) U_g^{-1}
    =
    U_{g\circ f} P^n \mathcal{O}(0) U_{g\circ f}^{-1}~.
\end{equation}
Since
\begin{equation}
    b_{g\circ f}(0)
    =
    b_g(f(0)) b_f(0) =  b_g(x_1)~,
\end{equation}
the composed transformation does not obey \eqref{bconstraint}. One can easily compensate by composing to it on the right a dilatation $U_D(\lambda)$ of parameter $\lambda=1/b_g(x_1)$, to find
\begin{equation}
\begin{split}
    O^{(n)}(g(x_1)|g(x_2)) = U_{g\circ f} \,U_D(\lambda)\, P^n \mathcal{O}(0) \,U_D(\lambda)^{-1}\, U_{g\circ f}^{-1} \\ =
    \frac{1}{b_g(x_1)^{\Delta+n}}  U_{g\circ f} P^n \mathcal{O}(0) U_{g\circ f}^{-1}~.
    \end{split}
\end{equation}
By replacing this equation in \eqref{Ugf}, we conclude that
\begin{equation} \label{bilocPrim}
    U_g \mathcal{O}^{(n)}(x_1|x_2) U_g^{-1}
    = b_g(x_1)^{\Delta+n}\, O^{(n)}(g(x_1)|g(x_2))~,
\end{equation}
which proves the (bilocal) primary transformation law.

For reference, the transformed translation operator appearing in \eqref{OnDef} is 
\begin{equation} \label{p'}
   P_f= U_f P U_f^{-1}
    = \frac{1}{x_{21}^2}
    \left(
        K-2x_2 D+x_2^2 P
    \right)~,
\end{equation}
as obtained from
\begin{equation}
    U_f= e^{x_1 P} 
    \circ e^{-(1/x_{21})K}~.
\end{equation}
By repeated applications of \eqref{p'}, one can construct all of the $O^{(n)}$. For instance,
\begin{equation}
    \mathcal{O}^{(2)}(x_1|x_2)=[P_f,[P_f, \mathcal{O}(x_1)]]=\left(\partial_{x_1}^2+\frac{2(1+2\Delta)}{x_1-x_2}\partial_{x_1}  + \frac{2 \Delta(1+2\Delta)}{(x_1-x_2)^2}\right)\mathcal{O}(x_1)~.
\end{equation}
One can directly check that these operators transform like bilocal primaries.

In a translational invariant CFT, the property \eqref{bilocPrim} amounts to a repackaging: a coordinate dependent linear combination of operators belonging to the same family can be designed to compensate for the inhomogeneous terms in the conformal transformation of each of them. As already remarked, on a cusped defect whose branches are arcs of circle, the value of $x_2$ is fixed as a function of $x_1$ and the curvature, hence NS quantization eigenstates give rise to local cusp operators. The construction in this section also explicitly shows that these operators can be constructed unambiguously, thanks to the normalization condition.

We conclude the section with a couple of instructive checks, which highlight the role of the bi-locality of the operators $\eqref{OnDef}.$

\subsection{Correlation functions of NS eigenstates}\label{3-pt-transform}
Consider first the three-point function
\begin{equation}
    \langle \mathcal{O}_1^{(1)}(x_1|x_S)\mathcal{O}_2(x_2)\mathcal{O}_3(x_3)\rangle ~,
\end{equation}
where $\mathcal{O}_{2}$ and $\mathcal{O}_3$ are local primary operators and $\mathcal{O}_1^{(1)}(x_1|x_S)$ is defined as in \eqref{eq:PcommaO} by acting on another local primary operator $\mathcal{O}_1$. 

Straightforwardly, one finds
\begin{equation}
\begin{split}
   &
   \braket{\mathcal{O}_1^{(1)}(x_1|x_S)\mathcal{O}_2(x_2)\mathcal{O}_3(x_3)}
   = - (\Delta_1 - \Delta_2 + \Delta_3 +2 \Delta_1 \zeta)\\
   &\times \frac{c_{\mathcal{O}_1 \mathcal{O}_2 \mathcal{O}_3}}{(x_2 - x_1)^{1 + \Delta_1 + \Delta_2 - \Delta_3} (x_3 - x_1)^{1 + \Delta_1 + \Delta_3 -\Delta_2} (x_3 - x_2)^{ \Delta_2 + \Delta_3-1 - \Delta_1 }} 
\end{split}
\end{equation}
where
\begin{equation}
    \zeta=\frac{(x_2-x_S)(x_3-x_1)}{(x_S-x_1)(x_3-x_2)}
\end{equation} 
is a conformal invariant, and $c_{\mathcal{O}_1 \mathcal{O}_2 \mathcal{O}_3}$ is the OPE coefficient of $\langle \mathcal{O}_1(x_1) \mathcal{O}_2(x_2) \mathcal{O}_3(x_3) \rangle$.
Because of the bi-local nature of the transformation law \eqref{bilocPrim}, the result is not fixed in terms of $x_1,\,x_2,\,x_3$ alone. Instead, the `coefficient of the three-point function' is position-dependent via the conformal invariant combination of the three insertion points with the south pole quantization point $x_S.$ This feature highlights why planar cusped loops with circular branches are special: in these cases, the value of $x_S$ is fixed in terms of the position of the cusps, and the cross ratio $\zeta$ is constant.

The same is true for an arbitrary number of cusps connected by planar arcs---see Section \ref{subsec:3loop}. To illustrate this point, 
consider a general 4-pt function of collinear points
\begin{align} \label{gen4}
\langle O_1(x_1) O_2(x_2) O_3(x_3) O_4(x_4)\rangle
=
\left(\frac{|x_{24}|}{|x_{14}|}\right)^{\Delta_{12}}
\left(\frac{|x_{14}|}{|x_{13}|}\right)^{\Delta_{34}}
\frac{G(\chi)}{|x_{12}|^{\Delta_1+\Delta_2}|x_{34}|^{\Delta_3+\Delta_4}}~,
\end{align}
with
\begin{equation}
\Delta_{ij} = \Delta_i - \Delta_j,
\qquad
\chi = \frac{|x_{12}||x_{34}|}{|x_{13}||x_{24}|}~.
\end{equation}

Without loss of generality, we assume $x_1 < x_2 < x_3 < x_4$. Applying the differential operator
$\partial_{x_1} - \frac{2\Delta_1}{x_2 - x_1} $ to the correlator above yields
\begin{align}
\nonumber
&\left(\partial_{x_1} - \frac{2\Delta_1}{x_2-x_1}\right)
\langle O_1(x_1) O_2(x_2) O_3(x_3) O_4(x_4)\rangle
\\[6pt] \nonumber
\nonumber
&=
(x_2-x_1)^{-\Delta_1-1-\Delta_2}
(x_4-x_3)^{-\Delta_3-\Delta_4}
\left(\frac{x_4-x_1}{x_3-x_1}\right)^{\Delta_3-\Delta_4}
\left(\frac{x_4-x_2}{x_4-x_1}\right)^{\Delta_1+1-\Delta_2}
\\[6pt] 
&\quad \times
\Big[ \left(
(\Delta_2-\Delta_1)
-
\chi(\Delta_4-\Delta_3)
 \right)G(\chi)
-\chi(1-\chi) G'(\chi) \Big]~,
\end{align}
where $'$  indicates derivative with respect to $\chi$. Comparing with \eqref{gen4}, the prefactors are those expected for a 4-pt function of primaries with dimensions
\begin{equation}
\{\Delta_1+1, \Delta_2, \Delta_3, \Delta_4\}~,
\end{equation}
and no further cross-ratios have been introduced. This is a consequence of having inserted the operator $O^{(1)}(x_1|x_2)$, where the south pole quantization point has been chosen to coincide with another insertion.

\section{A Cuspy Twist on Tauberian Analysis} \label{sec:bootsconf}


\begin{figure}[t]
\centering
\resizebox{0.98\linewidth}{!}{%
\begin{tikzpicture}[
    line cap=round,
    line join=round,
    scale=1,
    contour/.style={black, thick},
    aux/.style={black, thick, dash pattern=on 5pt off 4pt},
    angle mark/.style={black, semithick}
]


\coordinate (A0) at (0.0,0.0);      
\coordinate (A1) at (1.860370,1.076544);    
\coordinate (A2) at (1.860370,-1.076544);   
\coordinate (A4) at (5.189630,1.076544);    
\coordinate (A3) at (5.189630,-1.076544);   
\coordinate (A5) at (7.05,0.0);     

\draw[aux]
    (A0) .. controls (0.538440,0.498400) and (1.177310,0.857248) .. (A1);
\draw[contour]
    (A1) .. controls (2.933750,1.421152) and (4.116250,1.421152) .. (A4);
\draw[aux]
    (A4) .. controls (5.872690,0.857248) and (6.511560,0.498400) .. (A5);

\draw[aux]
    (A0) .. controls (0.538440,-0.498400) and (1.177310,-0.857248) .. (A2);
\draw[contour]
    (A2) .. controls (2.933750,-1.421152) and (4.116250,-1.421152) .. (A3);
\draw[aux]
    (A3) .. controls (5.872690,-0.857248) and (6.511560,-0.498400) .. (A5);

\draw[contour]
    (A1) .. controls (1.45,0.62) and (1.45,-0.62) .. (A2);
\draw[contour]
    (A4) .. controls (5.60,0.62) and (5.60,-0.62) .. (A3);

\path
    (A1) .. controls (2.933750,1.421152) and (4.116250,1.421152) ..
    coordinate[pos=0.1005552553] (A1topAngle)
    coordinate[pos=0.8994447447] (A4topAngle) (A4);
\path
    (A2) .. controls (2.933750,-1.421152) and (4.116250,-1.421152) ..
    coordinate[pos=0.1005552553] (A2bottomAngle)
    coordinate[pos=0.8994447447] (A3bottomAngle) (A3);
\path
    (A1) .. controls (1.45,0.62) and (1.45,-0.62) ..
    coordinate[pos=0.1696861589] (A1sideAngle)
    coordinate[pos=0.8303138411] (A2sideAngle) (A2);
\path
    (A4) .. controls (5.60,0.62) and (5.60,-0.62) ..
    coordinate[pos=0.1696861589] (A4sideAngle)
    coordinate[pos=0.8303138411] (A3sideAngle) (A3);
\path
    (A0) .. controls (0.538440,0.498400) and (1.177310,0.857248) ..
    coordinate[pos=0.1551303881] (A0topAngle) (A1);
\path
    (A0) .. controls (0.538440,-0.498400) and (1.177310,-0.857248) ..
    coordinate[pos=0.1551303881] (A0bottomAngle) (A2);

\draw[angle mark]
    (A0bottomAngle)
    arc[start angle=-40.7360988,end angle=40.7360988,radius=0.34];
\node at ($(A0)+(0:0.61)$) {$\varphi_s$};

\draw[angle mark]
    (A1sideAngle)
    arc[start angle=-120.6745062,end angle=15.9625523,radius=0.34];
\node at ($(A1)+(-52.3559769:0.61)$) {$\varphi_1$};

\draw[angle mark]
    (A2bottomAngle)
    arc[start angle=-15.9625523,end angle=120.6745062,radius=0.34];
\node at ($(A2)+(52.3559769:0.61)$) {$\varphi_2$};

\draw[angle mark]
    (A3sideAngle)
    arc[start angle=59.3254938,end angle=195.9625523,radius=0.34];
\node at ($(A3)+(127.6440231:0.61)$) {$\varphi_3$};

\draw[angle mark]
    (A4topAngle)
    arc[start angle=164.0374477,end angle=300.6745062,radius=0.34];
\node at ($(A4)+(232.3559769:0.61)$) {$\varphi_4$};

\fill (A0) circle (2.1pt);
\fill (A1) circle (2.1pt);
\fill (A2) circle (2.1pt);
\fill (A3) circle (2.1pt);
\fill (A4) circle (2.1pt);
\fill (A5) circle (2.1pt);

\node[left]        at (A0) {$x_s$};
\node[above left]  at (A1) {$x_1$};
\node[below left]  at (A2) {$x_2$};
\node[below right] at (A3) {$x_3$};
\node[above right] at (A4) {$x_4$};
\node[right]       at (A5) {$x_{\bar{s}}$};


\node at (7.95,0.0) {$=$};
\node at (8.85,-0.05) {$\displaystyle \sum_n$};


\begin{scope}[shift={(10.05,1.25)}]

\coordinate (B0) at (0.0,0.0);       
\coordinate (B4) at (3.217570,0.667457);     
\coordinate (B3) at (3.217570,-0.667457);    

\draw[contour]
    (B0) .. controls (0.858427,0.794592) and (2.128578,1.017078) .. (B4);
\draw[contour]
    (B0) .. controls (0.858427,-0.794592) and (2.128578,-1.017078) .. (B3);
\draw[contour]
    (B4) .. controls (3.472000,0.384400) and (3.472000,-0.384400) .. (B3);

\path
    (B0) .. controls (0.858427,0.794592) and (2.128578,1.017078) ..
    coordinate[pos=0.0603284875] (B0topAngle)
    coordinate[pos=0.9385495512] (B4topAngle) (B4);
\path
    (B0) .. controls (0.858427,-0.794592) and (2.128578,-1.017078) ..
    coordinate[pos=0.0603284875] (B0bottomAngle)
    coordinate[pos=0.9385495512] (B3bottomAngle) (B3);
\path
    (B4) .. controls (3.472000,0.384400) and (3.472000,-0.384400) ..
    coordinate[pos=0.1696861414] (B4sideAngle)
    coordinate[pos=0.8303138586] (B3sideAngle) (B3);

\draw[angle mark]
    (B0bottomAngle)
    arc[start angle=-40.7361022,end angle=40.7361022,radius=0.2108];
\node at ($(B0)+(0:0.50)$) {$\varphi_s$};

\draw[angle mark]
    (B4topAngle)
    arc[start angle=164.0374442,end angle=300.6745180,radius=0.2108];
\node at ($(B4)+(232.3559811:0.48)$) {$\varphi_4$};

\draw[angle mark]
    (B3sideAngle)
    arc[start angle=59.3254820,end angle=195.9625558,radius=0.2108];
\node at ($(B3)+(127.6440189:0.48)$) {$\varphi_3$};

\fill (B0) circle (2.1pt);
\fill (B4) circle (2.1pt);
\fill (B3) circle (2.1pt);

\node[left]        at (B0) {$x_s$};
\node[above right] at (B4) {$x_4$};
\node[below right] at (B3) {$x_3$};
\node at ($(B0)+(-0.15,0.38)$) {$n$};

\end{scope}

\node at (14.00,1.25) {$\times$};


\begin{scope}[shift={(14.732430,1.25)}]

\coordinate (C1) at (0.0,0.667457);      
\coordinate (C2) at (0.0,-0.667457);     
\coordinate (C5) at (3.217570,0.0);      

\draw[contour]
    (C1) .. controls (-0.254429,0.384400) and (-0.254429,-0.384400) .. (C2);
\draw[contour]
    (C1) .. controls (1.088993,1.017078) and (2.359143,0.794592) .. (C5);
\draw[contour]
    (C2) .. controls (1.088993,-1.017078) and (2.359143,-0.794592) .. (C5);

\path
    (C1) .. controls (-0.254429,0.384400) and (-0.254429,-0.384400) ..
    coordinate[pos=0.1696861589] (C1sideAngle)
    coordinate[pos=0.8303138411] (C2sideAngle) (C2);
\path
    (C1) .. controls (1.088993,1.017078) and (2.359143,0.794592) ..
    coordinate[pos=0.0614504435] (C1topAngle) (C5);
\path
    (C2) .. controls (1.088993,-1.017078) and (2.359143,-0.794592) ..
    coordinate[pos=0.0614504435] (C2bottomAngle) (C5);

\draw[angle mark]
    (C1sideAngle)
    arc[start angle=-120.6745062,end angle=15.9625546,radius=0.2108];
\node at ($(C1)+(-52.3559758:0.48)$) {$\varphi_1$};

\draw[angle mark]
    (C2bottomAngle)
    arc[start angle=-15.9625546,end angle=120.6745062,radius=0.2108];
\node at ($(C2)+(52.3559758:0.48)$) {$\varphi_2$};

\fill (C1) circle (2.1pt);
\fill (C2) circle (2.1pt);
\fill (C5) circle (2.1pt);

\node[above left] at (C1) {$x_1$};
\node[below left] at (C2) {$x_2$};
\node[right]      at (C5) {$x_{\bar{s}}$};
\node at ($(C5)+(0.20,0.38)$) {$n$};

\end{scope}


\draw[contour] (9.45,0.0)--(18.55,0.0);


\begin{scope}[shift={(11.525,-1.52)}]

\coordinate (D0) at (0.0,0.0);
\coordinate (D5) at (4.95,0.0);

\draw[contour]
    (D0) .. controls (1.35,1.25) and (3.60,1.25) .. (D5);
\draw[contour]
    (D0) .. controls (1.35,-1.25) and (3.60,-1.25) .. (D5);

\path
    (D0) .. controls (1.35,1.25) and (3.60,1.25) ..
    coordinate[pos=0.0473250595] (D0topAngle) (D5);
\path
    (D0) .. controls (1.35,-1.25) and (3.60,-1.25) ..
    coordinate[pos=0.0473250595] (D0bottomAngle) (D5);
\draw[angle mark]
    (D0bottomAngle)
    arc[start angle=-40.5619744,end angle=40.5619744,radius=0.26];
\node at ($(D0)+(0:0.55)$) {$\varphi_s$};

\fill (D0) circle (2.1pt);
\fill (D5) circle (2.1pt);

\node[left]  at (D0) {$x_s$};
\node at ($(D0)+(-0,0.38)$) {$n$};
\node[right] at (D5) {$x_{\bar{s}}$};
\node at ($(D5)+(-0,0.38)$) {$n$};

\end{scope}

\end{tikzpicture}%
}
\caption{$s$-channel COE for a 4-point function defined by arcs on a plane.
The decomposition on the RHS contains products of 3-cusp functions divided
by cusp 2-point functions, whose kinematics is fixed and takes the canonical
form \eqref{eq:cusp2pt},\eqref{eq:cusp3pt}.}
\label{fig:sboot}
\end{figure}
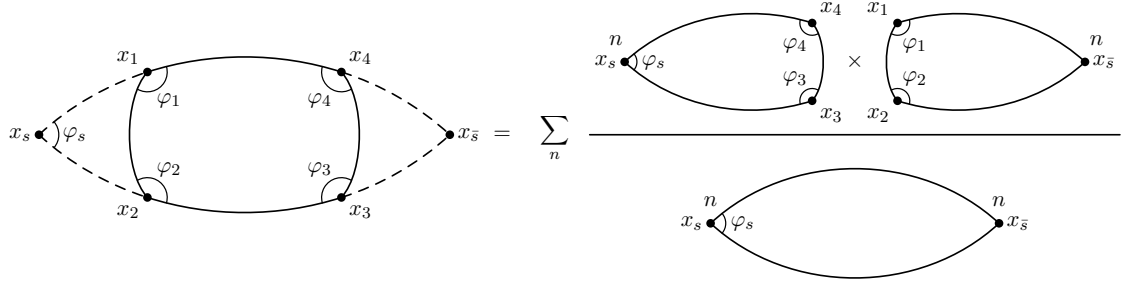


\begin{figure}[t]
\centering
\resizebox{0.7\linewidth}{!}{%
\begin{tikzpicture}[
    line cap=round,
    line join=round,
    contour/.style={black, thick},
    aux/.style={black, thick, dash pattern=on 5pt off 4pt},
    angle mark/.style={black, semithick}
]


\coordinate (A0) at (3.525,2.500);       
\coordinate (A1) at (1.860370,1.076544); 
\coordinate (A2) at (1.860370,-1.076544);
\coordinate (A3) at (5.189630,-1.076544);
\coordinate (A4) at (5.189630,1.076544); 
\coordinate (A5) at (3.525,-2.500);      

\draw[aux]
    (A0) arc[start angle=105.5756965,end angle=155.4930363,radius=2.595309];
\draw[contour]
    (A1) arc[start angle=155.4930363,end angle=204.5069637,radius=2.595309];
\draw[aux]
    (A2) arc[start angle=204.5069637,end angle=254.4243035,radius=2.595309];

\draw[aux]
    (A0) arc[start angle=74.4243035,end angle=24.5069637,radius=2.595309];
\draw[contour]
    (A4) arc[start angle=24.5069637,end angle=-24.5069637,radius=2.595309];
\draw[aux]
    (A3) arc[start angle=-24.5069637,end angle=-74.4243035,radius=2.595309];

\draw[contour]
    (A1) .. controls (2.933750,1.421152) and (4.116250,1.421152) .. (A4);
\draw[contour]
    (A2) .. controls (2.933750,-1.421152) and (4.116250,-1.421152) .. (A3);

\path
    (A1) .. controls (2.933750,1.421152) and (4.116250,1.421152) ..
    coordinate[pos=0.1005552553] (A1topAngle)
    coordinate[pos=0.8994447447] (A4topAngle) (A4);
\path
    (A2) .. controls (2.933750,-1.421152) and (4.116250,-1.421152) ..
    coordinate[pos=0.1005552553] (A2bottomAngle)
    coordinate[pos=0.8994447447] (A3bottomAngle) (A3);
\coordinate (A0leftAngle)  at ($(A0)+(-160.6685802:0.34)$);
\coordinate (A0rightAngle) at ($(A0)+(-19.3314198:0.34)$);
\coordinate (A1sideAngle)  at ($(A1)+(-110.7512404:0.34)$);
\coordinate (A2sideAngle)  at ($(A2)+(110.7512404:0.34)$);
\coordinate (A4sideAngle)  at ($(A4)+(-69.2487596:0.34)$);
\coordinate (A3sideAngle)  at ($(A3)+(69.2487596:0.34)$);

\draw[angle mark]
    (A0leftAngle)
    arc[start angle=-160.6685802,end angle=-19.3314198,radius=0.34];
\node at ($(A0)+(-90:0.62)$) {$\varphi_t$};

\draw[angle mark]
    (A1sideAngle)
    arc[start angle=-110.7512404,end angle=15.9625523,radius=0.34];
\node at ($(A1)+(-47.3943441:0.61)$) {$\varphi_1$};

\draw[angle mark]
    (A2bottomAngle)
    arc[start angle=-15.9625523,end angle=110.7512404,radius=0.34];
\node at ($(A2)+(47.3943441:0.61)$) {$\varphi_2$};

\draw[angle mark]
    (A3sideAngle)
    arc[start angle=69.2487596,end angle=195.9625523,radius=0.34];
\node at ($(A3)+(132.6056559:0.61)$) {$\varphi_3$};

\draw[angle mark]
    (A4topAngle)
    arc[start angle=164.0374477,end angle=290.7512404,radius=0.34];
\node at ($(A4)+(227.3943441:0.61)$) {$\varphi_4$};

\foreach \p in {A0,A1,A2,A3,A4,A5}
    \fill (\p) circle (2.1pt);

\node[above]       at (A0) {$x_{t}$};
\node[above left]  at (A1) {$x_1$};
\node[below left]  at (A2) {$x_2$};
\node[below right] at (A3) {$x_3$};
\node[above right] at (A4) {$x_4$};
\node[below]       at (A5) {$x_{\bar{t}}$};


\node at (6.70,0.0) {$=$};
\node at (7.45,-0.05) {$\displaystyle \sum_n$};


\begin{scope}[shift={(9.65,0.65)}]

\coordinate (B0) at (0.000000,1.609445);
\coordinate (B2) at (-0.749083,0.000000);
\coordinate (B3) at (0.749084,0.000000);

\draw[contour]
    (B0) arc[start angle=105.5756965,end angle=204.5069637,radius=1.167889];
\draw[contour]
    (B0) arc[start angle=74.4243035,end angle=-24.5069637,radius=1.167889];
\draw[contour]
    (B2) .. controls (-0.266063,-0.155074) and (0.266063,-0.155074) .. (B3);

\coordinate (B0leftAngle)  at ($(B0)+(-160.6685802:0.153)$);
\coordinate (B0rightAngle) at ($(B0)+(-19.3314198:0.153)$);
\coordinate (B2sideAngle)  at ($(B2)+(110.7512404:0.153)$);
\coordinate (B3sideAngle)  at ($(B3)+(69.2487596:0.153)$);
\path
    (B2) .. controls (-0.266063,-0.155074) and (0.266063,-0.155074) ..
    coordinate[pos=0.1005552553] (B2bottomAngle)
    coordinate[pos=0.8994447447] (B3bottomAngle) (B3);

\draw[angle mark]
    (B0leftAngle)
    arc[start angle=-160.6685802,end angle=-19.3314198,radius=0.153];
\node at ($(B0)+(-90:0.39)$) {$\varphi_t$};

\draw[angle mark]
    (B2bottomAngle)
    arc[start angle=-15.9625523,end angle=110.7512404,radius=0.153];
\node at ($(B2)+(47.3943441:0.40)$) {$\varphi_2$};

\draw[angle mark]
    (B3sideAngle)
    arc[start angle=69.2487596,end angle=195.9625523,radius=0.153];
\node at ($(B3)+(132.6056559:0.40)$) {$\varphi_3$};

\foreach \p in {B0,B2,B3}
    \fill (\p) circle (2.1pt);

\node[above]       at (B0) {$x_{t}$};
\node[below left]  at (B2) {$x_2$};
\node[below right] at (B3) {$x_3$};
\node at ($(B0)+(-0.40,0.05)$) {$n$};

\end{scope}

\node at (11.05,1.455) {$\times$};


\begin{scope}[shift={(12.45,2.259445)}]

\coordinate (C1) at (-0.749083,0.000000);
\coordinate (C4) at (0.749084,0.000000);
\coordinate (C5) at (0.000000,-1.609445);

\draw[contour]
    (C1) arc[start angle=155.4930363,end angle=254.4243035,radius=1.167889];
\draw[contour]
    (C4) arc[start angle=24.5069637,end angle=-74.4243035,radius=1.167889];
\draw[contour]
    (C1) .. controls (-0.266063,0.155074) and (0.266063,0.155074) .. (C4);

\coordinate (C1sideAngle) at ($(C1)+(-110.7512404:0.153)$);
\coordinate (C4sideAngle) at ($(C4)+(-69.2487596:0.153)$);
\path
    (C1) .. controls (-0.266063,0.155074) and (0.266063,0.155074) ..
    coordinate[pos=0.1005552553] (C1topAngle)
    coordinate[pos=0.8994447447] (C4topAngle) (C4);

\draw[angle mark]
    (C1sideAngle)
    arc[start angle=-110.7512404,end angle=15.9625523,radius=0.153];
\node at ($(C1)+(-47.3943441:0.40)$) {$\varphi_1$};

\draw[angle mark]
    (C4topAngle)
    arc[start angle=164.0374477,end angle=290.7512404,radius=0.153];
\node at ($(C4)+(227.3943441:0.40)$) {$\varphi_4$};

\foreach \p in {C1,C4,C5}
    \fill (\p) circle (2.1pt);

\node[above left]  at (C1) {$x_1$};
\node[above right] at (C4) {$x_4$};
\node[below]       at (C5) {$x_{\bar{t}}$};
\node at ($(C5)+(0.40,-0.02)$) {$n$};

\end{scope}

\draw[contour] (8.35,0.0)--(13.75,0.0);


\begin{scope}[shift={(11.05,-1.55)}]

\coordinate (D0) at (0.000000,1.000000);
\coordinate (D5) at (0.000000,-1.000000);

\draw[contour]
    (D0) arc[start angle=105.5756965,end angle=254.4243035,radius=1.038124];
\draw[contour]
    (D0) arc[start angle=74.4243035,end angle=-74.4243035,radius=1.038124];

\coordinate (D0leftAngle)  at ($(D0)+(-160.6685802:0.136)$);
\coordinate (D0rightAngle) at ($(D0)+(-19.3314198:0.136)$);

\draw[angle mark]
    (D0leftAngle)
    arc[start angle=-160.6685802,end angle=-19.3314198,radius=0.136];
\node at ($(D0)+(-90:0.36)$) {$\varphi_t$};

\fill (D0) circle (2.1pt);
\node at ($(D0)+(0.38,0)$) {$n$};
\fill (D5) circle (2.1pt);
\node at ($(D5)+(0.38,0)$) {$n$};
\node[above] at (D0) {$x_{t}$};
\node[below] at (D5) {$x_{\bar{t}}$};

\end{scope}

\end{tikzpicture}%
}
\caption{$t$-channel COE for a 4-point function defined by arcs on a plane.
The decomposition on the RHS contains products of 3-cusp functions divided
by cusp 2-point functions, whose kinematics is fixed and takes the canonical
form \eqref{eq:cusp2pt},\eqref{eq:cusp3pt}.}
\label{fig:tboot}
\end{figure}
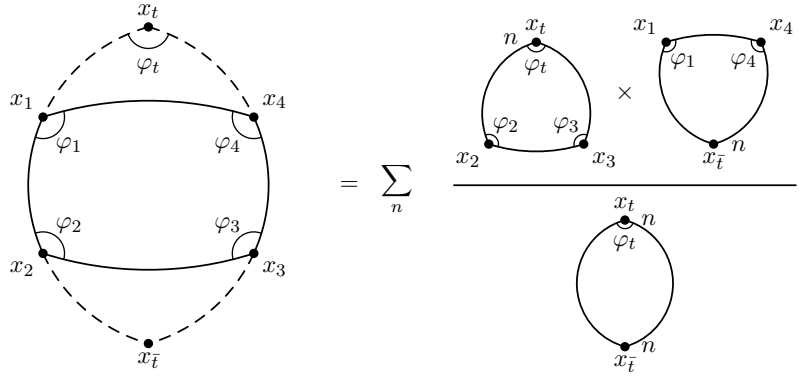

Given the existence of the COE expansion, it is a natural question to ask what constraints can be placed on the dynamical data of the cusp Hilbert space. 
In this section we discuss two setups in which 
 consistency conditions such as the equivalence of the COE in two channels (cf. Figures \ref{fig:sboot} and \ref{fig:tboot})  lead to  analytic constraints on the cusp data. Reflection positivity is requires for these results to be valid, so the defect types are assumed to be equal on all arcs (one can easily relax this requirement to picking them pairwise equal, which we do not do explicitly for notational simplicity).

\subsection{The asymptotc COE Density and Defect Fusion
}\label{sec:fusingcusps} 
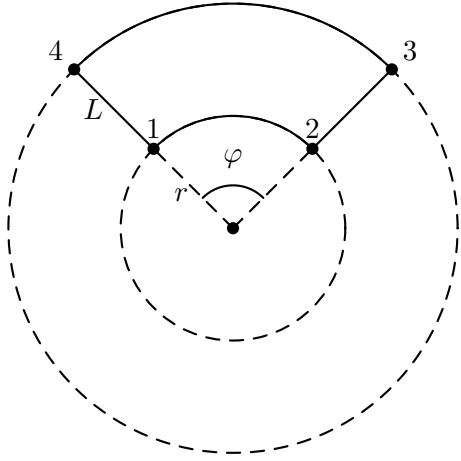
\begin{figure}[t]
\centering

\begin{tikzpicture}[
    scale=1.35,
  line cap=round,
    line join=round,
    contour/.style={black, thick},
    aux/.style={black, thick, dash pattern=on 5pt off 4pt}
]

\def\Rout{2.2}   
\def\Rin{1.1}    

\def\angL{135}
\def\angR{45}

\coordinate (O) at (0,0);

\coordinate (P1) at (\angL:\Rin);
\coordinate (P4) at (\angL:\Rout);

\coordinate (P2) at (\angR:\Rin);
\coordinate (P3) at (\angR:\Rout);


\draw[aux] (O) circle (\Rout);

\draw[aux] (O) circle (\Rin);

\draw[aux] (O) -- (P1);
\draw[aux] (O) -- (P2);


\draw[contour]
    (P1)
    arc[
        start angle=\angL,
        end angle=\angR,
        radius=\Rin
    ];

\draw[contour]
    (P2) -- (P3);

\draw[contour]
    (P3)
    arc[
        start angle=\angR,
        end angle=\angL,
        radius=\Rout
    ];

\draw[contour]
    (P4) -- (P1);

\draw[black,thick]
    (\angR:0.42)
    arc[
        start angle=\angR,
        end angle=\angL,
        radius=0.42
    ];

\node at (90:0.68) {$\varphi$};

\fill[black] (O)  circle (1.7pt);
\fill[black] (P1) circle (1.7pt);
\fill[black] (P2) circle (1.7pt);
\fill[black] (P3) circle (1.7pt);
\fill[black] (P4) circle (1.7pt);

\node at ($(P4)+(-0.18,0.18)$) {$4$};
\node at ($(P1)+(0,0.20)$) {$1$};

\node at ($(P2)+(0,0.20)$) {$2$};
\node at ($(P3)+(0.18,0.18)$) {$3$};

\node at ($(O)!0.52!(P1)+(-0.10,-0.08)$) {$r$};
\node at ($(P4)!0.5!(P1)+(-0.20,0.00)$) {$L$};

\end{tikzpicture}

\caption{Configuration of four cusps formed by two concentric circles of radii $r$ and $r+L$ respectively. For $L\rightarrow 0^+$, we expect a leading divergence determined by the Casimir energy. }
\label{fig:newcusp}
\end{figure}

\begin{figure}[t]
\centering

\begin{tikzpicture}[
    x=0.75pt,
    y=0.75pt,
    yscale=-1,
    xscale=1,
    line cap=round,
    line join=round,
    contour/.style={black,thick}
]



\coordinate (L1) at (129.63,113.34);
\coordinate (L2) at (133.19,179.38);
\coordinate (L3) at (170.79,191.07);
\coordinate (L4) at (165.59,96.61);

\draw[contour]
    (L3) .. controls (177.07,180.21) and (181,164.71) .. (181,147.5)
         .. controls (181,125.93) and (174.83,107.05) .. (L4);

\draw[contour]
    (L2) .. controls (137.39,171.73) and (140,160.95) .. (140,149)
         .. controls (140,133.95) and (135.86,120.76) .. (L1);

\draw[contour] (L2) -- (L3);
\draw[contour] (L1) -- (L4);


\coordinate (RfirstN) at (271,146);
\coordinate (R4) at (361.35,95.68);
\coordinate (R3) at (357.58,192.11);

\draw[contour] (RfirstN) -- (R4);
\draw[contour] (RfirstN) -- (R3);

\draw[contour]
    (R4) .. controls (369.03,107.28) and (373.86,122.83) .. (374.16,139.99)
         .. controls (374.53,161) and (368.02,179.84) .. (R3);


\coordinate (R1) at (406.09,107.23);
\coordinate (R2) at (410.18,183.12);
\coordinate (RsecondN) at (467.5,143);

\draw[contour] (R1) -- (RsecondN);
\draw[contour] (R2) -- (RsecondN);

\draw[contour]
    (R2) .. controls (402.75,173.41) and (398,158.59) .. (398,142)
         .. controls (398,128.7) and (401.05,116.54) .. (R1);


\draw (198,135.4) node [anchor=north west,inner sep=0.75pt,font=\large] {$=$};
\draw (236,134.4) node [anchor=north west,inner sep=0.75pt] {$\sum$};
\draw (239,153.4) node [anchor=north west,inner sep=0.75pt,font=\small] {$n$};

\draw (292.53,104.66) node
    [anchor=north west,inner sep=0.75pt,rotate=-332.23] {$\mathnormal{r+L}$};
\draw (442.99,104.53) node
    [anchor=north west,inner sep=0.75pt,rotate=-26.29] {$\mathnormal{r}$};


\foreach \p in {L1,L2,L3,L4,RfirstN,R4,R3,R1,R2,RsecondN}
    \fill (\p) circle (2.1pt);

\draw (267,149.4) node [anchor=north west,inner sep=0.75pt,font=\small] {$n$};
\draw (459,150.4) node [anchor=north west,inner sep=0.75pt,font=\small] {$n$};


\draw (119,175.4) node [anchor=north west,inner sep=0.75pt,font=\footnotesize] {$2$};
\draw (117,105.4) node [anchor=north west,inner sep=0.75pt,font=\footnotesize] {$1$};
\draw (171,86.4) node [anchor=north west,inner sep=0.75pt,font=\footnotesize] {$4$};
\draw (173.79,189.47) node [anchor=north west,inner sep=0.75pt,font=\footnotesize] {$3$};

\draw (364,85.4) node [anchor=north west,inner sep=0.75pt,font=\footnotesize] {$4$};
\draw (363,188.4) node [anchor=north west,inner sep=0.75pt,font=\footnotesize] {$3$};
\draw (397,179.4) node [anchor=north west,inner sep=0.75pt,font=\footnotesize] {$2$};
\draw (393,100.4) node [anchor=north west,inner sep=0.75pt,font=\footnotesize] {$1$};

\end{tikzpicture}

\caption{COE decomposition of the configuration at Figure \ref{fig:newcusp}.
For illustrative purposes, the $12$ COE has been inverted around the origin.}
\label{fig:ope}
\end{figure}
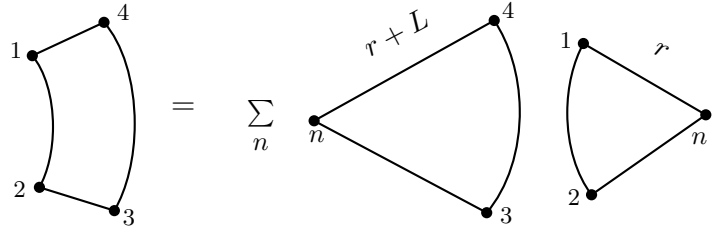

Consider the configuration of Figure \ref{fig:newcusp}. The cusp angles are all the same and equal to $\pi/2$ and we denote by $\varphi$  the angle formed at the cusp obtained by prolonging the arcs $14$ and $23$. The arcs $12$ and $34$, instead, do not admit such an intersection, so this configuration is different from the 4-point function illustrated in Figures \ref{fig:sboot} and \ref{fig:tboot}. 
 In this more singular configuration, the COE can be applied only in one channel, as shown in Figure \ref{fig:ope}. We will obtain information by comparing this expansion with the fusion limit $L\to 0$, where the defects on the inner and outer circles approach each other and should fuse to form a new defect~\cite{Kravchuk:2024qoh}. 
 This comparison will allow us to constrain the density of the cusp spectrum, weighted by the squared COE coefficients.


Applying the COE in the channel shown in Figure \ref{fig:ope}, we obtain that the 4-cusp function of the setup we are describing can be written as\footnote{Equation \eqref{ope1} is simply an application of the rule illustrated in Figure \ref{fig:sboot}, keeping into account the kinematics of our setup (in particular $x_{\bar{s}} = \infty$). }
\begin{equation}\label{ope1}
\text{4-pt} = \sum_n  \frac{a_n(\varphi)}{\abs{ 4 r (r+L)  \sin^2\frac{\varphi}{2} }^{2 \Gamma_0(\frac{\pi}{2} ) - \Gamma_n(\varphi) } (r+L)^{2 \Gamma_n(\varphi)}} ~,
\end{equation}
where we define for brevity
\begin{equation}
a_n(\varphi) \equiv C^2_{00n}\left(\frac{\pi}{2}, \frac{\pi}{2}, \varphi\right)~,
\end{equation}
in terms of the COE coefficients defined in \eqref{eq:cuspcuspope}. Introducing the spectral density 
\begin{equation}\label{spectraldensity1}
    \rho(\Delta) \equiv \sum_n\frac{a_n(\varphi)}{\abs{2 \sin\frac{\varphi}{2} }^{-2 \Gamma_n(\varphi)}  } \delta(\Delta-\Gamma_n(\varphi))~,
\end{equation} 
the 4-cusp function can be written as
\begin{equation}\label{4ptintegral}\text{4-pt} = 
   \abs{4 r^2 e^{\beta} \sin^2\frac{\varphi}{2} }^{-2\Gamma_0(\frac{\pi}{2})} \;
   \int d\Delta \, \rho(\Delta)  e^{-\beta \Delta } ~, 
\end{equation}
where we introduced the parameter \begin{equation}\label{eq:defbeta}
e^{\beta} \equiv 1 + \frac{L}{r}~.
\end{equation}
Now we would like to study the fusion limit $\beta \rightarrow 0^+$, corresponding to $L\rightarrow 0^+$ for fixed $r$, where we should find the divergence associated to the Casimir energy, i.e. $\sim \frac{1}{\beta^p}\text{exp}\left(\frac{\mathcal{E}}{e^{\beta}-1} \varphi + \dots \right)$,  where $\mathcal{E}$ is the Casimir energy.\footnote{The Casimir energy is a characteristic  of the theory, e.g. for  Wilson lines in conformal gauge theories it represents minus the 
 quark-antiquark potential. As explained in \cite{Correa:2012hh,Drukker:2012de}, with a conformal map to the cylinder one can relate it to a singular limit of the ground state cusp dimension for the same pair of defects, i.e. $\Gamma_0(\alpha) \sim \frac{\mathcal{E}}{\alpha}$ for $\alpha \rightarrow 0^+$.} This follows from defect-fusion effective theory~\cite{Bachas:2007td,Bachas:2013ora,Konechny:2015qla,Diatlyk:2024qpr,Diatlyk:2024zkk,Cuomo:2024psk,Kravchuk:2024qoh}, which describes the leading contribution to the fusion of nearby defects to be the coupling to the identity, producing a Casimir term proportional to the length \(\ell\) of the fused defects divided by the distance \(L\). The $1/\beta$ power that appears in front of the exponential accounts for the divergence of the fusion of cusps. It will become clear below that this power-law factor is subleading  with respect to the logarithmic growth of the integrated spectral density at large $\Delta$.

The core idea of this example is that the COE expansion has to reproduce the  exponential divergence associated with this Casimir term. 
The situation is not identical but reminiscent of the case of the standard crossing equation, where the leading OPE short-distance divergence in one channel needs to be reproduced by an infinite sum in the other channel, giving constraints on the spectral density, as first shown in \cite{Pappadopulo:2012jk}. Considering the annulus partition function, analogous  arguments were used to constrain a spectral density weighted with boundary  structure constants in terms of the Casimir energy in \cite{Diatlyk:2024qpr}. In our case, notice that each individual term of \eqref{ope1} is actually regular for $\beta \rightarrow 0^+$. 
Thus, from the \eqref{4ptintegral} and the fusion limit, we have 
\begin{equation} \label{eq1}
    \log\left[\int_{\Delta_{\text{min}}}^{\infty} \mathrm{d} \Delta' \; \rho(\Delta')  e^{-\beta\Delta'} \right]
\sim   \left( \frac{\mathcal{E} \varphi }{\beta}  
 + \dots  \right)~,
 \qquad \, \beta\to 0^+~.
\end{equation} 
Finding the large $\Delta$ asymptotics of a positive density from the small $\beta$ behavior of its Laplace transform is an example a class of problems addressed by Tauberian theorems. For this specific case, by Kohlbecker's Theorem \cite{Kohlbecker1958}, we have 
\begin{equation}
   \int_0^{\Delta} \mathrm{d}\Delta'\rho(\Delta') \sim \exp(2\sqrt{\mathcal{E} \varphi \Delta} + \dots ) ~, \qquad \Delta \to +\infty~,
\end{equation}
where $\dots$ stand for subleading terms in the limit. 
In the next subsection, we find the asymptotic density for different values of the angles, and show that the Cardy growth is replaced by a milder asymptotics.

\subsection{The asymptotic COE density for two cusps and a defect operator 
}\label{sec:tauberian}

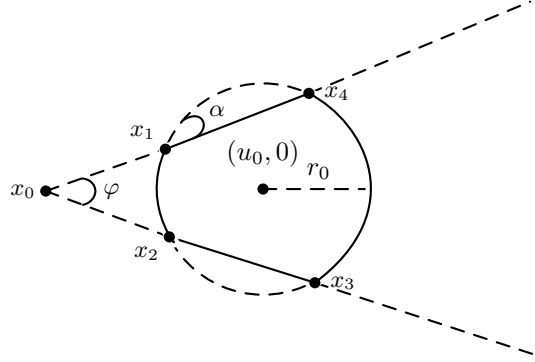
\begin{figure}[t]
\centering

\begin{tikzpicture}[
    x=0.75pt,
    y=0.75pt,
    yscale=-1,
    xscale=1,
    line cap=round,
    line join=round,
    aux/.style={
        black,
        thick,
        dash pattern=on 5pt off 4pt
    },
    contour/.style={
        black,
        thick,
        preaction={draw,white,line width=4pt}
    },
    angle mark/.style={black,thick}
]

\coordinate (X0)     at (206,200);
\coordinate (Center) at (315,199);
\coordinate (Z1)     at (266,179);
\coordinate (Z2)     at (268,223);
\coordinate (Z3)     at (341,246);
\coordinate (Z4)     at (338,151);

\coordinate (CircleLeft)   at (262,199);
\coordinate (CircleTop)    at (315,146);
\coordinate (CircleRight)  at (368,199);
\coordinate (CircleBottom) at (315,252);


\draw[aux]
    (CircleLeft)
    .. controls (262,169.73) and (285.73,146) .. (CircleTop)
    .. controls (344.27,146) and (368,169.73) .. (CircleRight)
    .. controls (368,228.27) and (344.27,252) .. (CircleBottom)
    .. controls (285.73,252) and (262,228.27) .. (CircleLeft)
    -- cycle;

\draw[aux] (X0) -- (Z1);
\draw[aux] (X0) -- (Z2);
\draw[aux] (Z4) -- (449,105);
\draw[aux] (Z3) -- (451,282);

\draw[aux] (Center) -- (CircleRight);


\draw[contour] (Z4) -- (Z1);
\draw[contour] (Z2) -- (Z3);

\draw[contour]
    (Z2) .. controls (277,233) and (252,210) .. (Z1);

\draw[contour]
    (Z3) .. controls (389,210) and (367,166) .. (Z4);


\path (X0) -- coordinate[pos=0.30] (PhiTop) (Z1);
\path (X0) -- coordinate[pos=0.30] (PhiBottom) (Z2);

\path
    (CircleLeft) .. controls (262,169.73) and (285.73,146) ..
    coordinate[pos=0.45] (DeltaCircle) (CircleTop);
\path (Z1) -- coordinate[pos=0.194] (DeltaChord) (Z4);


\draw[angle mark]
    (PhiTop)
    .. controls (227.91,193.74) and (230.93,196.69) .. (230.88,200.09)
    .. controls (230.83,203.58) and (227.55,206.50) .. (PhiBottom);

\draw[angle mark]
    (DeltaCircle)
    .. controls (280.51,160.95) and (284.13,162.55) .. (285.05,165.47)
    .. controls (286.00,168.46) and (283.82,171.95) .. (DeltaChord);


\foreach \p in {X0,Center,Z1,Z2,Z3,Z4}
    \fill[black] (\p) circle (2.1pt);


\node[font=\small] at (343,190) {$r_{0}$};
\node[font=\small] at (315,181) {$(u_{0},0)$};
\node[font=\footnotesize] at (239,200) {$\varphi$};
\node[font=\footnotesize] at (292,160) {$\alpha$};

\node[font=\footnotesize,anchor=east]       at (X0) {$x_{0}$};
\node[font=\footnotesize,anchor=south east] at (Z1) {$x_{1}$};
\node[font=\footnotesize,anchor=north east] at (Z2) {$x_{2}$};
\node[font=\footnotesize,anchor=west,xshift=2pt] at (Z3) {$x_{3}$};
\node[font=\footnotesize,anchor=west,xshift=2pt] at (Z4) {$x_{4}$};

\end{tikzpicture}

\caption{Configuration of four cusps (at $x_1$--$x_4$) located on a circle
centered at $(u_0,0)$ and intersected by two lines symmetric about the
$x$-axis. The exchanged cusps are at $x_0=0$ and $x_6=\infty$.}
\label{fig:fourc}
\end{figure}

A special case of the three-cusp function is obtained when one of the angles is equal to $\pi$. Then, one obtains a correlator of two cusps and a local defect operator. We will now find the asymptotics of the COE density associated to this correlators, i.e. the COE describing the fusion of two cusps with equal angles into local defect operators.
The key is 
to consider the setup depicted in Figure \ref{fig:fourc}.

The configuration we start from is a 4-point functions of cusps connected by planar arcs, chosen in such as a way that two of the non-adjacent arcs belong to the same circle. Then, while one of the two COE channels (the s-channel) will still admit a COE decomposition in terms of a cusp Hilbert space, in the other channel we are expanding two cusps in terms of states of the 1D  CFT living on the circular defect. 
This expansion is governed by the standard kinematics of the OPE in a 1D defect CFT. We will compare the two channels in a particular limit described explicitly below, {in which the four-point configuration defined by the cusp points  approaches a triangle while keeping the angle $\alpha$ fixed}.

Let us analyze the kinematics of the problem. We choose the concrete configuration shown in Figure \ref{fig:fourc}, where the arcs 12 and 34 belong to the same circle, and we are working in a frame where the arcs 14 and 23 are straight lines. Moreover,   we will choose the center of the circle on the symmetry axis of the 14 and 23 lines. Then, the four cusp angles are all equal, $\varphi_1 = \varphi_2 = \varphi_3 = \varphi_4 \equiv \pi - \alpha$. The other independent parameter is $\varphi$, the angle between the lines 14 and 23. The only independent conformal invariants are  $\varphi$ and $\alpha$. 
Explicitly, the intersection of the straight lines is placed at $x_0 = (0,0)$, and the positions of the four cusps are  $x_i \equiv (u_i,\pm m u_i)$, where 
$$m \equiv \tan\varphi/2 .
$$
Using a global rescaling, we can set $u_3=u_4=1$, and the positions of the remaining two points are then fixed by 
\begin{align}
    u_1&=u_2=\frac{u_0  - \sqrt{(1+m^2)r_0^2-(m u_0)^2}}{1+m^2} ~, 
\end{align}
where $r_0$ is the radius of the circle and $(u_0, 0)$ its center, given by
\begin{align}
  r_0&=\sqrt{1+m^2-2 u_0+u_0^2}~, \qquad 
    u_0 = \frac{1+m^2}{1+m \tan \alpha}~.
\end{align}
 Importantly, the cross ratio of the four points is 
\begin{equation} \label{cr1}
    z=\frac{\abs{x_{12}} \abs{x_{34}}}{\abs{x_{13}} \abs{x_{24}}} =  
     1-(1 + \tan^2\frac{\varphi}{2} )\sin^2\alpha ~,
\end{equation}
where the range of parameters we consider is $0<\varphi<\pi$ and $0< \alpha <\frac{\pi-\varphi}{2}$, guaranteeing $0<z<1$. 

Let us decompose the correlator in the two channels. The  s-channel COE gives, setting\footnote{The initial configuration is assumed to have cusps in their ground state at the four points  $x_i$. The generalization where they are all in the same excited state is obvious. } $\tilde\Gamma \equiv \Gamma_0(\pi -\alpha)$:
\begin{equation} \label{COEs}
\begin{split}
  \text{4-pt} = \frac{1}{\abs{x_{12}}^{2\tilde\Gamma} \abs{x_{34}}^{2\tilde\Gamma} }\sum_n c^2_{n}(\varphi, \alpha)\; \frac{1}{(1+m^2)^{\Gamma_n(\varphi)}}\abs{x_{12}}^{\Gamma_n(\varphi)} \abs{x_{34}}^{\Gamma_n(\varphi)}~,
\end{split} 
\end{equation}
with
$$
c_n^2(\varphi, \alpha)\equiv~C^2_{00n}(\pi - \alpha, \pi - \alpha, \varphi )~.
$$
In the t-channel, as we anticipated, we fuse two cusps to form point-like operators on the defect. This decomposition obeys the same rules as in a 1D CFT; in particular, it involves the standard 1D conformal blocks. It can be written as
\begin{equation}\label{eq:COEt}
 \text{4-pt} = \frac{G(1-z)}{\abs{x_{14}}^{2\tilde\Gamma} \abs{x_{23}}^{2\tilde\Gamma}} ~,
\end{equation} 
with 
\begin{equation}\label{eq:Gexp}
    G(z) = \sum_{\bar{\mathcal{O}}} \lambda_{c c \bar{\mathcal{O}}}^2(\alpha) g_{\Delta_{\bar{\mathcal{O}}}}(z), \hspace{1cm} g_{\Delta_{\bar{\mathcal{O}}}}(z)=z^{\Delta_{\bar{\mathcal{O}}}}  \hspace{2pt}_2F_1(\Delta_{\bar{\mathcal{O}}},\Delta_{\bar{\mathcal{O}}},2\Delta_{\bar{\mathcal{O}}}; z)~,
\end{equation}  
where the $\lambda_{cc\mathcal{O}}(\alpha)$'s are OPE coefficients for the fusion of two cusp operators to create a primary local  operator on the line defect.
 Notice that in the expansion \eqref{eq:Gexp} the $\varphi$ dependence is neatly confined to the cross ratio $z$ through \eqref{cr1}. 

Now, we would like to consider the limit $z \rightarrow 0^+$, which can be reached, e.g., keeping $\alpha$ fixed and varying $\varphi$ (or vice versa). In particular, \eqref{cr1} shows that $z=0$ corresponds to  $\varphi = \pi - 2 \alpha$, which describes the limit configuration where the circle touches the intersection of the two straight lines. 
 When we tune the values in this way, we do not expect anything singular to happen to the COE coefficients $c_n(\pi - 2 \alpha, \alpha)$. Thus, the most singular behaviour of the s-channel expansion for $z\rightarrow 0^+ $ is determined by the term exchanging the ground state in \eqref{COEs}.  Thus, we get that for $z\rightarrow 0^+$,
    \begin{equation}
     \frac{G(1-z)}{\abs{x_{14}}^{2\tilde\Gamma} \abs{x_{23}}^{2\tilde\Gamma}} \simeq  \frac{c_{0}^2}{\abs{x_{12}}^{2\tilde\Gamma-\Gamma_0} \abs{x_{34}}^{2\tilde\Gamma-\Gamma_0} } \;\left(\sin\alpha\right)^{2 \Gamma_0},
\end{equation} 
where we wrote for brevity $c_0 \equiv C_{000}(\pi - \alpha , \pi - \alpha , \pi - 2 \alpha )$, $\tilde\Gamma = \Gamma_0(\pi - \alpha)$ and $\Gamma_0\equiv \Gamma_0(\pi - 2 \alpha)$,  
which simplifies further to
 \begin{equation}\label{eq:schematic}
     \frac{G(1-z)}{(1-z)^{2\tilde\Gamma}} \simeq  \frac{c_{0}^2}{z^{2\tilde\Gamma -\Gamma_0}}, \;\;\; z\rightarrow 0^+ .
\end{equation} 
The treatment of this equation follows from the method of \cite{Qiao:2017xif}. The conformal blocks scale as
\begin{equation}
    g_{\Delta_{\bar{\mathcal{O}}}}(1-z) \approx  F(\Delta_{\bar{\mathcal{O}}}) K_0(2\sqrt{z}\Delta_{\bar{\mathcal{O}}}) , \hspace{1cm} F(\Delta)=4^{\Delta_{\bar{\mathcal{O}}}}\sqrt{\frac{\Delta_{\bar{\mathcal{O}}}}{\pi}}~,
\end{equation} 
valid for $\Delta_{\bar{\mathcal{O}}}\gg 1, \hspace{2pt} z\ll 1$ and $z\Delta_{\bar{\mathcal{O}}}\ll 1$. Therefore, the crossing equation in the $z\to 0$ limit leads to the following asymptotic constraint
\begin{equation}
    \int_0^\infty \mathrm{d}\Delta F(\Delta) p_{\alpha}(\Delta) K_0(2 \sqrt{z}\Delta) \sim \frac{c_{0}^2}{z^{2\bar{\Gamma} -\Gamma_0}}  \hspace{1cm}  (\text{as}  \hspace{2pt} z\to 0^+) ,
\end{equation} 
where $p_{\alpha}(\Delta) $ is the spectral density defined as follows: 
\begin{equation}
p_{\alpha}(\Delta) = \sum_{\bar{\mathcal{O}}} \lambda_{c c \bar{\mathcal{O}}}^2(\alpha) \; \delta( \Delta - \Delta_{\bar{\mathcal{O}}} ) ~, 
\end{equation}
where the sum is over all 1D CFT primaries. This can also be interpreted as a deformation of the standard spectral density of the 1D CFT. 

Following \cite{Qiao:2017xif}, the constraint above implies
\begin{equation} \label{g1}
\int_0^{\Delta_{\bar{\mathcal{O}}}}\mathrm{d}\Delta F(\Delta) p_{\alpha}(\Delta) \sim C_{000}^2(\pi -  \alpha, \pi - \alpha, \pi - 2 \alpha )( A  \gamma )^{-1} \Delta_{\bar{\mathcal{O}}}^{\gamma}\end{equation} 
where the coefficients $A$, $\gamma$ are also now dependent on $\alpha$ and read
\begin{equation}
    A \equiv  \frac{\Gamma_{\text{Euler}}(\gamma/2)^2}{4}, \hspace{1cm} \gamma \equiv 2(2 \tilde\Gamma -\Gamma_0) . 
\end{equation} 
This differs from the asymptotic result for the coefficients of three local operators by the  contribution of the ground state cusp operator to the definition of the exponent $\gamma$. In particular, we see that there is no Cardy-like behavior at large scaling dimension in this case: the contribution of defect primary operators to the expansion is exponentially suppressed, as in the standard (d)CFT case \cite{Pappadopulo:2012jk,Lauria:2017wav}.

Let us now consider what happens when we try to reach $z=1$. From \eqref{cr1} we have 
\begin{equation}
    \tan\frac{\varphi}{2}= \sqrt{\frac{\cos^2\alpha-z}{\sin^2\alpha}}
\end{equation}
which yields an upper bound for $z$
\begin{equation}
    z\leq \cos^2 \alpha.
\end{equation} 
Therefore, at fixed \(\alpha\), the \(s\)-channel COE cannot be continued within the space of real Euclidean cusp configurations all the way to \(z=1\); its physical kinematic domain terminates at \(z=\cos^2\alpha\). This is in contrast with the usual four-point kinematics of local operators on a line, for which the full Euclidean interval \(0<z<1\) is physically accessible.

\section{Conclusions and Outlook} \label{sec:conclu}

In the present work, we explored the features and constraints on cusped defects that are embedded in generic CFTs. We first described the construction of a Hilbert space around a cusped defect and, from basic principles, established the existence of a Cusp Operator Expansion, where a piece of contour enclosed by a sphere that it pierces orthogonally can be expanded in terms of states created by cusped lines. 
We discussed 
a basis of cusp eigenstates that diagonalizes the appropriate North-South-pole evolution  Hamiltonian. 
We provided a  general construction of cusp local operators corresponding to these states through the state-operator map, and proved the conformal covariance of the expectation value of an $n$-cusped contour made of circular arcs, where any excited cusp transforms as a primary local operator, as was observed previously in the context of $\mathcal{N}$=4 SYM.

We then proceeded to check these properties in the specific example of the Pinning defect embedded in the four-dimensional free scalar theory. After computing the explicit form of the two- and three-point functions, to confirm conformal covariance, we proceeded to the construction of the excited cusp states and showed that they are derived by operator insertions at the cusp. The overlap of the excited cusp states with states appearing in the COE of a smooth contour was also computed. An interesting feature of the operators that diagonalize the mixing problem of cusp eigenstates is their explicit dependence on spacetime coordinates other than the point where the cusp is. While this dependence can be expressed in terms of local data at the cusp, it does give rise to genuinely bi-local operators when the defect is absent. This provided a sanity check. Local operators in a CFT can be thought of as cusped scaling operators of a topological defect: consistently, for any local scaling operator one can construct a bi-local operator transforming like a primary.

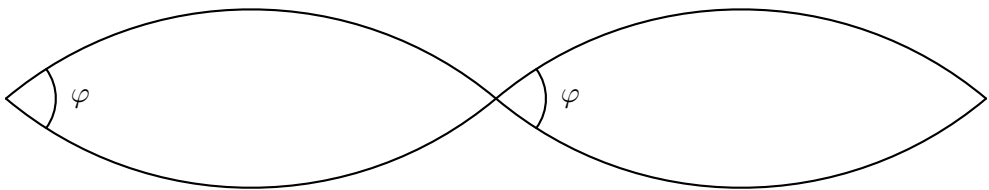
\begin{figure}[t]
\centering
\begin{tikzpicture}[
    scale=1.2,
    line cap=round,
    line join=round
]


\def\a{2.7}

\def\phiang{80}

\def\anglerad{0.55}

\pgfmathsetmacro{\alpha}{(180-\phiang)/2}
\pgfmathsetmacro{\h}{\a*tan(\alpha)}
\pgfmathsetmacro{\Rc}{sqrt(\a*\a+\h*\h)}

\pgfmathsetmacro{\beta}{
    acos(\anglerad/(2*\Rc))-\alpha
}


\coordinate (C1) at (-\a,0);
\coordinate (C2) at ( \a,0);

\coordinate (C3) at (3*\a,0);


\draw[black,thick]
    (C1)
    arc[
        start angle=180-\alpha,
        end angle=\alpha,
        radius=\Rc
    ];

\draw[black,thick]
    (C1)
    arc[
        start angle=180+\alpha,
        end angle=360-\alpha,
        radius=\Rc
    ];

\draw[black,thick]
    (C1) ++(-\beta:\anglerad)
    arc[
        start angle=-\beta,
        end angle=\beta,
        radius=\anglerad
    ];

\node at ($(C1)+(0.82,0)$) {$\varphi$};


\draw[black,thick]
    (C2)
    arc[
        start angle=180-\alpha,
        end angle=\alpha,
        radius=\Rc
    ];

\draw[black,thick]
    (C2)
    arc[
        start angle=180+\alpha,
        end angle=360-\alpha,
        radius=\Rc
    ];

\draw[black,thick]
    (C2) ++(-\beta:\anglerad)
    arc[
        start angle=-\beta,
        end angle=\beta,
        radius=\anglerad
    ];

\node at ($(C2)+(0.82,0)$) {$\varphi$};

\end{tikzpicture}
\caption{Fusion limit between two almonds leading to a so-called cross anomalous dimension.}
 \label{fig:crossanomdim}
\end{figure}

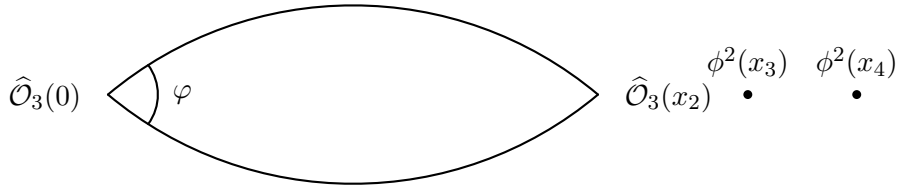
\begin{figure}[t]
\centering

\tikzset{every picture/.style={line width=0.75pt}} 
\begin{tikzpicture}[
    scale=1.2,
    line cap=round,
    line join=round
]


\def\a{2.7}
\def\phiang{80}
\def\anglerad{0.55}

\pgfmathsetmacro{\alpha}{(180-\phiang)/2}
\pgfmathsetmacro{\h}{\a*tan(\alpha)}
\pgfmathsetmacro{\Rc}{sqrt(\a*\a+\h*\h)}

\pgfmathsetmacro{\beta}{
    acos(\anglerad/(2*\Rc))-\alpha
}

\coordinate (C1) at (-\a,0);
\coordinate (C2) at ( \a,0);

\draw[black,thick]
    (C1)
    arc[
        start angle=180-\alpha,
        end angle=\alpha,
        radius=\Rc
    ];

\draw[black,thick]
    (C1)
    arc[
        start angle=180+\alpha,
        end angle=360-\alpha,
        radius=\Rc
    ];

\draw[black,thick]
    (C1) ++(-\beta:\anglerad)
    arc[
        start angle=-\beta,
        end angle=\beta,
        radius=\anglerad
    ];

\node at ($(C1)+(0.82,0)$) {$\varphi$};

\node[left=6pt]  at (C1) {$\widehat{\mathcal O}_3(0)$};
\node[right=6pt] at (C2) {$\widehat{\mathcal O}_3(x_2)$};

\coordinate (B1) at (4.35,0);
\fill (B1) circle (1.4pt);
\node[above=2pt] at (B1) {$\phi^2(x_3)$};

\coordinate (B2) at (5.55,0);
\fill (B2) circle (1.4pt);
\node[above=2pt] at (B2) {$\phi^2(x_4)$};

\end{tikzpicture}
\caption{Configuration with two cusps collinear to two local operators.}
 \label{fig:almlocloc}
\end{figure}


In the last part of the paper, we investigated some constraints that the existence of the COE places on the theory. In the first instance, the setup illustrated in Figure \ref{fig:newcusp} allowed us to constrain a spectral density of cusp states at generic angle by relating it to the Casimir energy describing the fusion limit of defects living on concentric circles. This, in particular, shows a Cardy-like growth for the spectrum of cusp states. The second configuration in Figure \ref{fig:fourc} allowed us to derive a generalization (depending on a  continuous parameter) of the Tauberian analysis for local operators, leading to an asymptotic result for a new spectral function defined on the spectrum of the 1D  defect CFT.

In future work, we would also like to obtain numerical (and hence potentially stronger) constraints on crossing equations involving cusps.  As the previous two examples show and was already pointed out in \cite{ladder}, the generic  configuration with four cusps connected by planar arcs does immediately yield `crossing' constraints, coming from the fact that the COE can be performed in two channels, see Figures \ref{fig:sboot} and \ref{fig:tboot}. 
In the case of local operators, analogous crossing constraints on 4-point functions allow to probe systematically the space of CFT data with the numerical conformal bootstrap~\cite{Rattazzi:2008pe}. Importantly, in the case of local operators a single crossing equation provides infinitely many constraints on the same dynamical data, which can roughly be obtained by evaluating the equation at different values of the cross ratios. 

In the case of four cusps at generic angles, as we anticipated in Section \ref{subsec:3loop} there is an important conceptual difference, due to the fact that dynamical data such as the dimensions and COE coefficients depend on  the kinematics of the problem. In fact, it is not possible to vary the cross ratios while keeping \emph{all} the relevant angles fixed. These angles include the four cusp angles and the two angles defining the cusp Hilbert spaces that enters the COE expansions, denoted as $\varphi_s$ (for the s-channel) and $\varphi_t$ (for the t-channel) in Figures \ref{fig:sboot} and \ref{fig:tboot}. Since COE coefficients and cusp dimensions depend on these angles, varying the location of the cusps 
necessarily also alters the dynamical data. 
Thus, the numerical exploration of parameter space  demands, at least partially, a rethinking of the problem. 

 The system of four cusps is not the only configuration involving cusp data which leads to crossing-type constraints.   Consider, for instance, the setup in Figure \ref{fig:cutalm}, where we assume that the almond can be truncated by appropriate defect-ending operators.  
Here, we can expand the chopped almond in two distinct ways. In one channel, one would use the COE for the chopped lines, in the quantization scheme defined by the almond, while in the other channel, one could collapse each pair of line-ending operators to create local operators, which would be naturally organized in terms of the two $SL(2, \mathbb{R}) $ symmetries corresponding to the branches of the original almond (one of these symmetries is twisted by a rotation with respect to the other). 

 Another setup  one can consider is  the system of two almonds along a line, where in one instance the almonds are fused at their tip, creating states in a ``cross'' Hilbert space\footnote{The cross anomalous dimension, which would be the ground state of the natural NS Hamiltonian on this four-punctured Hilbert space, was studied for example in \cite{Korchemsky:1993hr,Korchemskaya:1994qp,Munkler:2018cvu}.} (as in Figure \ref{fig:crossanomdim}), whereas in the other channel each almond is separately expanded in terms of local operators. 
Another interesting mixed system could involve  correlators where we have an almond and local operators placed on the line (see Figure \ref{fig:almlocloc}). Clearly, there are also other possibilities worth exploring. We hope to come back to some of these problems in the future. 

Finally, there exist additional natural generalizations to the work presented in this manuscript. The first concerns non-coplanar cusped contours, and more generally higher co-dimension defects. Another application includes the derivation of Ward identities and integrated constraints for cusps in a similar spirit to what has already been done for flat defects \cite{Cavaglia:2022yvv,Gabai:2025zcs,Girault:2025kzt,Drukker:2025dfm}. Finally, it would be a particularly instructive to compute in perturbation theory ($d=4-\varepsilon$, large $N$) the spectrum of the leading excited cusps states. This would provide crucial data to benchmark against in future numerical bootstrap studies.

\acknowledgments

We are grateful to  Gabriel Cuomo, Gregory Korchemsky, Ryan Lanzetta and Lorenzo Magnea   for  discussions, and to Nikolay Gromov and Fedor Levkovich-Maslyuk for inspiring collaboration on related topics.
LB, ADG, SRK and MM are partially supported by the INFN “Iniziativa Specifica” STEFI. AC is partially supported by the INFN “Iniziativa Specifica” SFT. SRK and MM are supported by the Italian Ministry of University and Research (MUR) under the FIS grant BootBeyond (CUP: D53C24005470001). The authors  of this work participate in the Marie Sk\l odowska-Curie Action (MSCA) High energy Intelligence (HORIZON-MSCA-2023-SE-01-101182937-HeI). LB's research is partially supported
by the MUR PRIN contract 2022N9CTAE ``Constraining strongly coupled quantum field theories using symmetry''.

\appendix 

\section{Conjugacy Classes of $PSL(2,\mathbb{C)}$ and its Flow Lines} \label{Conju}  
\begin{figure}[ht]
\centering
\begin{tikzpicture}[
    scale=1.15,
    line cap=round,
    line join=round
]

\def\k{0.16}       
\def\A{0.50}       
\def\tmin{-25}     
\def\tmax{9}       
\def\L{2.7}        
\def\phiA{0.55}    
\def\phiB{-0.55}   

\draw[thin, gray!70, ->]
    (-\L,0) -- (\L,0)
    node[right] {$x_1$};

\draw[thin, gray!70, ->]
    (0,-\L) -- (0,\L)
    node[above] {$x_2$};

\draw[black, thick,
      domain=\tmin:\tmax,
      samples=450,
      variable=\t]
  plot (
    {\A*exp(\k*\t)*cos((\t+\phiA) r)},
    {\A*exp(\k*\t)*sin((\t+\phiA) r)}
  );

\draw[black, thick,
      domain=\tmin:\tmax,
      samples=450,
      variable=\t]
  plot (
    {\A*exp(\k*\t)*cos((\t+\phiB) r)},
    {\A*exp(\k*\t)*sin((\t+\phiB) r)}
  );


\end{tikzpicture}

\caption{Two loxodromic orbits of $H=D+\omega M_{12}$.
Their closures meet at $0$, but the cusp angle is not finite.}
\label{fig:loxodromic}
\end{figure}
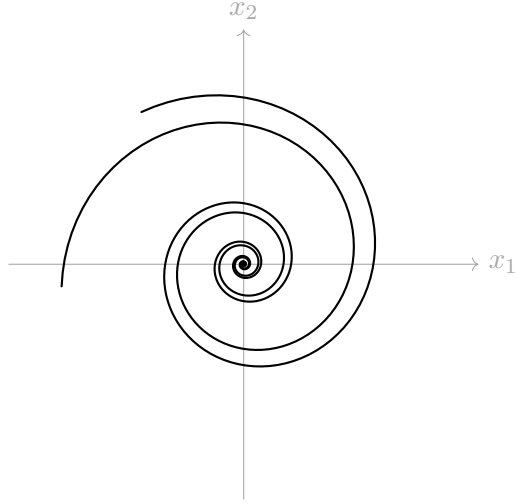

As mentioned in Subsection \ref{subsec:gener}, M\"obius transformations are classified into four conjugacy classes: elliptic, hyperbolic, loxodromic, and parabolic. By using a conjugacy transformation ($g = h \circ f \circ h^{-1}$), any transformation can be brought to a representative of one of these families. This algebraic classification reveals the map's fundamental action. Elliptic maps represent pure rotations, hyperbolic maps represent pure dilations, loxodromic maps combine the previous two into a twisted spiral, and parabolic maps represent translations. 

For a vector field $H=v^\mu(x) \partial_\mu$, the flow is the family of maps $h_s$ defined by
\begin{equation}
    h_s(x_0)=x(s),
\end{equation}
where $x(s)$ is obtained by solving
\begin{equation} \label{floweq}
    \frac{\mathrm{d}x^\mu(s)}{\mathrm{d}s}
    =
    v^\mu(x(s)),
    \qquad
    x^\mu(0)=x_0^\mu 
\end{equation}
for $s\in \mathbb{R}$. A complete  $H$-orbit through $x_0$ is defined as
\begin{equation}
    O_{x_0}=\{h_s(x_0): s\in \mathbb{R}\}. 
\end{equation}

Above, we defined the cusped defects as defects formed by the union of two complete $H$-orbits whose closures meet with a finite angle at one or more finite fixed points of $H$. This implies that the defect is invariant under the flow generated by $H$. In this section, we show that only the hyperbolic Hamiltonians can give rise to such cusped defects. 
\begin{itemize}
    \item Let us start by studying the elliptic flows. One representative Hamiltonian is 
    \begin{equation}
        H=M_{12}=x_1 \partial_2 -x_2 \partial_1=\partial_\theta
    \end{equation}
    by replacing $x_1=\rho \cos\theta$ and $x_2=\rho \sin\theta $. From equation \eqref{floweq}, the flow is 
    \begin{equation}
        \theta(s)=\theta_0 +s, \hspace{0.5cm} \rho(s)=\rho_0
    \end{equation}
    The non-trivial orbits are concentric circles around $x_1=x_2=0$, which corresponds to a fixed point in $d=2$. Notice, however, that disjoint circles cannot form a cusp. Thus, we conclude that elliptic flows cannot give rise to cusped defects. 

    \item In the same way, we can study the parabolic flows, which correspond to translations, for which we can have the following representative Hamiltonian $H=\partial_1$. Note that, in $\mathbb{R}^2$, it has no fixed point. The orbits of $H$ are parallel lines $x_1(s)=x_{0,1}+s$ with $x_i(s)=x_{0,i}$ and, as in the previous case, they cannot form cusped defects since their closures do not meet at any finite fixed point. 

    \item Now, let us consider the hyperbolic flow. As we saw previously, the standard representative is the dilatation generator $H=D=x^\mu \partial_\mu= r\partial_r$ on a plane. The nontrivial orbits are open rays from the origin; on the compactification, their closures contain both the origin and the point at infinity. In general, the cusped defect $\mathcal{C}$ in $\mathbb{R}^2$ is given by 
    \begin{equation}
        \mathcal{C}=\{0\}\cup \{r n_1: r>0\}\cup \{r n_2: r>0\}
    \end{equation} 
    with $n_{1,2}$ unit vectors defining the cusp angle through $\cos\varphi= n_1\cdot n_2$. Under conjugations of $D$, the cusped defect $g(\mathcal{C})$ represents a conformal transformation of $\mathcal{C}$. As seen in Appendix \ref{sec:conf}, a conformal map of $\mathcal{C}$ leads to defects like that in Figure \ref{fig:nonzerochi}.
    \item Finally, let us consider the combination of the hyperbolic and elliptic flows, which gives rise to the loxodromic flow. In polar coordinates, a standard representative vector field is $H=D+\omega M_{12}= r\partial_r +\omega \partial_\theta$. Under compactification, this Hamiltonian has two fixed points, the origin and infinity. The flow equation \eqref{floweq} leads to the following flow 
    \begin{equation}
        r(s)=e^s r_0, \hspace{0.5cm}\theta(s)=\theta_0+\omega s
    \end{equation}
    The $H-$orbits are logarithmic spirals
    \begin{equation}
        \theta(r)=\theta_0 +\omega \log\frac{r}{r_0}
    \end{equation}
    The closures of two distinct $H$-orbits meet at the origin and at infinity, as seen in Figure \ref{fig:loxodromic}. However, as we approach the origin, $r\rightarrow 0$, there is no limiting tangent line and no finite cusp angle. For this reason, we discard this kind of invariant curve in our study.
\end{itemize}
In conclusion, only hyperbolic flows can give rise to cusped defects. In practical terms, this means that the only Hamiltonian available compatible with symmetric cusped defects is the dilatation operator and its possible conjugations, such as the NS Hamiltonian.

\section{Describing the Almond in the Plane} \label{sec:conf}

Let us derive the parametrization \eqref{xpm} and \eqref{zeta} that we use throughout this work. The starting point is the cusped defect with cusp angle $\varphi$ and with upper and lower branches given respectively by 
\begin{equation} \label{initcon}
\begin{split}
    z_+(\tau)&=e^\tau e^{i\varphi} \\
    z_-(\tau)&=e^\tau .
\end{split}
\end{equation}

We introduce a M\"obius transformation that maps the origin to itself and brings the point at infinity to $1$
\begin{equation} \label{conmap}
    f(z)=\frac{z}{z+a}
\end{equation}
where both $z$ and $a$ are complex numbers. 
We conveniently define $a= |a| e^{i\delta}$.
Note that we can start from any configuration in \eqref{initcon}, since any other can be reached by a rotation parametrized by $\delta$, we will see how this works below. 
Under the conformal map \eqref{conmap}, for instance, the (+) ray is mapped to
\begin{equation}
        f(z_+(\tau))=\frac{1}{1+ e^{-s} e^{i(\chi-\varphi)/2}}  
\end{equation}
where we have replaced $e^{-s}= |a| e^{-\tau}$  and $\delta=\frac{\chi+\varphi}{2}$ above. Likewise, with the replacement $e^s=\abs{a} e^{-\tau}$, since by convention we have chosen the parameter $s$ in the lower arc to increase in the opposite direction from the one in the upper arc, we have 
\begin{equation}
   f_-(s):= f(z_-(\tau))=\frac{1}{1+e^{s}e^{i(\chi+\varphi)/2}}.
\end{equation}
Finally, we send $0\rightarrow w_1$ and $1\rightarrow w_2$ by an affine transformation
\begin{equation}
    \zeta_{\pm}(s)= w_1+(w_2-w_1) f_{\pm}(s)
\end{equation}
and recover our parametrization of the almond. See Figure \ref{fig:nonzerochi} for an illustration of an almond with (and without) $\chi=0$.

\section{Chopping the Almond}\label{app:chopped} 
\begin{figure}[ht]
    \centering

\begin{tikzpicture}[
    x=0.75pt,
    y=0.75pt,
    yscale=-1,
    xscale=1,
    line cap=round,
    line join=round
]

\draw[black,thick]
    (291.7,141.38)
    .. controls (305.67,127.97) and (336.79,118.49) .. (373.01,118.23)
    .. controls (409.75,117.97) and (441.39,127.27) .. (455.25,140.79);

\draw[black,thick]
    (455.9,162.88)
    .. controls (441.95,176.31) and (410.84,185.84) .. (374.62,186.15)
    .. controls (337.88,186.47) and (306.23,177.22) .. (292.35,163.73);

\draw (278,119.4)
    node [anchor=north west][inner sep=0.75pt]
    [font=\scriptsize]
    {$-\Lambda_{1}$};

\draw (445,171.4)
    node [anchor=north west][inner sep=0.75pt]
    [font=\scriptsize]
    {$-\Lambda_{4}$};

\draw (289,173.4)
    node [anchor=north west][inner sep=0.75pt]
    [font=\scriptsize]
    {$\Lambda_{2}$};

\draw (447,120.4)
    node [anchor=north west][inner sep=0.75pt]
    [font=\scriptsize]
    {$\Lambda_{3}$};

\end{tikzpicture}

\caption{Chopped almond.}
\label{fig:cutalm}
\end{figure}
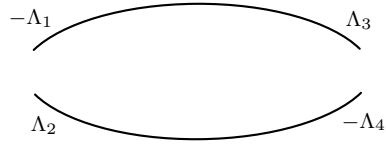

Here we outline a construction for the spectrum of excited cusp states which draws some initial intuition from \cite{ladder}. In that work, the cusped almond was regulated by chopping off the arcs before they reach the cusp points, as shown in Figure \ref{fig:cutalm}. The positions at which the lines are chopped off are taken to be in general different from each other. As in equation \eqref{biglambda}, we take 
\begin{equation}\label{repL}
    \Lambda_i= \log\left(\frac{x_{12}}{\epsilon_i}\right).
\end{equation}
Then, the chopped defect, as seen in Figure \ref{fig:cutalm}, is defined as follows 
\begin{equation}\label{eq:regulated}
    D_{p,\text{cut}}=\exp\left(-\lambda \int_{-\Lambda_1}^{\Lambda_3}\mathrm{d}s \hspace{2pt} J^+(s) \phi(s) -\lambda\int_{-\Lambda_4}^{\Lambda_2}\mathrm{d}s \hspace{2pt} J^-(s) \phi(s)\right)
\end{equation} 
with the $+ $ and $-$ superscripts denoting the upper and lower curves, respectively. Note that for brevity, we write $\phi(s):= \phi(x(s))$. We have also denoted 
\begin{equation}
    J^\pm (s)= \abs{\frac{d x_{\pm}}{ds}}.
\end{equation} 
Already at the level of \eqref{eq:regulated} it becomes apparent that taking derivatives with respect to the regulators $\Lambda_i$ brings down powers of the field $\phi$. While it is intuitive that additional powers of the field $\phi$ lead to excited states, we will make this more precise by explicitly evaluating the expression in \eqref{eq:regulated} and showing that combinations of derivatives with respect to the regulators indeed reproduce excited states and project out other states.

Let us state the result for this expectation value of the defect on the cut almond, denoted as $D_{p,\text{cut}}$
\begin{equation}
\begin{split}
    \langle D_{p,\text{cut}} \rangle&= e^{- \Gamma y}\Bigg (1+ 2\lambda^2 \left(\cosh\eta +\cosh \delta\right) e^{-y} \\
    &+\left[\lambda^2 (\cosh 2\eta +\cos\varphi \cosh 2 \delta )+2 \lambda^4(\cosh\eta +\cosh \delta)^2\right]e^{-2y}  +\cdots\Bigg) 
\end{split}
\end{equation}  
where 
\begin{align}
    y&=\dfrac{\Lambda_1+\Lambda_2+\Lambda_3+\Lambda_4}{2} \\
    \eta&=\frac{\Lambda_{12}-\Lambda_{43}}{2}\\ 
    \delta&= \frac{\Lambda_{12}+\Lambda_{43}}{2}, \hspace{1cm} \Lambda_{ij}=\Lambda_i-\Lambda_j.
\end{align}
The expression above indeed reveals an expansion in terms of a tower of states with integer separation. This integer spacing, however, is just an artifact of the free theory, and we don't expect it to hold in interacting theories.
It is easy to verify that the differential operator\footnote{These operators are dimensionless derivatives along the curves.} $-(\partial_{\Lambda_1}+\partial_{\Lambda_2}+\Gamma)$ projects out the ground state, in agreement with our previous statements above. Concretely, 
\begin{equation}\label{eq:overlap}
\begin{split}
&-(\partial_{\Lambda_1}+\partial_{\Lambda_2}+\Gamma) \langle D_{p,\text{cut}} \rangle = e^{- \Gamma y}\Bigg ( 2\lambda^2 \left(\cosh\eta +\cosh \delta\right) e^{-y} \\
&+ { 2}\left[\lambda^2 (\cosh 2\eta +\cos\varphi \cosh 2 \delta )+2 \lambda^4(\cosh\eta +\cosh \delta)^2\right]e^{-2y}  +\ldots\Bigg)
\end{split}
\end{equation}
which in the limit $\Lambda_i=\Lambda=\log\left( \dfrac{x_{12}}{\epsilon}\right)$ as $\epsilon\rightarrow 0$ reduces to
\begin{equation}\label{projected}
     -(\partial_{\Lambda_1}+\partial_{\Lambda_2}+\Gamma) \langle D_{p,\text{cut}} \rangle = 4\lambda^2 \left(\frac{\epsilon}{x_{12}}\right)^{2 (\Gamma+1)} + \ldots
\end{equation}
where $\ldots$ denotes terms of higher order in $\epsilon$. It is now evident that the ground state contribution has been projected out, since the vacuum term $\sim (\frac{\epsilon}{x_{12}})^{2\Gamma}$ is indeed absent in \eqref{projected}. The action of the differential operator should be understood as an insertion of the first excited operator, of dimension $\Gamma+1$, at the endpoints parametrized by $\Lambda_1$ and $\Lambda_2$. Normalizing both the excited operator and the ground state by $\epsilon^{\Gamma+1}$ and $\epsilon^\Gamma$, respectively, yields
\begin{equation}
-\frac{(\partial_{\Lambda_1}+\partial_{\Lambda_2}+\Gamma) \langle D_{p,\text{cut}} \rangle}{\epsilon^{\Gamma+1}\epsilon^\Gamma}= 4 \lambda^2 \frac{\epsilon}{x_{12}^{2(\Gamma+1)}}
\end{equation}
which in the limit $\epsilon \rightarrow 0$ is zero
\begin{equation}\label{eq:gammagamma1}
    -\frac{(\partial_{\Lambda_1}+\partial_{\Lambda_2}+\Gamma) \langle D_{p,\text{cut}} \rangle}{\epsilon^{\Gamma+1}\epsilon^\Gamma} |_{\epsilon\rightarrow0}=0 \sim \langle \Gamma+1 | \Gamma\rangle.
\end{equation}

At this point, we encourage the reader to recall our comments about the importance of normalizing states with the correct powers of $\epsilon$ below \eqref{eq:2ptfunc}. The RHS of \eqref{eq:gammagamma1} states the orthogonality between the ground state and the first excited state. If we apply this operation on both sides and divide by the appropriate $\epsilon^{\Gamma+1}$ for each insertion, we find
\begin{equation}
     \frac{(\partial_{\Lambda_3}+\partial_{\Lambda_4}+\Gamma)(\partial_{\Lambda_1}+\partial_{\Lambda_2}+\Gamma) \langle D_{p,\text{cut}} \rangle}{\epsilon^{\Gamma+1}\epsilon^{\Gamma+1}}\underset{\epsilon\rightarrow0} = 4\lambda^2 \left(\frac{1}{x_{12}} \right)^{2 (\Gamma+1)} \sim \langle \Gamma+1|\Gamma+1 \rangle
\end{equation} 
which represents the two-point function of the first excited  cusp state.
In this way, excited states are built by successive applications of differential operators. In general, the $n-$th excitation of the ground state is given by 
\begin{equation}
\widehat{\mathcal{O}}_{n} = \frac{(-1)^n}{n!} \prod_{i=0}^{n-1} \frac{\partial_{\Lambda_1}+\partial_{\Lambda_2}+\Gamma_i}{\Gamma_{i+1}-\Gamma_i}.
\end{equation}

We would now like to convert this intuition coming from differential operators to the standard picture of building excited states in terms of composite operators. Thus, to see what these operators look like in terms of bulk operators, we apply this differential operator to the defect $D_{p,\text{cut}}$. Then, for instance, applying the first differential operator (forgetting about the overall factor), we have
\begin{equation}
   -(\partial_{\Lambda_1}+\partial_{\Lambda_2}+\Gamma) D_{p,\text{cut}} = \lambda \left( J^+(-\Lambda_1) \phi(-\Lambda_1) + J^-(\Lambda_2) \phi(\Lambda_2) + \frac{\Gamma}{\lambda} \right) D_{p, \text{cut}}.
\end{equation} 

As suggested earlier, the action of the differential operator $\widehat{\mathcal{O}}_{1}$ is to insert the first excited operator at the endpoints. For this reason, we allow ourselves to abuse the notation and label both the differential operators and the insertions themselves as $\widehat{\mathcal{O}}_{n}$. In the limit $\epsilon\rightarrow0$, $\widehat{\mathcal{O}}_{1}$ becomes 
\begin{equation} \label{oc1}
    \widehat{\mathcal{O}}_{1}(x_1)= \left (2 \phi(x_1) +\frac{\Gamma}{\epsilon\lambda} \right).
\end{equation} 
Thus, the first excited operator on the cusp is created by inserting the bulk operator $\phi$ mixed with the identity. Notice that $\phi$ here is indeed allowed to mix with a constant term, since our regulator is dimensionful. In particular, the general pattern which we study more systematically in Section \ref{sec:CompositeOps} is that the presence of the cusp results in mixing with the vacuum due to the $1/\epsilon^a$ singularity arising from the cusp that needs to be subtracted off. The value of $a$ depends on how excited the state is, for the first excited state above, we had $a=1$. 

Let us now see what the operator $\mathcal{O}_{c,2}$ at dimension 2 looks like.  Applying the differential operator, we have 
\begin{equation}
\begin{aligned}
&(\partial_{\Lambda_1}+\partial_{\Lambda_2}+\delta_2)
(\partial_{\Lambda_1}+\partial_{\Lambda_2}+\delta_1)D_{p,\mathrm{cut}} \nonumber \\
&=\Bigg[
\delta_1\delta_2
-\lambda(\delta_1+\delta_2)\Big(J^+(-\Lambda_1)\phi(-\Lambda_1)+J^-(\Lambda_2)\phi(\Lambda_2)\Big)
\\
&
+\lambda\,\partial_s\!\big(J^+(s)\phi(s)\big)\Big|_{s=-\Lambda_1}
-\lambda\,\partial_s\!\big(J^-(s)\phi(s)\big)\Big|_{s=\Lambda_2}\\
&+\lambda^2\Big(J^+(-\Lambda_1)\phi(-\Lambda_1)+J^-(\Lambda_2)\phi(\Lambda_2)\Big)^2
\Bigg]
D_{p,\mathrm{cut}}
\end{aligned}
\end{equation} 
with $\delta_1=\Gamma$ and $\delta_2=\Gamma +1$. By inspecting the expression in brackets, we notice that as $\epsilon\rightarrow 0$ this becomes a linear combination of the identity, $\epsilon \phi$,  $(\epsilon \phi)^2$ , and a derivative term which deserves a few extra words. In terms of derivatives in $x$ and $y$, i.e. the two Euclidean coordinates, as $\epsilon\rightarrow0$, the derivative becomes
\begin{equation}
\begin{split}\label{tangled}
&\partial_s \big( J^+(s)\phi(s) \big) \Big|_{s=-\Lambda_1} - \partial_s \big( J^-(s)\phi(s) \big) \Big|_{s=\Lambda_2} 
=2 \epsilon \phi(x_1) \\
&+ 4 \epsilon^2 \sin\left(\frac{\phi}{2}\right) \left[ \frac{2 \cos\left(\frac{\chi}{2}\right) \phi(x_1)}{x_1 - x_2} - \sin\left(\frac{\chi}{2}\right) \partial_y \phi(x_1) + \cos\left(\frac{\chi}{2}\right) \partial_x \phi(x_1) \right] +\mathcal{O}(\epsilon^3)
\end{split}
\end{equation}
where, for simplicity, we have constrained the points $x_1$ and $x_2$ to lie on the $y=0$ line. The parameter $\chi$, as noted in Subsection \ref{subsec:twocuspf}, is a parameter that bends the arcs without changing the cusp angle. 

While the approach outlined in this appendix is intuitive, once one starts studying operators at dimension 2 and above, where there are multiple scaling operators at each dimension, it becomes cumbersome to disentangle/distinguish the scaling operators from each other in equations such as the RHS of \eqref{tangled}. This is the reason why we pursued a different route in Section \ref{sec:CompositeOps}.

\section{Partition Function at Finite Temperature} \label{sec:parti}

The partition function in the presence of the defect $W$ \eqref{dp} is defined as 
\begin{equation}
    Z_{W}=\int \mathcal{D}\phi \hspace{2pt} e^{-S_{\text{free}}(\phi)} W
\end{equation}
where $S_{\text{free}(\phi)}$ denotes the action of the free bulk scalar field in flat space. On an $S^1_\beta \cross S^3$ background, the defect is written as 
\begin{equation}
    W=\exp \left( -\lambda \int_0^\beta \mathrm{d}\tau \phi(\tau, \hat{n}_1) -\lambda\int_\beta^0 \mathrm{d}\tau \phi(\tau ,\hat{n}_2)\right)
\end{equation}
where the field $\phi$ is understood to be defined on the cylinder with compactified direction $\tau\rightarrow\tau+\beta$. In the expression above $\hat{n}_{1,2}$ denote unit vectors on $S^3$. The propagator for bosonic operators at finite temperature is given by 
\begin{equation}
    G_{\beta}(\Delta \tau,\varphi)=\sum_{n \in \mathbb{Z}} G_{\text{cyl}}(\Delta\tau +n\beta,\varphi) 
\end{equation}
where
\begin{equation}
    G_{\text{cyl}}(\Delta\tau,\varphi)=\left(\frac{1}{2(\cosh(\Delta\tau)-\cos\varphi )} \right)^\Delta
\end{equation}
with $\Delta$ being the scaling dimension of the scalar operator and $\varphi$ the cusp angle defined by $\cos\varphi =\hat{n}_1\cdot \hat{n}_2$.

The expectation value of the defect is given by

\begin{equation} \label{zdp}
    \langle W \rangle_\beta =\frac{Z_{W}}{Z_0}
\end{equation}
where $Z_0$ denotes the partition function of the free scalar field on $S^1_\beta \cross S^3$. By use of the Poisson resummation formula, which states that for a periodic function $f(t)$ with period $\beta$
\begin{equation}
    \sum_{n\in \mathbb{Z}} f(t+n\beta)=\frac{1}{\beta} \sum_m e^{i\omega_m t} \tilde{f}(\omega_m), \hspace{1cm} \omega_m=\frac{2\pi m }{\beta}
\end{equation}
where $\tilde{f}(\omega)$ is the Fourier transform of $f(t)$, we can show that 
\begin{equation} \label{expdp}
    \langle W \rangle_\beta=x^\Gamma, \hspace{1cm} \Gamma=\lambda^2\left(1-\frac{\pi -\varphi}{\sin\varphi}\right)
\end{equation}
with $x=e^{-\beta}$. On the other hand 
\begin{equation}
    Z_0= \exp \left[\sum_{k=1}^\infty (-k^2) \log(1-x^k)\right]
\end{equation}
where $k$ labels a sum over spherical harmonic eigenstates.
Thus, from \eqref{zdp} and \eqref{expdp} one concludes that the action of the defect is to shift the scaling dimension of the free theory spectrum by the cusp anomalous dimension $\Gamma$, while the operator degeneracies at each level stay the same. 

Explicitly, expanding in $x$, we find 
\begin{equation}
    Z_{W}=x^\Gamma(1+x+5x^2+14x^3+\cdots) 
\end{equation}
which, by comparison to \eqref{zcount}, tells us that there is one operator with $\Delta=\Gamma$, one with $\Delta=\Gamma+1$ and 5 with $\Delta=\Gamma+2$, just as in a free scalar CFT at $d=4$ spacetime dimensions.

\section{Two-Cusp Functions for the Pinning Defect} \label{sec:matrix} 

Below we give the matrix $G$ of correlators defined in \eqref{gmn} in terms of bare operators of \eqref{basisop} up to dimension 2. In practice, we choose to normalize the matrix elements by the expectation value of the defect $W$
\begin{equation} \label{GN}
    G^N_{mn}= \frac{ \langle \hat{\Psi}_m(x_2) W \hat{\Psi}_n(0)\rangle}{\langle W\rangle}
\end{equation}
where the superscript $N$ denotes a normalized matrix element, and $\langle W\rangle$ is given by \eqref{expealm}.

It is important to emphasize that the matrix elements are scheme-dependent. In particular, as a consequence, note that the constants $c_{1,2}$ in \eqref{diagx1} are scheme-dependent. Below, these are computed in a scheme in which the operators are inserted directly at the position of the cusps. Then, the chopped almond in Figure \ref{fig:cutalm} approaches the almond as the cutoff is taken to zero ($\epsilon \rightarrow 0$). For simplicity, we have set $\chi=0$, and the matrix elements have the following form

\begin{equation}
\label{eq:Gmatrix}
G_{mn}^N
=
\begin{pmatrix}
1
&
\epsilon\delta
&
0
&
-\epsilon^2 q
&
\epsilon^2\delta^2
\\[2mm]
\epsilon\delta
&
\epsilon^2\left(\dfrac{1}{x_2^2}+\delta^2\right)
&
0
&
\epsilon^3\left(\dfrac{2}{x_2^3}-\delta q\right)
&
\epsilon^3\left(\delta^3+\dfrac{2\delta}{x_2^2}\right)
\\[3mm]
0
&
0
&
\dfrac{2\epsilon^4}{x_2^4}
&
0
&
0
\\[3mm]
\epsilon^2 q
&
\epsilon^3\left(-\dfrac{2}{x_2^3}+\delta q\right)
&
0
&
\epsilon^4\left(-\dfrac{6}{x_2^4}-q^2\right)
&
\epsilon^4\left(q\delta^2-\dfrac{4\delta}{x_2^3}\right)
\\[3mm]
\epsilon^2\delta^2
&
\epsilon^3\left(\delta^3+\dfrac{2\delta}{x_2^2}\right)
&
0
&
\epsilon^4\left(-q\delta^2+\dfrac{4\delta}{x_2^3}\right)
&
\epsilon^4\left(
\delta^4+\dfrac{4\delta^2}{x_2^2}
+\dfrac{2}{x_2^4}
\right)
\end{pmatrix}~,
\end{equation}
with
\begin{align}
\label{eq:m_definition}
\delta&=-\frac{2\lambda}{\epsilon}
+\frac{2\lambda\epsilon}{x_2^2} \\
q&=
\lambda\left[
\frac{2\cos\frac{\varphi}{2}}{\epsilon^2}
+\frac{4}{x_2\epsilon}
-\frac{4\epsilon}{x_2^3}
-\frac{2\epsilon^2}{x_2^4}
\cos\frac{\varphi}{2} +\cdots
\right]~.
\end{align}
The elements above should be expanded up to order $\epsilon^4$. This is the highest power that will contribute in a non-vanishing way to two-point functions involving operators up to dimension 2 in the $\epsilon \rightarrow0$ limit. Recall that the renormalized scaling operators are obtained after stripping the expected factor \(\epsilon^{\Gamma_n}\) and taking \(\epsilon\to0\); their two-point functions then remain finite (see the discussion around \ref{eq:2ptfunc}). Diagonalization of this matrix up to order $\epsilon^4$ leads to the basis in equations \eqref{diagx1} and \eqref{b2}. 

\bibliography{bib}

@article{Dorn:2019yms,
    author = "Dorn, Harald",
    title = "{More on Wilson loops for two touching circles}",
    eprint = "1905.01101",
    archivePrefix = "arXiv",
    primaryClass = "hep-th",
    reportNumber = "HU-EP-19/11",
    doi = "10.1007/JHEP07(2019)088",
    journal = "JHEP",
    volume = "07",
    pages = "088",
    year = "2019"
}

@article{Makeenko:1979pb,
    author = "Makeenko, Yu. M. and Migdal, Alexander A.",
    title = "{Exact Equation for the Loop Average in Multicolor QCD}",
    reportNumber = "ITEP-86-1979",
    doi = "10.1016/0370-2693(79)90131-X",
    journal = "Phys. Lett. B",
    volume = "88",
    pages = "135",
    year = "1979",
    note = "[Erratum: Phys.Lett.B 89, 437 (1980)]"
}

@article{Girault:2025kzt,
    author = "Girault, Bastien and Paulos, Miguel F. and van Vliet, Philine",
    title = "{Consequences of symmetry-breaking on conformal defect data}",
    eprint = "2509.26561",
    archivePrefix = "arXiv",
    primaryClass = "hep-th",
    month = "9",
    year = "2025"
}

@article{Gabai:2025zcs,
    author = "Gabai, Barak and Sever, Amit and Zhong, De-liang",
    title = "{Universal constraints for conformal line defects}",
    eprint = "2501.06900",
    archivePrefix = "arXiv",
    primaryClass = "hep-th",
    doi = "10.1103/gsfg-wrps",
    journal = "Phys. Rev. D",
    volume = "112",
    number = "6",
    pages = "065004",
    year = "2025"
}

@article{Kravchuk:2024qoh,
    author = "Kravchuk, Petr and Radcliffe, Alex and Sinha, Ritam",
    title = "{Effective theory for fusion of conformal defects}",
    eprint = "2406.04561",
    archivePrefix = "arXiv",
    primaryClass = "hep-th",
    doi = "10.1088/1751-8121/ae14c5",
    journal = "J. Phys. A",
    volume = "58",
    number = "46",
    pages = "465402",
    year = "2025"
}

@article{Cavaglia:2020hdb,
    author = "Cavaglia, Andrea and Grabner, David and Gromov, Nikolay and Sever, Amit",
    title = "{Colour-twist operators. Part I. Spectrum and wave functions}",
    eprint = "2001.07259",
    archivePrefix = "arXiv",
    primaryClass = "hep-th",
    reportNumber = "CERN-TH-2020-012",
    doi = "10.1007/JHEP06(2020)092",
    journal = "JHEP",
    volume = "06",
    pages = "092",
    year = "2020"
}

@article{Rattazzi:2008pe,
    author = "Rattazzi, Riccardo and Rychkov, Vyacheslav S. and Tonni, Erik and Vichi, Alessandro",
    title = "{Bounding scalar operator dimensions in 4D CFT}",
    eprint = "0807.0004",
    archivePrefix = "arXiv",
    primaryClass = "hep-th",
    doi = "10.1088/1126-6708/2008/12/031",
    journal = "JHEP",
    volume = "12",
    pages = "031",
    year = "2008"
}

@Book{Collins:1984xc,
  author    = {Collins, John C.},
  publisher = {Cambridge University Press},
  title     = {{Renormalization: An Introduction to Renormalization, the Renormalization Group and the Operator-Product Expansion}},
  year      = {1984},
  address   = {Cambridge},
  isbn      = {978-0-521-31177-9, 978-0-511-86739-2, 978-1-009-40180-7, 978-1-009-40176-0, 978-1-009-40179-1},
  series    = {Cambridge Monographs on Mathematical Physics},
  volume    = {26},
  doi       = {10.1017/9781009401807},
}

@article{Brown:1979pq,
    author = "Brown, Lowell S.",
    title = "{Dimensional regularization of composite operators in scalar field theory}",
    reportNumber = "Print-79-0945 (IAS,PRINCETON)",
    doi = "10.1016/0003-4916(80)90377-2",
    journal = "Annals Phys.",
    volume = "126",
    pages = "135",
    year = "1980"
}

@Article{Brezin:1974zr,
  author  = {Brezin, E. and De Dominicis, C. and Zinn-Justin, J.},
  journal = {Lett. Nuovo Cim.},
  title   = {{Anomalous dimensions of higher-order operators in the $\varphi^4$-theory}},
  year    = {1974},
  pages   = {483--486},
  volume  = {9S2},
  doi     = {10.1007/BF02819916},
}

@article{Cuomo:2024psk, 
    author = "Cuomo, Gabriel and He, Yin-Chen and Komargodski, Zohar",
    title = "{Impurities with a cusp: general theory and 3d Ising}",
    eprint = "2406.10186",
    archivePrefix = "arXiv",
    primaryClass = "hep-th",
    doi = "10.1007/JHEP11(2024)061",
    journal = "JHEP",
    volume = "11",
    pages = "061",
    year = "2024"
}

@inproceedings{Simmons-Duffin:2016gjk,
    author = "Simmons-Duffin, David",
    title = "{The Conformal Bootstrap}",
    booktitle = "{Theoretical Advanced Study Institute in Elementary Particle Physics}: {New Frontiers in Fields and Strings}",
    eprint = "1602.07982",
    archivePrefix = "arXiv",
    primaryClass = "hep-th",
    doi = "10.1142/9789813149441_0001",
    pages = "1--74",
    year = "2017"
}

@article{ladder,
    author = "Cavagli{\`a}, Andrea and Gromov, Nikolay and Levkovich-Maslyuk, Fedor",
    title = "{Quantum spectral curve and structure constants in $ \mathcal{N}=4 $ SYM: cusps in the ladder limit}",
    eprint = "1802.04237",
    archivePrefix = "arXiv",
    primaryClass = "hep-th",
    doi = "10.1007/JHEP10(2018)060",
    journal = "JHEP",
    volume = "10",
    pages = "060",
    year = "2018"
}

@article{Correa:2012hh,
    author = "Correa, Diego and Maldacena, Juan and Sever, Amit",
    title = "{The Quark Anti-Quark Potential and the Cusp Anomalous Dimension from a TBA Equation}",
    eprint = "1203.1913",
    archivePrefix = "arXiv",
    primaryClass = "hep-th",
    doi = "10.1007/JHEP08(2012)134",
    journal = "JHEP",
    volume = "08",
    pages = "134",
    year = "2012"
}

@article{Diatlyk:2024qpr,
    author = "Diatlyk, Oleksandr and Khanchandani, Himanshu and Popov, Fedor K. and Wang, Yifan",
    title = "{Effective Field Theory of Conformal Boundaries}",
    eprint = "2406.01550",
    archivePrefix = "arXiv",
    primaryClass = "hep-th",
    doi = "10.1103/PhysRevLett.133.261601",
    journal = "Phys. Rev. Lett.",
    volume = "133",
    number = "26",
    pages = "261601",
    year = "2024"
}

@article{Diatlyk:2024zkk,
    author = "Diatlyk, Oleksandr and Khanchandani, Himanshu and Popov, Fedor K. and Wang, Yifan",
    title = "{Defect fusion and Casimir energy in higher dimensions}",
    eprint = "2404.05815",
    archivePrefix = "arXiv",
    primaryClass = "hep-th",
    doi = "10.1007/JHEP09(2024)006",
    journal = "JHEP",
    volume = "09",
    pages = "006",
    year = "2024"
}

@article{Qiao:2017xif,
    author = "Qiao, Jiaxin and Rychkov, Slava",
    title = "{A tauberian theorem for the conformal bootstrap}",
    eprint = "1709.00008",
    archivePrefix = "arXiv",
    primaryClass = "hep-th",
    reportNumber = "CERN-TH-2017-176",
    doi = "10.1007/JHEP12(2017)119",
    journal = "JHEP",
    volume = "12",
    pages = "119",
    year = "2017"
}

@article{LanzettaMoultWang2,
  author  = "Lanzetta, Ryan A. and Moult, Ian and Wang, Yifan",
  title   = "{Eye opening bounds on Cusps}",
  year    = {2026},
  note    = {To appear}
}

@article{LanzettaMoultWang1,
  author  = "Lanzetta, Ryan A. and Moult, Ian and Wang, Yifan",
  title   = "{Cutting corners: exciting and magical bounds from the cusp bootstrap}",
  year    = {2026},
  note    = {To appear}
}

@article{Falcioni:2019nxk,
    author = "Falcioni, Giulio and Gardi, Einan and Milloy, Calum",
    title = "{Relating amplitude and PDF factorisation through Wilson-line geometries}",
    eprint = "1909.00697",
    archivePrefix = "arXiv",
    primaryClass = "hep-ph",
    doi = "10.1007/JHEP11(2019)100",
    journal = "JHEP",
    volume = "11",
    pages = "100",
    year = "2019"
}

@article{Dorn:2020meb,
    author = "Dorn, Harald",
    title = "{On anomalous conformal Ward identities for Wilson loops on polygon-like contours with circular edges}",
    eprint = "2001.03391",
    archivePrefix = "arXiv",
    primaryClass = "hep-th",
    reportNumber = "HU-EP-20/01",
    doi = "10.1007/JHEP03(2020)166",
    journal = "JHEP",
    volume = "03",
    pages = "166",
    year = "2020"
}

@article{Soderberg:2021kne,
    author = {S{\"o}derberg, Alexander},
    title = "{Fusion of conformal defects in four dimensions}",
    eprint = "2102.00718",
    archivePrefix = "arXiv",
    primaryClass = "hep-th",
    reportNumber = "UUITP-07/21",
    doi = "10.1007/JHEP04(2021)087",
    journal = "JHEP",
    volume = "04",
    pages = "087",
    year = "2021"
}

@inproceedings{Cardy:2008jc,
    author = "Cardy, John",
    title = "{Conformal Field Theory and Statistical Mechanics}",
    booktitle = "{Les Houches Summer School: Session 89: Exacts Methods in Low-Dimensional Statistical Physics and Quantum Computing}",
    eprint = "0807.3472",
    archivePrefix = "arXiv",
    primaryClass = "cond-mat.stat-mech",
    month = "7",
    year = "2008"
}

@inproceedings{Ginsparg:1988ui,
    author = "Ginsparg, Paul H.",
    title = "{APPLIED CONFORMAL FIELD THEORY}",
    booktitle = "{Les Houches Summer School in Theoretical Physics: Fields, Strings, Critical Phenomena}",
    eprint = "hep-th/9108028",
    archivePrefix = "arXiv",
    reportNumber = "HUTP-88-A054",
    month = "9",
    year = "1988"
}

@article{Polyakov:1980ca,
    author = "Polyakov, Alexander M.",
    title = "{Gauge Fields as Rings of Glue}",
    doi = "10.1016/0550-3213(80)90507-6",
    journal = "Nucl. Phys. B",
    volume = "164",
    pages = "171--188",
    year = "1980"
}

@article{Brandt:1981kf,
    author = "Brandt, Richard A. and Neri, Filippo and Sato, Masa-aki",
    title = "{Renormalization of Loop Functions for All Loops}",
    reportNumber = "NYU/TR2/81",
    doi = "10.1103/PhysRevD.24.879",
    journal = "Phys. Rev. D",
    volume = "24",
    pages = "879",
    year = "1981"
}

@article{Korchemsky:1987wg,
    author = "Korchemsky, G. P. and Radyushkin, A. V.",
    title = "{Renormalization of the Wilson Loops Beyond the Leading Order}",
    doi = "10.1016/0550-3213(87)90277-X",
    journal = "Nucl. Phys. B",
    volume = "283",
    pages = "342--364",
    year = "1987"
}

@article{Korchemskaya:1994qp,
    author = "Korchemskaya, I. A. and Korchemsky, G. P.",
    title = "{High-energy scattering in QCD and cross singularities of Wilson loops}",
    eprint = "hep-ph/9409446",
    archivePrefix = "arXiv",
    reportNumber = "ITP-SB-94-42",
    doi = "10.1016/0550-3213(94)00553-Q",
    journal = "Nucl. Phys. B",
    volume = "437",
    pages = "127--162",
    year = "1995"
}

@article{Korchemsky:1993hr,
    author = "Korchemsky, Gregory P.",
    title = "{On Near forward high-energy scattering in QCD}",
    eprint = "hep-ph/9311294",
    archivePrefix = "arXiv",
    reportNumber = "ITP-SB-93-73",
    doi = "10.1016/0370-2693(94)90040-X",
    journal = "Phys. Lett. B",
    volume = "325",
    pages = "459--466",
    year = "1994"
}

@article{Correa:2012nk,
    author = "Correa, Diego and Henn, Johannes and Maldacena, Juan and Sever, Amit",
    title = "{The cusp anomalous dimension at three loops and beyond}",
    eprint = "1203.1019",
    archivePrefix = "arXiv",
    primaryClass = "hep-th",
    doi = "10.1007/JHEP05(2012)098",
    journal = "JHEP",
    volume = "05",
    pages = "098",
    year = "2012"
}

@article{Munkler:2018cvu,
    author = {M{\"u}nkler, Hagen},
    title = "{The Cross Anomalous Dimension in Maximally Supersymmetric Yang-Mills Theory}",
    eprint = "1805.06448",
    archivePrefix = "arXiv",
    primaryClass = "hep-th",
    reportNumber = "HU-EP-18/15, HU-EP-18-15",
    doi = "10.1007/JHEP10(2018)162",
    journal = "JHEP",
    volume = "10",
    pages = "162",
    year = "2018"
}

@article{Giombi:2025evu,
    author = "Giombi, Simone and Pendse, Anurag",
    title = "{Line Defects with a Cusp in Fermionic CFTs}",
    eprint = "2511.08547",
    archivePrefix = "arXiv",
    primaryClass = "hep-th",
    month = "11",
    year = "2025"
}

@article{Sun:2024qhv,
    author = "Sun, Xinyu and Jian, Shao-Kai",
    title = "{Holographic dual of defect conformal field theory with corner contributions}",
    eprint = "2407.19003",
    archivePrefix = "arXiv",
    primaryClass = "hep-th",
    doi = "10.1103/kqk3-lc64",
    journal = "Phys. Rev. D",
    volume = "112",
    number = "4",
    pages = "L041902",
    year = "2025"
}

@article{Billo:2016cpy,
    author = "Bill{\`o}, Marco and Gon{\c{c}}alves, Vasco and Lauria, Edoardo and Meineri, Marco",
    title = "{Defects in conformal field theory}",
    eprint = "1601.02883",
    archivePrefix = "arXiv",
    primaryClass = "hep-th",
    doi = "10.1007/JHEP04(2016)091",
    journal = "JHEP",
    volume = "04",
    pages = "091",
    year = "2016"
}

@article{Wilson:1974sk,
    author = "Wilson, Kenneth G.",
    title = "{Confinement of Quarks}",
    journal = "Phys. Rev. D",
    volume = "10",
    pages = "2445--2459",
    year = "1974",
    doi = "10.1103/PhysRevD.10.2445"
}

@article{tHooft:1977nqb,
    author = "'t Hooft, Gerard",
    title = "{On the Phase Transition Towards Permanent Quark Confinement}",
    journal = "Nucl. Phys. B",
    volume = "138",
    pages = "1--25",
    year = "1978",
    doi = "10.1016/0550-3213(78)90153-0"
}

@article{Gaiotto:2014kfa,
    author = "Gaiotto, Davide and Kapustin, Anton and Seiberg, Nathan and Willett, Brian",
    title = "{Generalized Global Symmetries}",
    eprint = "1412.5148",
    archivePrefix = "arXiv",
    primaryClass = "hep-th",
    doi = "10.1007/JHEP02(2015)172",
    journal = "JHEP",
    volume = "02",
    pages = "172",
    year = "2015"
}

@article{Cardy:1984bb,
    author = "Cardy, John L.",
    title = "{Conformal Invariance and Surface Critical Behavior}",
    doi = "10.1016/0550-3213(84)90241-4",
    journal = "Nucl. Phys. B",
    volume = "240",
    pages = "514--532",
    year = "1984"
}

@article{McAvity:1995zd,
    author = "McAvity, D. M. and Osborn, H.",
    title = "{Conformal field theories near a boundary in general dimensions}",
    eprint = "cond-mat/9505127",
    archivePrefix = "arXiv",
    reportNumber = "DAMTP-95-1, UBC-TP-95-002",
    doi = "10.1016/0550-3213(95)00476-9",
    journal = "Nucl. Phys. B",
    volume = "455",
    pages = "522--576",
    year = "1995"
}

@article{Bachas:2007td,
    author = "Bachas, C. and Brunner, I.",
    title = "{Fusion of conformal interfaces}",
    eprint = "0712.0076",
    archivePrefix = "arXiv",
    primaryClass = "hep-th",
    doi = "10.1088/1126-6708/2008/02/085",
    journal = "JHEP",
    volume = "02",
    pages = "085",
    year = "2008"
}

@article{Dorn:2018srz,
    author = "Dorn, Harald",
    title = "{On Wilson loops for two touching circles with opposite orientation}",
    eprint = "1811.00799",
    archivePrefix = "arXiv",
    primaryClass = "hep-th",
    reportNumber = "HU-EP-18/33",
    doi = "10.1088/1751-8121/ab0003",
    journal = "J. Phys. A",
    volume = "52",
    number = "9",
    pages = "095401",
    year = "2019"
}

@article{Dorn:2015bfa,
    author = "Dorn, Harald",
    title = "{Wilson loops at strong coupling for curved contours with cusps}",
    eprint = "1509.00222",
    archivePrefix = "arXiv",
    primaryClass = "hep-th",
    reportNumber = "HU-EP-15-38",
    doi = "10.1088/1751-8113/49/14/145402",
    journal = "J. Phys. A",
    volume = "49",
    number = "14",
    pages = "145402",
    year = "2016"
}

@article{Drukker:2011za,
    author = "Drukker, Nadav and Forini, Valentina",
    title = "{Generalized quark-antiquark potential at weak and strong coupling}",
    eprint = "1105.5144",
    archivePrefix = "arXiv",
    primaryClass = "hep-th",
    reportNumber = "IMPERIAL-TP-2011-ND-02, NSF-KITP-11-073, AEI-2011-027",
    doi = "10.1007/JHEP06(2011)131",
    journal = "JHEP",
    volume = "06",
    pages = "131",
    year = "2011"
}

@article{Bachas:2013ora,
    author = "Bachas, C. and Brunner, I. and Roggenkamp, D.",
    title = "{Fusion of Critical Defect Lines in the 2D Ising Model}",
    eprint = "1303.3616",
    archivePrefix = "arXiv",
    primaryClass = "cond-mat.stat-mech",
    reportNumber = "LMU-ASC-13-13, LPTENS-13-06",
    doi = "10.1088/1742-5468/2013/08/P08008",
    journal = "J. Stat. Mech.",
    volume = "1308",
    pages = "P08008",
    year = "2013"
}

@article{Lewkowycz:2013laa,
    author = "Lewkowycz, Aitor and Maldacena, Juan",
    title = "{Exact results for the entanglement entropy and the energy radiated by a quark}",
    eprint = "1312.5682",
    archivePrefix = "arXiv",
    primaryClass = "hep-th",
    doi = "10.1007/JHEP05(2014)025",
    journal = "JHEP",
    volume = "05",
    pages = "025",
    year = "2014"
}

@article{Konechny:2015qla,
    author = "Konechny, Anatoly",
    title = "{Fusion of conformal interfaces and bulk induced boundary RG flows}",
    eprint = "1509.07787",
    archivePrefix = "arXiv",
    primaryClass = "hep-th",
    doi = "10.1007/JHEP12(2015)114",
    journal = "JHEP",
    volume = "12",
    pages = "114",
    year = "2015"
}

@article{Grozin:2014hna,
    author = "Grozin, Andrey and Henn, Johannes M. and Korchemsky, Gregory P. and Marquard, Peter",
    title = "{Three Loop Cusp Anomalous Dimension in QCD}",
    eprint = "1409.0023",
    archivePrefix = "arXiv",
    primaryClass = "hep-ph",
    reportNumber = "IPHT-T14-111, DESY-14-148, SFB-CPP-14-64, LPN14-104",
    doi = "10.1103/PhysRevLett.114.062006",
    journal = "Phys. Rev. Lett.",
    volume = "114",
    number = "6",
    pages = "062006",
    year = "2015"
}

@article{Bruser:2018jnc,
    author = {Br{\"u}ser, Robin and Caron-Huot, Simon and Henn, Johannes M.},
    title = "{Subleading Regge limit from a soft anomalous dimension}",
    eprint = "1802.02524",
    archivePrefix = "arXiv",
    primaryClass = "hep-th",
    reportNumber = "MITP-18-004",
    doi = "10.1007/JHEP04(2018)047",
    journal = "JHEP",
    volume = "04",
    pages = "047",
    year = "2018"
}

@article{Griguolo:2012iq,
    author = "Griguolo, Luca and Marmiroli, Daniele and Martelloni, Gabriele and Seminara, Domenico",
    title = "{The generalized cusp in ABJ(M) N = 6 Super Chern-Simons theories}",
    eprint = "1208.5766",
    archivePrefix = "arXiv",
    primaryClass = "hep-th",
    doi = "10.1007/JHEP05(2013)113",
    journal = "JHEP",
    volume = "05",
    pages = "113",
    year = "2013"
}

@article{Bonini:2016fnc,
    author = "Bonini, Marisa and Griguolo, Luca and Preti, Michelangelo and Seminara, Domenico",
    title = "{Surprises from the resummation of ladders in the ABJ(M) cusp anomalous dimension}",
    eprint = "1603.00541",
    archivePrefix = "arXiv",
    primaryClass = "hep-th",
    doi = "10.1007/JHEP05(2016)180",
    journal = "JHEP",
    volume = "05",
    pages = "180",
    year = "2016"
}

@article{Bykov:2012sc,
    author = "Bykov, D. and Zarembo, K.",
    title = "{Ladders for Wilson Loops Beyond Leading Order}",
    eprint = "1206.7117",
    archivePrefix = "arXiv",
    primaryClass = "hep-th",
    reportNumber = "NORDITA-2012-49, UUITP-18-12",
    doi = "10.1007/JHEP09(2012)057",
    journal = "JHEP",
    volume = "09",
    pages = "057",
    year = "2012"
}

@article{Bianchi:2018scb,
    author = "Bianchi, Lorenzo and Preti, Michelangelo and Vescovi, Edoardo",
    title = "{Exact Bremsstrahlung functions in ABJM theory}",
    eprint = "1802.07726",
    archivePrefix = "arXiv",
    primaryClass = "hep-th",
    doi = "10.1007/JHEP07(2018)060",
    journal = "JHEP",
    volume = "07",
    pages = "060",
    year = "2018"
}

@article{McGovern:2019sdd,
    author = "McGovern, Joseph",
    title = "{Scalar insertions in cusped Wilson loops in the ladders limit of planar $ \mathcal{N} $ = 4 SYM}",
    eprint = "1912.00499",
    archivePrefix = "arXiv",
    primaryClass = "hep-th",
    doi = "10.1007/JHEP05(2020)062",
    journal = "JHEP",
    volume = "05",
    pages = "062",
    year = "2020"
}

@article{Fiol:2015spa,
    author = "Fiol, Bartomeu and Gerchkovitz, Efrat and Komargodski, Zohar",
    title = "{Exact Bremsstrahlung Function in $N=2$ Superconformal Field Theories}",
    eprint = "1510.01332",
    archivePrefix = "arXiv",
    primaryClass = "hep-th",
    reportNumber = "WIS-08-15-NOV-DPPA",
    doi = "10.1103/PhysRevLett.116.081601",
    journal = "Phys. Rev. Lett.",
    volume = "116",
    number = "8",
    pages = "081601",
    year = "2016"
}

@article{Cooke:2017qgm,
    author = "Cooke, Michael and Dekel, Amit and Drukker, Nadav",
    title = "{The Wilson loop CFT: Insertion dimensions and structure constants from wavy lines}",
    eprint = "1703.03812",
    archivePrefix = "arXiv",
    primaryClass = "hep-th",
    reportNumber = "NORDITA-2017-22",
    doi = "10.1088/1751-8121/aa7db4",
    journal = "J. Phys. A",
    volume = "50",
    number = "33",
    pages = "335401",
    year = "2017"
}

@article{Drukker:2012de,
    author = "Drukker, Nadav",
    title = "{Integrable Wilson Loops}",
    eprint = "1203.1617",
    archivePrefix = "arXiv",
    primaryClass = "hep-th",
    doi = "10.1007/JHEP10(2013)135",
    journal = "JHEP",
    volume = "10",
    pages = "135",
    year = "2013"
}

@article{Chernikov:2026lcv,
    author = "Chernikov, Filipp and Gromov, Nikolay and Sever, Amit",
    title = "{Quark Anti-Quark Fusion and Walking RG Flows}",
    eprint = "2607.01328",
    archivePrefix = "arXiv",
    primaryClass = "hep-th",
    month = "7",
    year = "2026"
}

@article{Cavaglia:2022yvv,
    author = "Cavagli{\`a}, Andrea and Gromov, Nikolay and Julius, Julius and Preti, Michelangelo",
    title = "{Integrated correlators from integrability: Maldacena-Wilson line in $ \mathcal{N} $ = 4 SYM}",
    eprint = "2211.03203",
    archivePrefix = "arXiv",
    primaryClass = "hep-th",
    doi = "10.1007/JHEP04(2023)026",
    journal = "JHEP",
    volume = "04",
    pages = "026",
    year = "2023"
}

@article{Grabner:2020nis,
    author = "Grabner, David and Gromov, Nikolay and Julius, Julius",
    title = "{Excited States of One-Dimensional Defect CFTs from the Quantum Spectral Curve}",
    eprint = "2001.11039",
    archivePrefix = "arXiv",
    primaryClass = "hep-th",
    doi = "10.1007/JHEP07(2020)042",
    journal = "JHEP",
    volume = "07",
    pages = "042",
    year = "2020"
}

@article{Gromov:2016rrp,
    author = "Gromov, Nikolay and Levkovich-Maslyuk, Fedor",
    title = "{Quark-anti-quark potential in $ \mathcal{N} =$ 4 SYM}",
    eprint = "1601.05679",
    archivePrefix = "arXiv",
    primaryClass = "hep-th",
    reportNumber = "NORDITA-2016-134",
    doi = "10.1007/JHEP12(2016)122",
    journal = "JHEP",
    volume = "12",
    pages = "122",
    year = "2016"
}

@article{Gromov:2015dfa,
    author = "Gromov, Nikolay and Levkovich-Maslyuk, Fedor",
    title = "{Quantum Spectral Curve for a Cusped Wilson Line in $\mathcal{N}=4$ SYM}",
    eprint = "1510.02098",
    archivePrefix = "arXiv",
    primaryClass = "hep-th",
    doi = "10.1007/JHEP04(2016)134",
    journal = "JHEP",
    volume = "04",
    pages = "134",
    year = "2016"
}

@article{McAvity:1993ue,
  author = {McAvity, D. M. and Osborn, H.},
  title = {Energy Momentum Tensor in Conformal Field Theories Near a Boundary},
  eprint = {hep-th/9302068},
  journal = {Nucl. Phys. B},
  volume = {406},
  pages = {655--680},
  year = {1993},
  doi = {10.1016/0550-3213(93)90005-A}
}

@article{Witten:2025ayw,
    author = "Witten, Edward",
    title = "{Bras and kets in Euclidean path integrals}",
    eprint = "2503.12771",
    archivePrefix = "arXiv",
    primaryClass = "hep-th",
    doi = "10.4310/bpam.260113013520",
    journal = "Beijing J. Pure Appl. Math.",
    volume = "3",
    number = "1",
    pages = "1--34",
    year = "2026"
}

@article{Affleck:1995ge,
    author = "Affleck, Ian",
    title = "{Conformal Field Theory Approach to the Kondo Effect}",
    eprint = "cond-mat/9512099",
    archivePrefix = "arXiv",
    doi = "10.48550/arXiv.cond-mat/9512099",
    year = "1995"
}

@article{Correa:2012at,
    author = "Correa, Diego and Henn, Johannes and Maldacena, Juan and Sever, Amit",
    title = "{An exact formula for the radiation of a moving quark in N=4 super Yang Mills}",
    eprint = "1202.4455",
    archivePrefix = "arXiv",
    primaryClass = "hep-th",
    doi = "10.1007/JHEP06(2012)048",
    journal = "JHEP",
    volume = "06",
    pages = "048",
    year = "2012"
}

@book{needham1997visual,
title     = {Visual Complex Analysis},
author    = {Needham, Tristan},
year      = {1997},
publisher = {Clarendon Press / Oxford University Press},
address   = {Oxford, UK},
isbn      = {978-0198534464}
}

@article{Dorn:2023qbj,
    author = "Dorn, Harald",
    title = "{Remarks on conformal invariants for piecewise smooth curves and Wilson loops}",
    eprint = "2301.01513",
    archivePrefix = "arXiv",
    primaryClass = "hep-th",
    reportNumber = "HU-EP-23/01",
    month = "1",
    year = "2023"
}

@article{Dorn:2020vzj,
    author = "Dorn, Harald",
    title = "{Wilson loops for triangular contours with circular edges}",
    eprint = "2010.14822",
    archivePrefix = "arXiv",
    primaryClass = "hep-th",
    reportNumber = "HU-EP-20/29",
    doi = "10.1088/1751-8121/abe311",
    journal = "J. Phys. A",
    volume = "54",
    number = "22",
    pages = "225402",
    year = "2021"
}

@article{Lanzetta:2025xfw,
    author = "Lanzetta, Ryan A. and Liu, Shang and Metlitski, Max A.",
    title = "{The beginning of the endpoint bootstrap for conformal line defects}",
    eprint = "2508.14964",
    archivePrefix = "arXiv",
    primaryClass = "cond-mat.str-el",
    month = "8",
    year = "2025"
}

@article{Meineri:2023mps,
    author = "Meineri, Marco and Penedones, Joao and Spirig, Taro",
    title = "{Renormalization group flows in AdS and the bootstrap program}",
    eprint = "2305.11209",
    archivePrefix = "arXiv",
    primaryClass = "hep-th",
    doi = "10.1007/JHEP07(2024)229",
    journal = "JHEP",
    volume = "07",
    pages = "229",
    year = "2024"
}

@article{Osterwalder:1973,
	author = {Osterwalder, Konrad and Schrader, Robert},
	date = {1973/06/01},
	doi = {10.1007/BF01645738},
	id = {Osterwalder1973},
	isbn = {1432-0916},
	journal = {Communications in Mathematical Physics},
	number = {2},
	pages = {83--112},
	title = {Axioms for Euclidean Green's functions},
	url = {https://doi.org/10.1007/BF01645738},
	volume = {31},
	year = {1973}}

@article{Kravchuk:2021kwe,
    author = "Kravchuk, Petr and Qiao, Jiaxin and Rychkov, Slava",
    title = "{Distributions in CFT. Part II. Minkowski space}",
    eprint = "2104.02090",
    archivePrefix = "arXiv",
    primaryClass = "hep-th",
    doi = "10.1007/JHEP08(2021)094",
    journal = "JHEP",
    volume = "08",
    pages = "094",
    year = "2021"
}

@article{Pappadopulo:2012jk,
    author = "Pappadopulo, Duccio and Rychkov, Slava and Espin, Johnny and Rattazzi, Riccardo",
    title = "{OPE Convergence in Conformal Field Theory}",
    eprint = "1208.6449",
    archivePrefix = "arXiv",
    primaryClass = "hep-th",
    reportNumber = "LPTENS-12-31",
    doi = "10.1103/PhysRevD.86.105043",
    journal = "Phys. Rev. D",
    volume = "86",
    pages = "105043",
    year = "2012"
}

@article{Diehl:1996kd,
    author = "Diehl, H. W.",
    title = "{The Theory of boundary critical phenomena}",
    eprint = "cond-mat/9610143",
    archivePrefix = "arXiv",
    doi = "10.1142/S0217979297001751",
    journal = "Int. J. Mod. Phys. B",
    volume = "11",
    pages = "3503--3523",
    year = "1997"
}

@article{kondo,
    author = {Kondo, Jun},
    title = {Resistance Minimum in Dilute Magnetic Alloys},
    journal = {Progress of Theoretical Physics},
    volume = {32},
    number = {1},
    pages = {37-49},
    year = {1964},
    month = {07},
    issn = {0033-068X},
    doi = {10.1143/PTP.32.37},
    url = {https://doi.org/10.1143/PTP.32.37},
    eprint = {https://academic.oup.com/ptp/article-pdf/32/1/37/5193092/32-1-37.pdf},
}

@article{WilsonKondo,
  title = {The renormalization group: Critical phenomena and the Kondo problem},
  author = {Wilson, Kenneth G.},
  journal = {Rev. Mod. Phys.},
  volume = {47},
  issue = {4},
  pages = {773--840},
  numpages = {0},
  year = {1975},
  month = {Oct},
  publisher = {American Physical Society},
  doi = {10.1103/RevModPhys.47.773},
  url = {https://link.aps.org/doi/10.1103/RevModPhys.47.773}
}

@article{Bauer:2001yt,
    author = "Bauer, Christian W. and Pirjol, Dan and Stewart, Iain W.",
    title = "{Soft collinear factorization in effective field theory}",
    eprint = "hep-ph/0109045",
    archivePrefix = "arXiv",
    reportNumber = "UCSD-PTH-01-15",
    doi = "10.1103/PhysRevD.65.054022",
    journal = "Phys. Rev. D",
    volume = "65",
    pages = "054022",
    year = "2002"
}

@article{Bianchi:2015liz,
    author = "Bianchi, Lorenzo and Meineri, Marco and Myers, Robert C. and Smolkin, Michael",
    title = "{R{\'e}nyi entropy and conformal defects}",
    eprint = "1511.06713",
    archivePrefix = "arXiv",
    primaryClass = "hep-th",
    reportNumber = "DESY-15-229",
    doi = "10.1007/JHEP07(2016)076",
    journal = "JHEP",
    volume = "07",
    pages = "076",
    year = "2016"
}

@article{Calabrese:2004eu,
    author = "Calabrese, Pasquale and Cardy, John L.",
    title = "{Entanglement entropy and quantum field theory}",
    eprint = "hep-th/0405152",
    archivePrefix = "arXiv",
    doi = "10.1088/1742-5468/2004/06/P06002",
    journal = "J. Stat. Mech.",
    volume = "0406",
    pages = "P06002",
    year = "2004"
}

@article{Solodukhin:2008dh,
    author = "Solodukhin, Sergey N.",
    title = "{Entanglement entropy, conformal invariance and extrinsic geometry}",
    eprint = "0802.3117",
    archivePrefix = "arXiv",
    primaryClass = "hep-th",
    doi = "10.1016/j.physletb.2008.05.071",
    journal = "Phys. Lett. B",
    volume = "665",
    pages = "305--309",
    year = "2008"
}

@article{Allais:2014ata,
    author = "Allais, Andrea and Mezei, M{\'a}rk",
    title = "{Some results on the shape dependence of entanglement and R{\'e}nyi entropies}",
    eprint = "1407.7249",
    archivePrefix = "arXiv",
    primaryClass = "hep-th",
    reportNumber = "MIT-CTP-4569",
    doi = "10.1103/PhysRevD.91.046002",
    journal = "Phys. Rev. D",
    volume = "91",
    number = "4",
    pages = "046002",
    year = "2015"
}

@article{Graham:1999pm,
    author = "Graham, C. Robin and Witten, Edward",
    title = "{Conformal anomaly of submanifold observables in AdS / CFT correspondence}",
    eprint = "hep-th/9901021",
    archivePrefix = "arXiv",
    doi = "10.1016/S0550-3213(99)00055-3",
    journal = "Nucl. Phys. B",
    volume = "546",
    pages = "52--64",
    year = "1999"
}

@article{Bueno:2015rda,
    author = "Bueno, Pablo and Myers, Robert C. and Witczak-Krempa, William",
    title = "{Universality of corner entanglement in conformal field theories}",
    eprint = "1505.04804",
    archivePrefix = "arXiv",
    primaryClass = "hep-th",
    doi = "10.1103/PhysRevLett.115.021602",
    journal = "Phys. Rev. Lett.",
    volume = "115",
    pages = "021602",
    year = "2015"
}

@article{Bianchi:2018zpb,
    author = "Bianchi, Lorenzo and Lemos, Madalena and Meineri, Marco",
    title = "{Line Defects and Radiation in $\mathcal{N}=2$ Conformal Theories}",
    eprint = "1805.04111",
    archivePrefix = "arXiv",
    primaryClass = "hep-th",
    reportNumber = "DESY-18-071",
    doi = "10.1103/PhysRevLett.121.141601",
    journal = "Phys. Rev. Lett.",
    volume = "121",
    number = "14",
    pages = "141601",
    year = "2018"
}

@article{Drukker:2025dfm,
    author = "Drukker, Nadav and Kong, Ziwen and Kravchuk, Petr",
    title = "{Nonlinearly Realised Defect Symmetries and Anomalies}",
    eprint = "2512.15913",
    archivePrefix = "arXiv",
    primaryClass = "hep-th",
    month = "12",
    year = "2025"
}

@article{Cuomo:2026mop,
    author = "Cuomo, Gabriel and Giombi, Simone and Tizzano, Luigi",
    title = "{Impurities Near the Light Cone}",
    eprint = "2608.03704",
    archivePrefix = "arXiv",
    primaryClass = "hep-th",
    reportNumber = "CERN-TH-2026-186",
    month = "8",
    year = "2026"
}

@article{Kohlbecker1958,
  author  = {Kohlbecker, E. E.},
  title   = {Weak Asymptotic Properties of Partitions},
  journal = {Transactions of the American Mathematical Society},
  volume  = {88},
  number  = {2},
  pages   = {346--365},
  year    = {1958}
}

@article{Sudakov:1956,
  author  = {Sudakov, V. V.},
  title   = {Vertex Parts at Very High Energies in Quantum Electrodynamics},
  journal = {Sov. Phys. JETP},
  volume  = {3},
  pages   = {65--71},
  year    = {1956},
  note    = {Original Russian version: Zh. Eksp. Teor. Fiz. 30 (1956) 87--95},
  url     = {https://www.jetp.ras.ru/cgi-bin/dn/e_003_01_0065.pdf}
}

@article{Alday:2025pmg,
    author = {Alday, Luis F. and Armanini, Elisabetta and H{\"a}ring, Kelian and Zhiboedov, Alexander},
    title = "{From Partons to Strings: Scattering on the Coulomb Branch of $\mathcal{N}=4$ SYM}",
    eprint = "2510.19909",
    archivePrefix = "arXiv",
    primaryClass = "hep-th",
    reportNumber = "CERN-TH-2025-202",
    month = "10",
    year = "2025"
}

@article{Lauria:2017wav,
    author = "Lauria, Edoardo and Meineri, Marco and Trevisani, Emilio",
    title = "{Radial coordinates for defect CFTs}",
    eprint = "1712.07668",
    archivePrefix = "arXiv",
    primaryClass = "hep-th",
    doi = "10.1007/JHEP11(2018)148",
    journal = "JHEP",
    volume = "11",
    pages = "148",
    year = "2018"
}
\bibliographystyle{JHEP}
\end{document}